\documentclass[aps,prd,twocolumn,amsmath,superscriptaddress,preprintnumbers,amssymb,showpacs,floatfix,nofootinbib,longbibliography]{revtex4-1}

\usepackage{graphicx,dcolumn,hyperref,xcolor,bm,natbib}
\usepackage{calligra}
\usepackage{egothic}
\usepackage[T1]{fontenc}
\newfont{\rsfsten}{rsfs10 scaled 1200}
\newfont{\rsfsseven}{rsfs10 scaled 1200}
\newfont{\rsfsfive}{rsfs10 scaled 1200}
\usepackage{epsfig}
\usepackage{units}
\usepackage[utf8]{inputenc}

\newcommand{\aap}{{Astron.~Astrophys.}}

\newcommand{\apjs}{{Astrophys.~J.~Supp.}}

\newcommand{\mnras}{{Mon.~Not.~R.~Astron.~Soc.}}

\newcommand{\pasa}{{Publ.~Astron.~Soc.~Australia}}
\newcommand{\ssr}{{Space~Science~Rev.}}
\newcommand{\apss}{{Astrophys.~and~Space~Science}}
\newcommand{\jcap}{{Journ. of Cosmology and Astroparticle Physics}}

\newcommand{\be}{\begin{equation}}
\newcommand{\ee}{\end{equation}}
\newcommand{\bea}{\begin{eqnarray}}
\newcommand{\eea}{\end{eqnarray}}

\DeclareMathOperator{\GeV}{GeV}
\newcommand{\cm}{{\rm\; cm}}
\newcommand{\s}{{\rm\;s}}

\def\lsim{\mathrel{\raise.3ex\hbox{$<$\kern-.75em\lower1ex\hbox{$\sim$}}}}
\def\gsim{\mathrel{\raise.3ex\hbox{$>$\kern-.75em\lower1ex\hbox{$\sim$}}}}

\newcommand{\sigmav}{\langle\sigma_Av\rangle}

\usepackage{fontawesome}
\newcommand{\linkmaster}{\href{https://zenodo.org/record/6423495\#.Yk92zNPMKu4}{\scriptsize	 \faDownload}}

\begin{document}


\title{
The Return of the Templates: Revisiting the Galactic Center Excess with Multi-Messenger Observations
}

\author{Ilias Cholis}
\affiliation{Department of Physics, Oakland University, Rochester, Michigan, 48309, USA}
\author{Yi-Ming Zhong}
\affiliation{Kavli Institute for Cosmological Physics, University of Chicago, Chicago, IL 60637, USA}
\author{Samuel D.~McDermott}
\affiliation{Fermi National Accelerator Laboratory, Batavia, Illinois, 60510, USA}
\author{Joseph P.~Surdutovich}
\affiliation{Carleton College Physics Department, Northfield, Minnesota, 55057, USA}

\date{\today}

\begin{abstract}
The Galactic center excess (GCE) remains one of the most intriguing discoveries from the \textit{Fermi} Large Area Telescope (LAT)  observations. We revisit the characteristics of the GCE by first producing a new set of high-resolution galactic diffuse gamma-ray emission templates. This diffuse emission, which accounts for the bulk of the observed gamma rays, is ultimately due to cosmic-ray interactions with the interstellar medium. Using recent high-precision cosmic-ray observations, in addition to the continuing \textit{Fermi}-LAT observations and observations from lower energy photons, we constrain the properties of the galactic diffuse emission. We describe a large set of diffuse gamma-ray emission templates which account for a very wide range of initial assumptions on the physical conditions in the inner galaxy. The broad properties of the GCE that we find in this work are qualitatively unchanged despite the introduction of this new set of templates, though its quantitative features appear mildly different than those obtained in previous analyses. In particular, we find a high-energy tail at higher significance than previously reported. This tail is very prominent in the northern hemisphere, and less so in the southern hemisphere. This strongly affects one prominent interpretation of the excess: known millisecond pulsars are incapable of producing this high-energy emission, even in the relatively softer southern hemisphere, and are therefore disfavored as the sole explanation of the GCE. The annihilation of dark matter particles of mass $40^{+10}_{-7}$ GeV (95$\%$ CL) to $b$ quarks with a cross-section of $\sigmav = 1.4^{+0.6}_{-0.3} \times 10^{-26}$ cm$^{3}$s$^{-1}$  provides a good fit to the excess especially in the relatively cleaner southern sky. Dark matter of the same mass range annihilating to $b$ quarks or heavier dark matter particles annihilating to heavier Standard Model bosons can combine with millisecond pulsars to provide a good fit to the southern hemisphere emission as well, as can a broken power-law spectrum which would be related to recent cosmic-ray burst activity. As part of this paper, we make publicly available all of our templates and the data covariance matrix we have generated to account for systematic uncertainties.~\linkmaster 
\end{abstract}

\preprint{FERMILAB-PUB-21-709-T}

\maketitle
\tableofcontents

\section{Introduction}
\label{sec:introduction}

One of the most conspicuous unsolved mysteries of \textit{Fermi} Large Area Telescope (\textit{Fermi}-LAT) \cite{Gehrels:1999ri, fermiURL} observations remains the Galactic Center Excess (GCE) emission, which accounts for $\sim \mathcal O(10\%)$ of the gamma-ray emission in the inner few degrees of the Milky Way at GeV energies \cite{Goodenough:2009gk, Hooper:2010mq, Abazajian:2010zy, Hooper:2011ti, Hooper:2013rwa, Gordon:2013vta, Abazajian:2014fta, Daylan:2014rsa, Calore:2014xka, Zhou:2014lva, TheFermi-LAT:2015kwa, Huang:2015rlu, Linden:2016rcf, DiMauro:2021raz}. Since its first detection, several hypotheses have been proposed regarding its nature: it may be a signature of dark matter particles annihilating in the center of the Milky Way \cite{Goodenough:2009gk, Gordon:2013vta, Daylan:2014rsa, Calore:2014nla, Agrawal:2014oha, Berlin:2015wwa, Karwin:2016tsw, TheFermi-LAT:2017vmf, Leane:2019xiy}, it may come from an astrophysically novel population of point sources \cite{Abazajian:2012pn, Abazajian:2014fta, Hooper:2013nhl, Lee:2015fea, Petrovic:2014xra, Cholis:2014lta, OLeary:2015qpx, Bartels:2017vsx, Buschmann:2020adf,Gautam:2021wqn}, it may be related to activity from the supermassive black hole at the very center of our galaxy \cite{Petrovic:2014uda, Carlson:2014cwa, Cholis:2015dea}, or it may be an artifact of erroneous assumptions on the galactic diffuse emission that have led to significant mismodeling of astrophysical processes \cite{Gaggero:2015nsa, Macias:2016nev, Horiuchi:2016zwu}.

Each of these proposed explanations of the GCE makes other predictions. If the GCE is indeed a dark matter annihilation signal, one would expect to observe gamma-ray emission from observations toward known dwarf galaxies \cite{Strigari:2013iaa, Calore:2014nla, Charles:2016pgz, DeAngelis:2017gra} from which we currently have only limits \cite{Cholis:2012am, Geringer-Sameth:2014qqa, Drlica-Wagner:2015xua, Ackermann:2015zua, Fermi-LAT:2016uux} (see however \cite{Geringer-Sameth:2015lua, Hooper:2015ula}), from the Large Magellanic Cloud \cite{Buckley:2015doa}, from other galaxies \cite{Lisanti:2017qlb}, in the extragalactic isotropic emission \cite{Cholis:2013ena, Ackermann:2015tah}, in lower energy gamma-ray observations of the inner galaxy \cite{Bartels:2017dpb}, in observations at microwave wavelengths \cite{Aghanim:2018eyx, Elor:2015bho}, and also in antimatter cosmic-rays produced through the same annihilations \cite{Bringmann:2014lpa, Hooper:2014ysa, Cirelli:2014lwa, Evoli:2015vaa, Chen:2015cqa, Korsmeier:2017xzj}. Recently, \cite{Cuoco:2016eej, Cholis:2019ejx, Cuoco:2019kuu} have found a robust excess in the cosmic-ray antiproton-to-proton ratio measured by the Alpha Magnetic Spectrometer (\textit{AMS-02}) onboard the International Space Station \cite{Aguilar:2016kjl} which is consistent with the interpretation of the GCE as due to dark matter annihilation; given further assumptions on the dark matter particle properties, we also expect observable anti-deuteron and anti-helium cosmic-ray fluxes \cite{Cholis:2020twh}. Including astrophysical uncertainties, these observations are collectively consistent with a dark matter particle in the mass range $m_{\rm DM}$ of $10-200$ GeV and an annihilation cross-section $\langle \sigma v \rangle$ in the range of $(0.2-8) \times 10^{-26}$ cm$^{3}$/s depending on its annihilation products \cite{Calore:2014nla}. For $\text{DM DM} \to b \bar{b}$, for example, fits to dark matter properties have previously been found to favor the dark matter particle mass $m_{\rm DM}$ in the range $38-61$ GeV and $\langle \sigma v \rangle$ in the range $(1.2-2.3) \times 10^{-26}$ cm$^{3}$/s, consistent with being a thermal relic \cite{Steigman:2012nb}. Using the entirely new set of galactic diffuse background templates described in much more detail in this work, we confirm these values.

A population of point sources in the inner galaxy can explain the GCE if these sources have a spatial distribution that fits the nearly spherically symmetric morphology of the observed emission \cite{Daylan:2014rsa, Calore:2014xka}. Furthermore, this population must give the observed spectrum, which is similar to but not in perfect agreement with the observed gamma-ray spectra from millisecond pulsars \cite{Hooper:2013nhl, Cholis:2014lta, Cholis:2014noa}. Moreover, it would require that the GCE is the result of sources whose distribution of luminosity permits them to emit mostly
below the threshold of detection by the \textit{Fermi}-LAT instrument \cite{Lee:2015fea, Zhong:2019ycb}. With continuing observations and the detection of more point sources across the sky, more members of that population should be detected \cite{Bartels:2015aea, Lee:2015fea, Charles:2016pgz}. Several recent point source analyses toward the inner galaxy suggest that regular MSPs are unable to account for the entirety of the observed GCE emission \cite{Zhong:2019ycb, Leane:2020nmi, Leane:2020pfc, List:2020mzd, List:2021aer, Mishra-Sharma:2021oxe}. 
Such a population of point sources could give signals in X-rays \cite{Cholis:2014lta} and future radio observations \cite{Calore:2015bsx, Bhakta:2017mpk}.

If the GCE is a signal of cosmic-ray burst activity associated with the supermassive black hole, it may also be related to the \textit{Fermi} Bubbles observed in gamma-rays \cite{Dobler:2009xz, Su:2010qj, Fermi-LAT:2014sfa, Balaji:2018rwz} and at microwave energies \cite{Dobler:2007wv, Ade:2012nxf}. Recently, structures related to the Bubbles have been detected in X-rays at low latitudes \cite{Predehl:2020kyq} while upper limits on Bubbles-related flux have been reported at high energies \cite{Moulin:2021mug}.

The main challenge in moving forward our understanding of the GCE is to reduce the uncertainty on the dominant galactic diffuse emission, which ultimately arises from cosmic-ray interactions in the Milky Way. These uncertainties are associated with our imperfect modeling of the complex astrophysical processes at the center of our galaxy. By comparison, the statistical errors associated with the number of detected gamma rays are smaller \cite{Calore:2014xka}. Most gamma-ray data analyses so far use available templates \cite{Dobler:2007wv, Ackermann:2012pya, Calore:2014xka, fermiTemplates}, a technique that historically relies heavily on microwave observations \cite{Gold:2010fm, wmapTemplates, Adam:2015wua, Akrami:2018mcd, planckTemplates}. New models for the total galactic gamma-ray emission can now be developed which rely instead on our significantly improved understanding of galactic cosmic rays. The cosmic-ray observations from the \textit{AMS-02} \cite{2002NIMPA.478..119A} and \textit{Voyager 1} \cite{1977SSRv...21..355S}, which since 2012 has entered the interstellar medium (ISM), allow us to more accurately model the ISM conditions and also the impact of solar modulation on cosmic-rays once they enter the Heliosphere. Adding to the  continuing gamma-ray observations from \textit{Fermi}-LAT, in this work we attain true multi-messenger models of the gamma-ray sky connected to a multitude of observations.  

In this work, we use \textit{AMS-02} observations of cosmic-ray hydrogen, helium, carbon, beryllium, boron, and oxygen to update our models of the nearby ISM. We take those ISM propagation models as starting points for gamma-ray maps of the inner galaxy. By relaxing a sequence of assumptions on the ISM conditions at the inner galaxy we then test a variety of astrophysical uncertainties in that region. Most of the cosmic rays we observe with \textit{AMS-02} and \textit{Voyager 1} are produced within only a few kiloparsecs (kpc) of the Sun. Because the GCE emission originates considerably closer to the center of the Milky Way, local cosmic rays and gamma rays do not probe the exact same conditions of cosmic-ray propagation and ISM conditions as those from the inner galaxy. Thus, appropriate freedom, accounting also for the uncertain distribution of cosmic-ray sources, must be included when constructing gamma-ray maps of the inner galaxy. Throughout this work, we focus on building galactic diffuse emission template models and testing those models against gamma-ray observations.

The rest of this paper is organized as follows. In Sec.~\ref{sec:cosmicrays}, we describe our assumptions for the cosmic-ray propagation and how 
those are constrained by multi-messenger observations. In Sec.~\ref{sec:GDGRE} we discuss how we build 
templates describing our galactic diffuse emission models.
In Sec.~\ref{sec:method} we show fits to data and describe the full range of template 
models tested. In Sec.~\ref{sec:results} we discuss our results for the GCE emission, its spectrum,
its morphology, and its consistency in different subregions of interest. In Sec.~\ref{sec:Uncertainties}, we go to greater lengths to characterize the significance of our detection of the GCE and discuss systematic uncertainties related to this detection. Then, in Sec.~\ref{sec:Interpretations} we discuss interpretations of the observed GCE and some models that may give rise to it, especially focusing on implications for 
the case of annihilating dark matter and a population of gamma-ray point sources. In Sec.~\ref{sec:conclusions} we give our conclusions. In several appendices, we give further details on certain aspects of our central results.

\section{Cosmic Rays}
\label{sec:cosmicrays}

All astrophysical diffuse gamma-ray emission comes from cosmic rays interacting with the interstellar medium. Thus, the 
distribution of cosmic-ray sources, the spectral properties of propagating cosmic rays, and the conditions
in the ISM affect the total predicted diffuse gamma-ray emission. In this section, we 
present models that are good fits to recent measurements of local cosmic-ray properties as measured by
\textit{AMS-02}. These models are then used as the starting point in Sec.~\ref{sec:GDGRE} to evaluate 
the expected diffuse gamma-ray emission from the Milky Way. Most of the cosmic rays observed 
with local instruments originate within a few kpcs of Earth, which is located approximately 8.5 kpc from the
Galactic center. Given the fact that the gamma rays we observe can be produced within a larger volume, we allow for wider
freedom in the values of physical parameters defining our cosmic-ray models in the inner part of our galaxy, accounting for relevant uncertainties in conditions there.

To test the injection and propagation properties of cosmic rays in our galaxy, we use the publicly available 
\texttt{GALPROP v56} \cite{NEWGALPROP}. \texttt{GALPROP} solves numerically the cosmic-ray transport 
equation to evaluate the expected cosmic-ray flux and distribution in energy and in space within the galaxy
~\cite{GALPROPSite, Strong:2015zva, galprop, Moskalenko:2001ya}.  Using \texttt{GALPROP} we take 
cosmic rays to propagate within a disk of radius 20 kpc and height of 2$z_{L}$ and solve the propagation 
equation in cylindrical coordinates assuming azimuthal symmetry. The galactic disk is located at the center 
of that volume, thus $z_{L}$ is the maximum height above the disk that cosmic rays propagate; beyond 
this box it is assumed that they escape to the intergalactic medium. For more details on the numerical 
assumptions pertaining to the solution of the propagation equation of galactic cosmic rays see~\cite{GALPROPSite, 
Moskalenko:2001ya}.

For all cosmic-ray species in the GeV energy range, one of the most important modeling assumptions 
is their diffusion timescale inside the galactic disk and 
the energy dependence of this diffusion timescale. We assume isotropic 
and homogeneous diffusion of cosmic rays described by a single diffusion coefficient, $D_{xx}(R)$, that depends only on 
their rigidity $R \equiv p/q$ (where $p$ and $q$ are the particle's momentum and charge respectively), 
\begin{equation}
D_{xx}(R) = \beta D_{0} (R/4 \,{\rm GV})^{\delta}, 
\label{eq:Diff}
\end{equation}
where $\delta$ is known as the diffusion index related to the spectral index of magnetohydrodynamic turbulence 
in the ISM and $\beta \equiv v/c$. The $D_{0}$ coefficient sets the normalization of $D_{xx}$, but can also 
act as the breaking point for the diffusion index $\delta$. Some of our models  allow such a break to exist. We
take $\delta$ in the range of 0.3 to 0.5 that has been shown to be in agreement with past cosmic-ray data as 
in Ref.~\cite{Trotta:2010mx}. At hundreds of gigavolts (GV) in rigidity, the power-law-index $\delta$ of $D_{xx}(R)$ may change again
(see e.g. \cite{Genolini:2017dfb}). For the cosmic-ray 
and gamma-ray energies of interest here, 
a rigidity break occurring at $\simeq 200$ GV is of minor impact.

First order Fermi acceleration predicts that cosmic rays have at injection in the ISM a power-law spectrum 
$dN/dR \propto R^{-\alpha}$, with $\alpha \sim 2.0$. Since we observe the contribution of many sources 
separated from us at different distances and of a variety of types, including pulsars as well as supernova remnants of different ages, 
chemical composition, and environments, we take the following parameterization for 
the averaged injected cosmic-ray spectra,  
\begin{eqnarray}
 dN/dR \propto \begin{cases} R^{-\alpha_{1}}, \,\,\,\,\,\,\,
     \textrm{for} \; R<R_{\rm{br}_{1}}\\  
     R^{-\alpha_{2}}, \,\,\,\,\,\,\, \textrm{for} \; R_{\rm{br}_{1}}<R<R_{\rm{br}_{2}}, \\
     R^{-\alpha_{3}}, \,\,\,\,\,\,\, \textrm{for} \; R>R_{\rm{br}_{2}}. \\
     \end{cases}
\label{eq:Inj}
\end{eqnarray}
We allow for some small differences between species in the values of the injection indices $\alpha_{1}$, $\alpha_{2}$ 
and $\alpha_{3}$ and their location in rigidity space $R_{\rm{br}_{1}}$ and $R_{\rm{br}_{2}}$.
However, we find that apart from cosmic-ray electrons most cosmic-ray species that we care about can explain 
the observed fluxes with similar injection assumptions between them.  Moreover, we note that the density of 
cosmic-ray nuclei beyond helium in the ISM is small enough that the exact assumptions on these more massive 
cosmic-ray nuclei does not affect the derived diffuse gamma-ray emission maps. 

In addition to the above assumptions, of some relevance to the observed cosmic-ray nuclei spectral properties 
are the assumptions on the presence of convective winds in the Milky Way and of diffusive reacceleration (i.e.,
diffusion of cosmic rays in momentum space). We rely on the \texttt{GALPROP} parameterization taking the 
convective winds to start from the disk with zero initial velocity at the plane and increase linearly with
height above the disk, $z$:
\begin{equation}
v_{c} = \frac{dv_{c}}{d|z|} |z|.
\label{eq:Conv}
\end{equation}
The diffusive reacceleration is also parametrized via a rigidity-dependent diffusion coefficient in momentum 
space $D_{pp}(R)$, which also depends on the value of the diffusion coefficient in physical space $D_{xx}(R)$ 
\cite{1994ApJ...431..705S},
\begin{equation}
D_{pp}(R) \propto \frac{R^{2}v_{A}^{2}}{D_{xx}(R)},
\label{eq:DiffReAcc}
\end{equation}
where $v_{A}$ is the Alfv$\acute{\textrm{e}}$n speed.

\begin{figure}
\begin{centering}
\hspace{-0.2in}
\includegraphics[width=3.65in,angle=0]{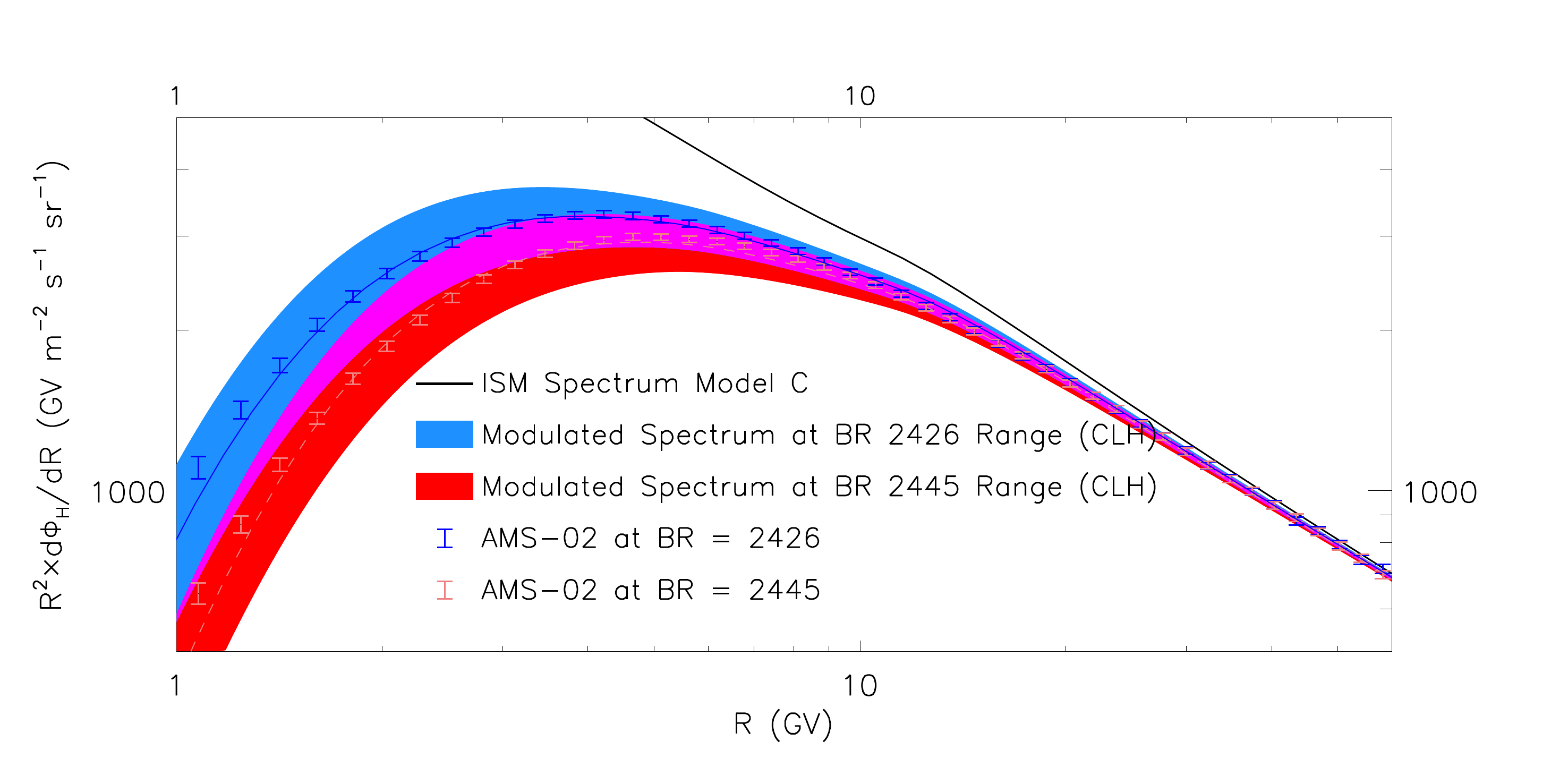} 
\end{centering}
\vspace{0.05in}
\caption{The hydrogen cosmic-ray spectrum at the local ISM (solid black line) 
for model C of Tab.~\ref{tab:ISM_models} and its modulated measurement by the 
\textit{AMS-02} at two different times: at Bartels' Rotation (BR) 2426 (blue data points) and BR 2445
(red data points). 
The blue and red bands show the predicted ranges of the modulated spectra for the equivalent times
of observation \cite{Cholis:2020tpi}. Where those bands overlap 
their color appears magenta. The thin blue and red lines give the expected hydrogen 
spectra assuming just the best fit modulation parameters of \cite{Cholis:2020tpi}.}
\label{fig:CR_hydrogen}
\end{figure}

\begin{figure}
\hspace{-0.18in}
\vspace{-0.33in}
\includegraphics[width=3.65in,angle=0]{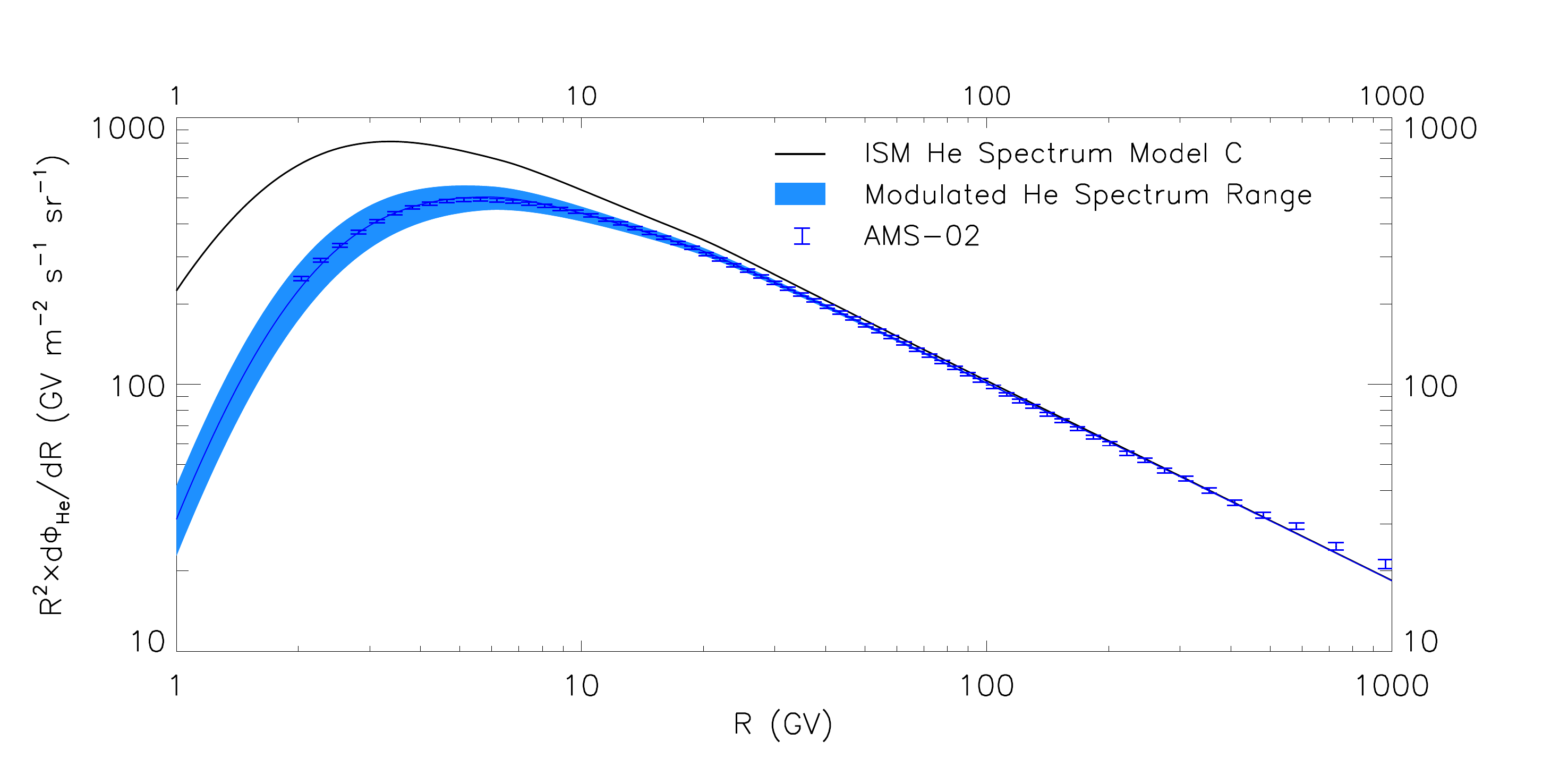} \\
\hspace{-0.18in}
\vspace{-0.33in}
\includegraphics[width=3.65in,angle=0]{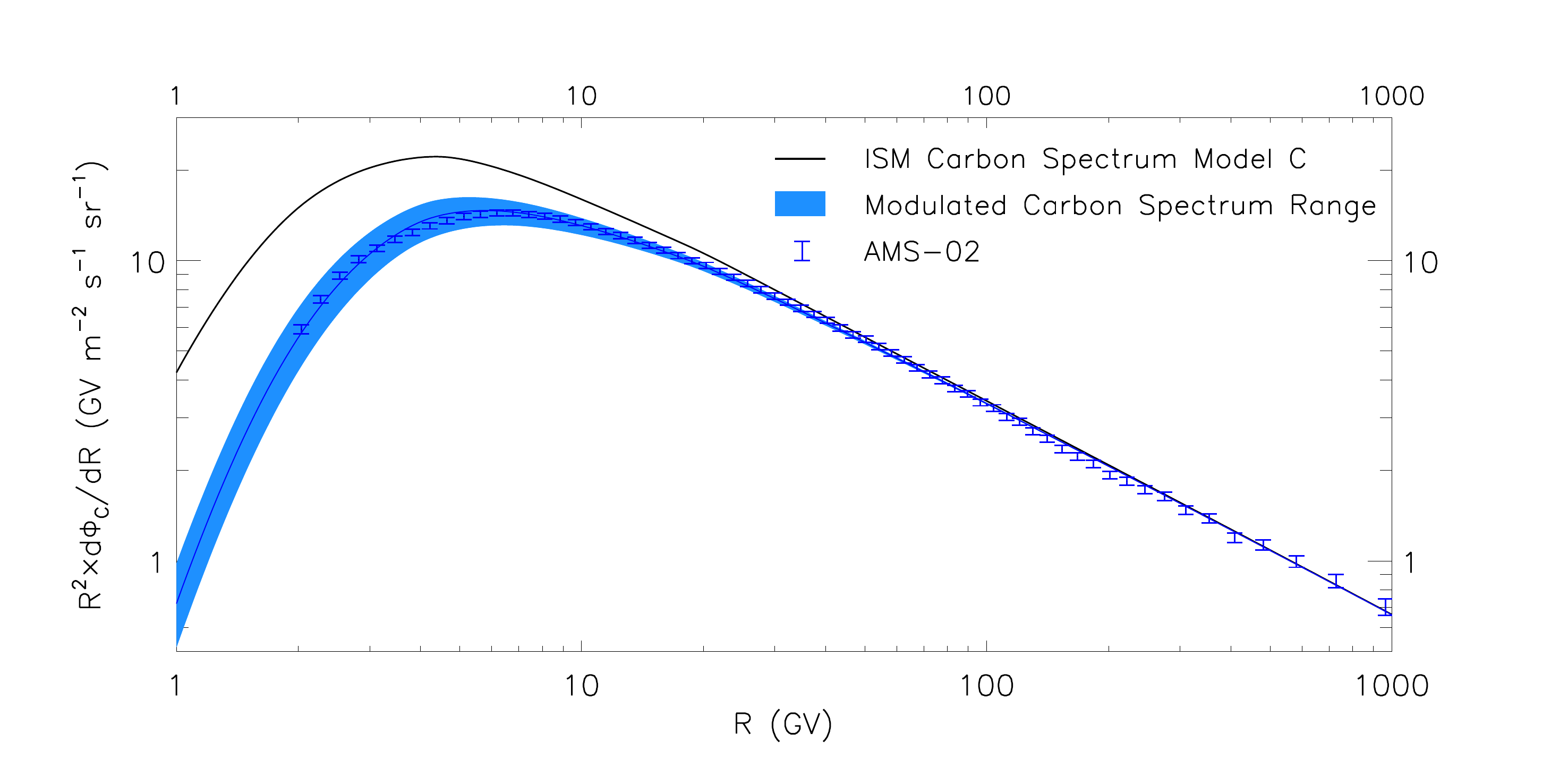} \\
\hspace{-0.18in}
\vspace{-0.33in}
\includegraphics[width=3.65in,angle=0]{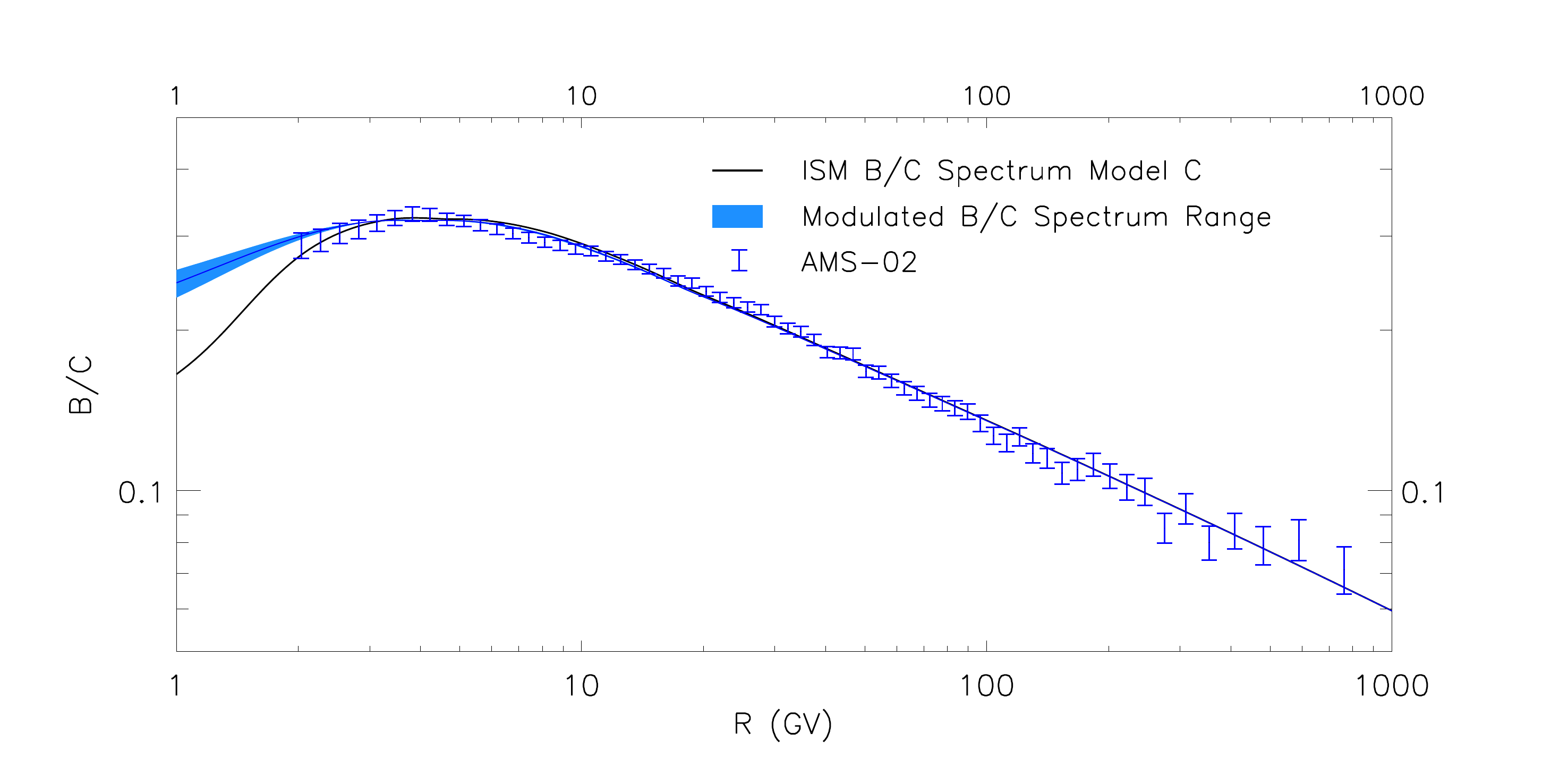} \\
\hspace{-0.18in}
\vspace{-0.33in}
\includegraphics[width=3.65in,angle=0]{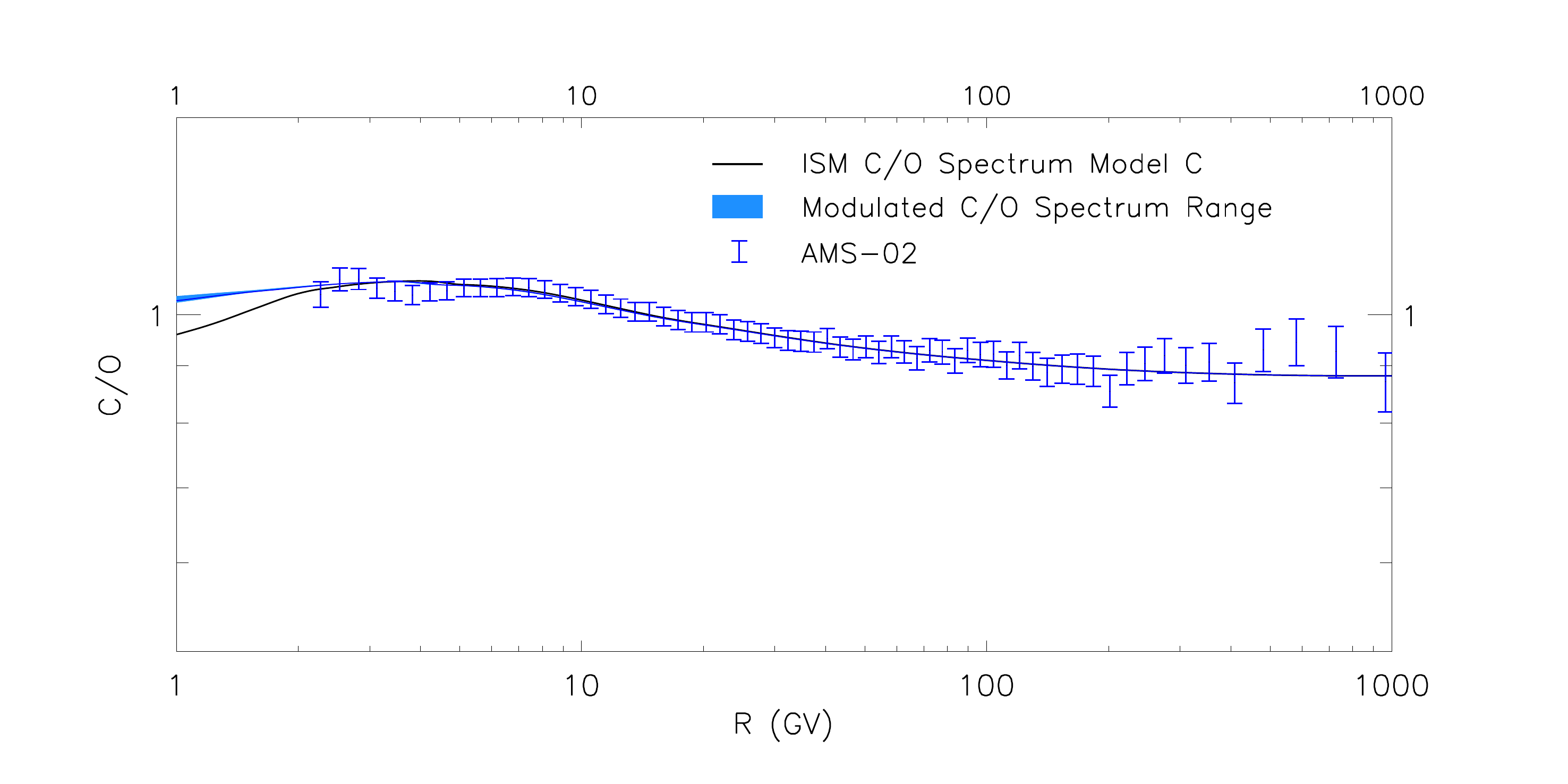} \\
\hspace{-0.18in}
\vspace{-0.33in}
\includegraphics[width=3.65in,angle=0]{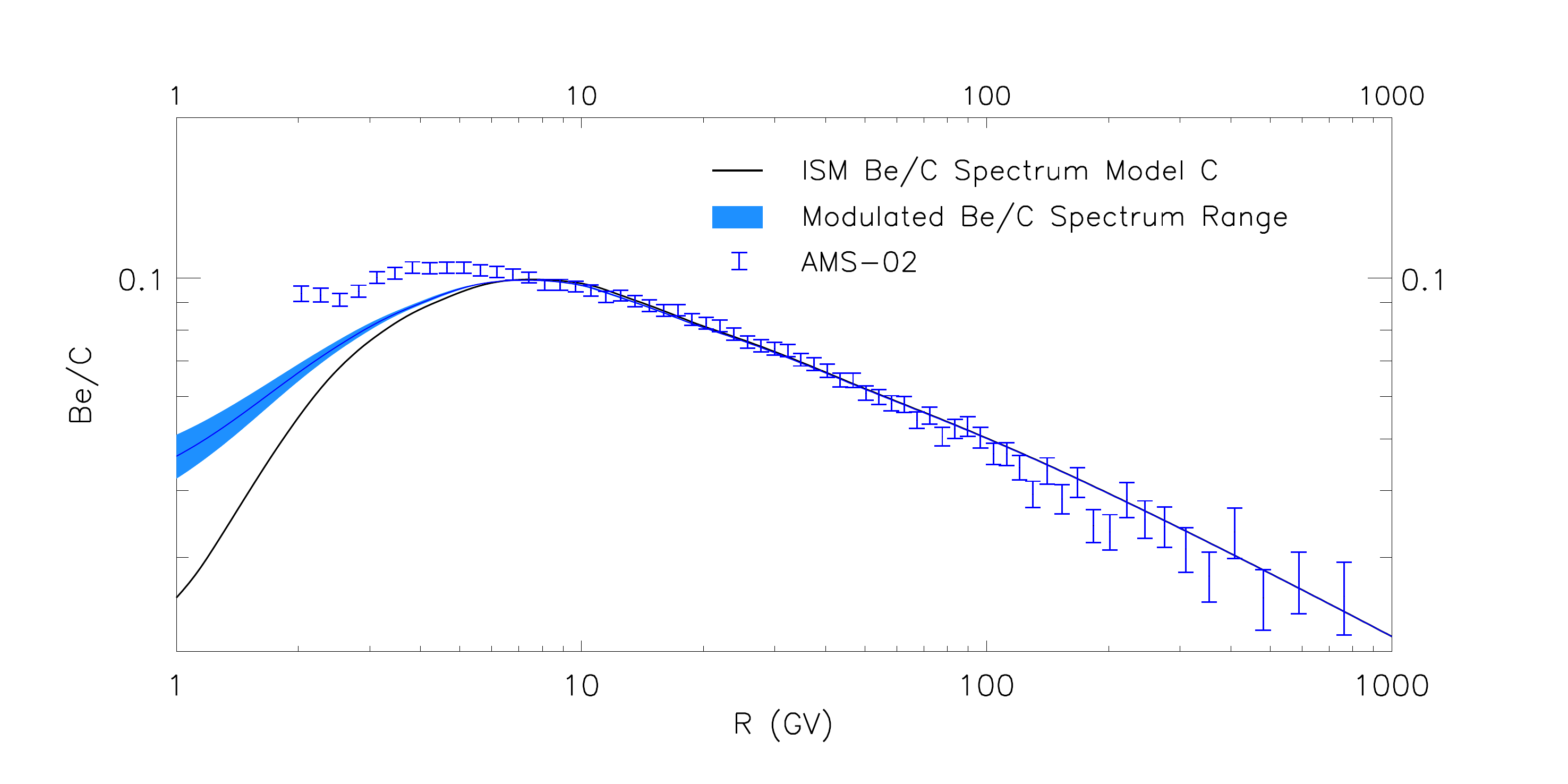}
\vspace{0.00in}
\caption{The measured \textit{AMS-02} cosmic-ray spectra for He (\textit{top}), 
carbon (\textit{second}), boron-to-carbon ratio (\textit{third}), carbon-to-oxygen 
ratio (\textit{forth}) and beryllium-to-carbon ratio  (\textit{bottom}). The black 
solid line gives the prediction from ISM model C, while the blue line and band show the 
expected modulated spectrum for the relevant periods of observation. As with 
Fig.~\ref{fig:CR_hydrogen}, the blue line (band) shows the best-fit to (allowed range for) the modulation 
parameters \cite{Cholis:2020tpi}.}
\label{fig:CR_nuclei}
\end{figure}

To account for systematic uncertainties on the ISM we first test six distinct cosmic-ray models that act as an envelope
on different uncertainties on the assumptions of the local diffusion index $\delta$ of Eq.~\ref{eq:Diff} and 
the height of the diffusion zone $z_{L}$. These models test the local galactic medium assumptions and give different values 
on the injection, convection, and diffusive cosmic-ray reacceleration properties of Eqs.~\ref{eq:Inj}--\ref{eq:DiffReAcc}. 
For gamma-ray templates we take these first six models as starting points on the conditions in the inner galaxy and further relax them. 
These models are given in Tab.~\ref{tab:ISM_models}. They all provide good fits to the \textit{AMS-02} cosmic-ray data \cite{Aguilar:2018wmi}, 
as we show in Figs.~\ref{fig:CR_hydrogen} and \ref{fig:CR_nuclei}. In these figures, we show low-energy \textit{Voyager 1} proton data \cite{Cummings:2015qvx, 
Potgieter:2013mcc}, and also a variety of cosmic-ray nuclei from recent measurements of 
\textit{AMS-02}  \cite{Aguilar:2015ctt,  Aguilar:2017hno, Aguilar:2018njt} including the He and C spectra  and the well-measured
boron-to-carbon (B/C), carbon-to-oxygen (C/O) and beryllium-to-carbon (Be/C) ratios. Apart from rigidities lower than 8  GV
in the Be/C ratio, every other spectrum is fitted very well throughout the observed rigidity range. 

\begin{table*}[t]
    \begin{tabular}{ccccccccccc}
         \hline
           CR Model & $\delta$ & $z_{L}$ & $D_{0} \times 10^{28}$ & $v_{A}$ & $dv_{c}/d|z|$ 
              & $\alpha_{1}$ & $R_{br_{1}}$ & $\alpha_{2}$ & $R_{br_{2}}$  & $\alpha_{3}$\\
              &  &  (kpc) & (cm$^2$/s) & (km/s) & (km/s/kpc) & H/He & H/He (GV) & H/He & H/He (GV) & H/He\\
            \hline \hline
            A  & 0.33 & 5.7 & 6.70 & 30.0 & 0 & 1.74/1.70 & 6.0/7.4 & 2.04/2.16 & 14.0/21.5 & 2.41/2.39 \\
            B &  0.37 & 5.5 & 5.50 & 30.0 & 2 & 1.72/1.74 & 6.0/8.0 & 2.00/2.14 & 12.4/21.0 & 2.38/2.375 \\
            C &  0.40 & 5.6 & 4.85 & 24.0 & 1 & 1.69/1.65 & 6.0/6.7 & 2.00/2.13 & 12.4/20 & 2.38/2.355 \\
            D &  0.45 & 5.7 & 3.90 & 24.0 & 5.5 & 1.69/1.68 & 6.0/7.0 & 1.99/2.12 & 12.4/18.7 & 2.355/2.34 \\ 
            E &  0.50 & 6.0 & 3.10 & 23.0 & 9 & 1.71/1.68 & 6.0/7.2 & 2.02/2.14 & 11.2/17.5 & 2.38/2.33 \\
            F &  0.43 & 3.0 & 1.85 & 20.0 & 2 & 1.68/1.74 & 6.0/10.5 & 2.08/2.09 & 13.0/21.0 & 2.41/2.33 \\
            \hline
        \end{tabular}
    \caption{The cosmic-ray propagation assumptions (CR Model), determined by the diffusion index 
    $\delta$, the diffusion scale height $z_L$, the normalization of the diffusion co-efficient $D_{0}$, the
Alfv$\acute{\textrm{e}}$n velocity $v_{A}$, the galactic convection gradient $dv_{c}/dv$, 
the injection indices $\alpha_{1}$, $\alpha_{2}$, $\alpha_{3}$, and the rigidity breaks 
$R_{br_{1}}$ and $R_{br_{2}}$ for cosmic-ray hydrogen and helium isotopes. 
In the last five columns, the first values refer to hydrogen injection properties and the second 
values to helium.}
    \label{tab:ISM_models}
\end{table*}

Cosmic-ray models with similar parameters have also been shown to provide good descriptions of the secondary
to primary ratio of cosmic-ray antiprotons to protons $\bar{p}/p$ measured by \textit{AMS-02} \cite{Aguilar:2016kjl},
apart from very high energies and an $\mathcal O(0.1)$ bump centered at $5-20$ GeV \cite{Cholis:2017qlb, Cholis:2019ejx}. 
We do not show the cosmic-ray electrons or positrons here, since we know that electrons and positrons in these energies 
experience fast energy losses, and as a result their spectral properties depend on the local galactic magnetic field amplitude 
and amplitude of the ISM radiation field. However, we do include these uncertainties when we produce our gamma-ray
templates. In addition, the observed properties of cosmic-ray electrons may also depend  on the distribution of local pulsars,
which we will revisit later when we discuss the gamma-rays from the inner 
galaxy. In any case, fitting the observed cosmic-ray electron observations will not provide strong constraints 
on the conditions of sub-TeV cosmic-ray propagation that we are interested in. Several
works have shown that models with relatively similar assumptions to those of Table ~\ref{tab:ISM_models} are in 
agreement with the observed cosmic-ray electrons and positrons ({\it e.g.}, \cite{Cholis:2018izy, Hooper:2017gtd, 
DiMauro:2014iia, Zhao-Dong:2019fez}). 

Cosmic rays lose energy as they  enter the volume of the Heliosphere and propagate inwards to the Earth's position 
in the solar system. How much energy they lose (predominantly via adiabatic energy losses) depends on their charge, 
their initial rigidity, and the amplitude, polarity, and morphology of the heliospheric magnetic field (HMF) they propagate through.
During the time of observations by \textit{AMS-02}, the polarity 
was negative $A<0$ up to the end of 2012; by the summer of 2014, it had stabilized to $A>0$.  For a 
negative polarity, negatively charged particles reach the Earth from the outer parts of the Heliosphere through the poles of 
the HMF. It takes them a typical travel time of a few months. At the same time, positively charged particles 
will travel predominantly through the Heliospheric Current Sheet: the area of the HMF that separates the north from the 
south magnetic hemispheres. Such particles travel much more slowly; they experience stronger energy losses, so
their arrival depends strongly on their initial rigidity at entrance \cite{2012Ap&SS.339..223S, Potgieter:2013pdj, Cholis:2015gna,
Kuhlen:2019hqb, Cholis:2020tpi}. The HMF changes polarity $A$ every approximately 11 years.
When the polarity of the HMF flips, the type of paths that oppositely charged particles follow 
through the volume of the Heliosphere also flips. The end result of these processes is that the observed cosmic-ray spectra 
are modulated compared to the local ISM ones before entrance into the Heliosphere. This is known as solar modulation 
\cite{1968ApJ...154.1011G}. The quantitative amount by which the cosmic-ray spectra are modulated is described by the 
modulation potential. In this work we follow the analytical model of \cite{Cholis:2015gna} to evaluate the charge-, time-, and 
rigidity-dependent modulation potential whose properties have most recently been evaluated in \cite{Cholis:2020tpi}. Since 
the properties of the amplitude of solar modulation potential is not perfectly constrained, for each cosmic-ray spectrum (or 
ratio of spectra) presented in Figs.~\ref{fig:CR_hydrogen} and~\ref{fig:CR_nuclei} we show a blue band that encompasses 
the $3\sigma$ relevant ranges.

\section{Diffuse Gamma-Ray Emission}
\label{sec:GDGRE}

In this work we develop and test templates to model the gamma-ray emission due to galactic cosmic rays
interacting with the ISM gas, the interstellar radiation field (ISRF), and the galactic magnetic field.
We also include the presence of galactic winds and the impact of turbulence in the galactic magnetic field. 
Moreover, we study different spatial distributions for the cosmic-ray sources, as well as varying the 
assumptions on the exact nature of of their injected primaries. 
In addition, we include the isotropic gamma-ray emission component 
associated with the combined emission from unresolved extragalactic sources and the {\it Fermi} Bubbles, as discussed below.

\subsection{Emission from Cosmic-Ray Interactions with the ISM}
\label{sec:CRtoGR}

As cosmic rays propagate they may interact with the interstellar medium, emitting gamma rays. The type of interaction 
determines the gamma-ray spectrum and morphology from these cosmic rays. This emission is broken into three basic components:
the $\pi^{0}$-emission component (Pi0), the bremsstrahlung emission (Bremss), and the inverse Compton scattering 
component (ICS). 

\begin{figure*}[t]
\hspace{-0.07in}
\includegraphics[width=2.32in,angle=0]{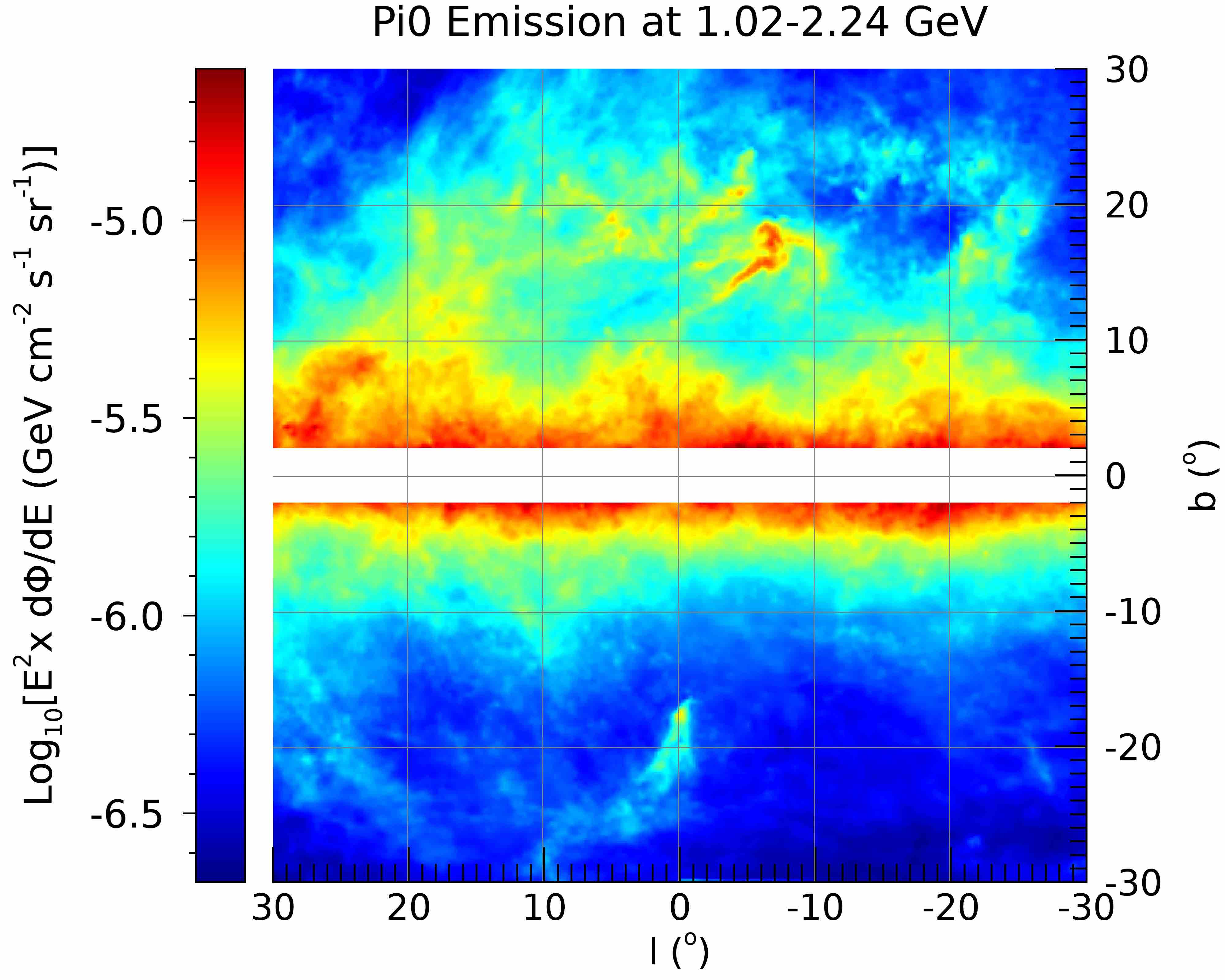} 
\hspace{-0.03in}
\includegraphics[width=2.32in,angle=0]{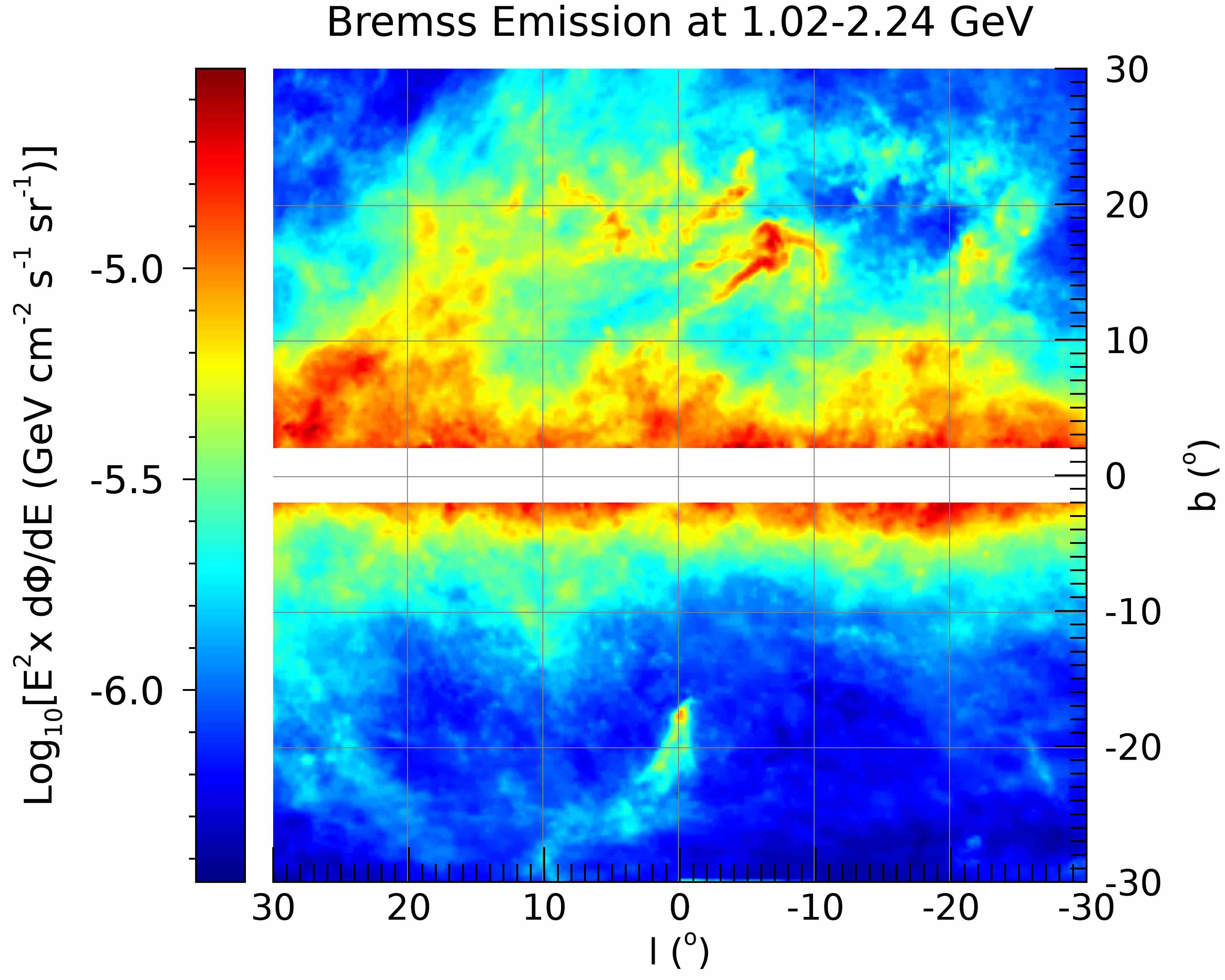} 
\hspace{-0.03in}
\includegraphics[width=2.32in,angle=0]{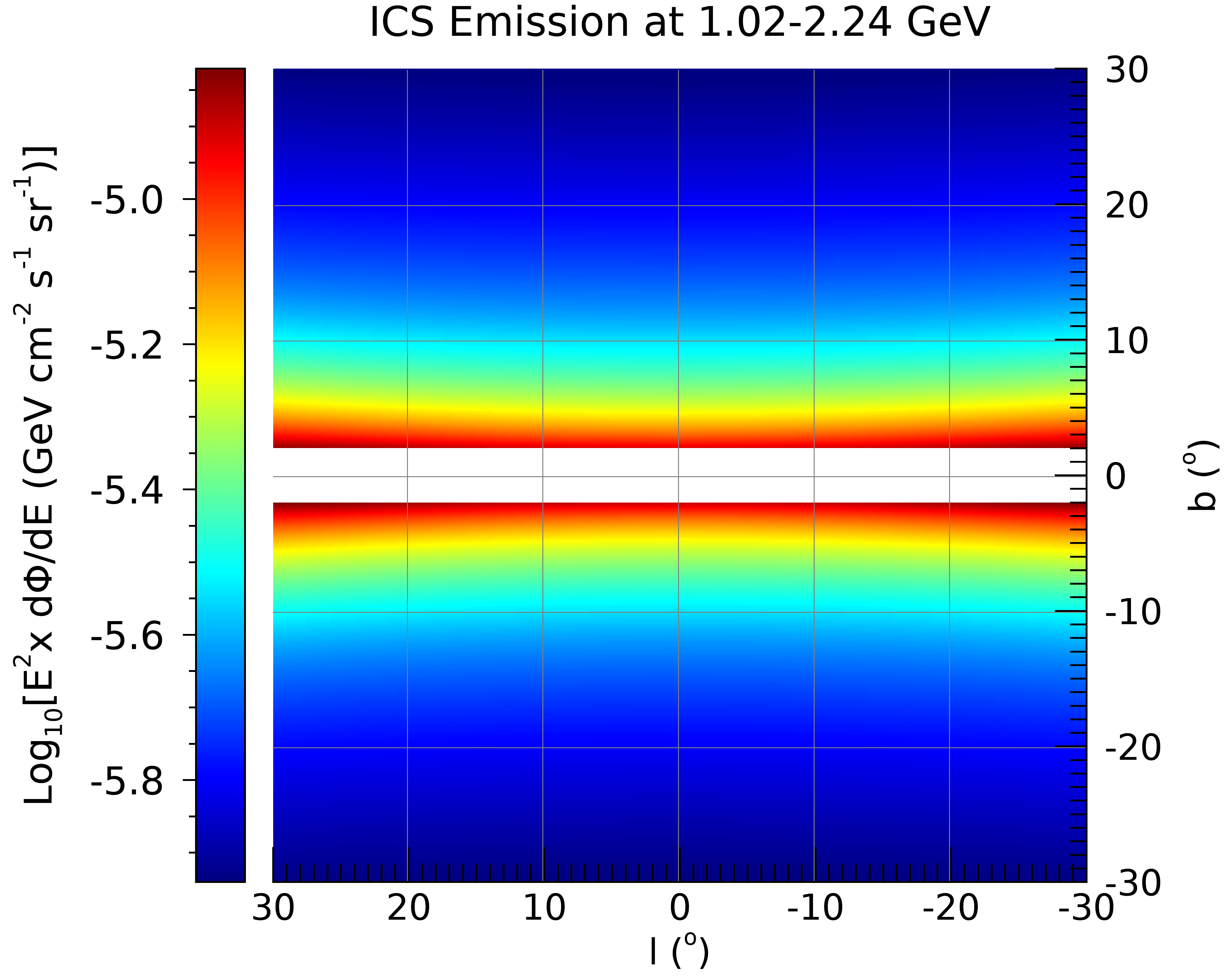}
\vspace{0.00in}
\caption{The morphology of the three major diffuse emission components for energies 
of 1.02-2.24 GeV in the inner $60^{\circ} \times 60^{\circ}$ region. \textit{Left}, Pi0, \textit{middle}, bremsstrahlung 
and \textit{right}, ICS. We use a resolution of 
$0.1^{\circ} \times 0.1^{\circ}$ pixels in a Cartesian grid. We do not include in these maps 
    the PSF smoothing of the \textit{Fermi}-LAT instrument, which emphasizes the differences 
in the morphology between the Pi0 and the bremsstrahlung emission components. We show the differential flux multiplied 
by the energy squared ($E^{2} \times d\Phi/dE$). In the map we mask the galactic disk ($|b| \leq 2^{\circ}$) as this region does not enter our fits.}
\label{fig:DiffComps}
\end{figure*}  

The Pi0 emission comes from inelastic collisions of cosmic-ray protons and more massive nuclei with the ISM molecular, 
atomic, and ionized hydrogen gas (H2, HI and HII). The ISM gases include an appreciable fraction of helium targets as well. 
Boosted mesons, including $\pi^{0}$ and $\eta$, are produced in these interactions. The decays of these mesons give significant 
amounts of high energy gamma rays \cite{2018PhRvD..98c0001T}, whose spectral properties are inherited from their 
parent cosmic-ray nuclei. Because cosmic-ray nuclei have a relatively soft spatial gradient, the morphology on the sky of 
the Pi0 component mostly resembles the density of target ISM gas nuclei integrated along the line of sight, but using 
\texttt{GALPROP} enables us to produce models that assume different conditions in the interstellar medium and which result in cosmic-ray 
nuclei having different spectral and spatial distributions in the galaxy. Those result in a model-dependent Pi0 morphology
and energy spectrum. Moreover, as we will describe in detail in Sec.~\ref{sec:TemplateModels}, we test a variety of assumptions
on the spatial distributions of H2, HI and HII gases.  An example of a predicted Pi0 emission morphology is presented in
Fig.~\ref{fig:DiffComps} (left panel) for the inner $60^{\circ} \times 60^{\circ}$ at gamma-ray energies of 1.02 to 2.24 GeV.
In this plot we ignore the effects of the \textit{Fermi}-LAT detector such as its non-isotropic sky scan strategy and its 
imperfect angular resolution, which, as we discuss in more detail below, is characterized by a non-zero point spread function (PSF). 

Cosmic-ray electrons may also interact with the ISM gases, leading to gamma rays via bremsstrahlung radiation
\cite{1970RvMP...42..237B}. This component acquires its energy spectrum from the parent electrons and its morphology 
mostly from the ISM gas distribution. In the middle panel of Fig.~\ref{fig:DiffComps}, we show the predicted
bremsstrahlung emission for the same set of ISM conditions at the same energy and part of sky as was shown for the Pi0 component. 
In dense gas environments, such as at low galactic latitudes, the combination of Pi0 and bremsstrahlung is responsible for most of 
the galactic diffuse emission. Because Pi0 and Bremss emission morphologies are both primarily determined by the ISM gas,
we fix their relative normalizations when we fit the total gamma-ray emission, a process we describe in much more detail below.

In addition to the Bremss emission, highly boosted electrons can up-scatter low energy photons into the gamma-ray range; this is the inverse Compton 
scattering (ICS) \cite{1970RvMP...42..237B}. The resulting ICS component has a spectrum that is inherited from the combination 
of the energy spectrum of the cosmic-ray electrons and the energy spectrum of the target photons, and is typically harder 
than the Pi0 and Bremss components. It significance on the sky depends on the gamma-ray energy and direction. 
We include as target photons both the interstellar radiation field spanning from the 
far infrared to the ultraviolet and the lower-energy cosmic microwave background photons (CMB). For the rest of this 
paper we will simply refer to all the target photons, including the CMB, as ISRF. In our combined gamma-ray templates, 
we allow for different assumptions between models on the amplitude of the infrared ISRF photons emitted 
from the ISM gas and dust and of the amplitude of the near IR-visible-UV photons emitted by stars, while keeping the CMB 
fixed. 

Since cosmic-ray electrons above a few GeV in energy lose significant amounts of their energy via ICS, the 
ISRF assumptions affect the prediction of both the ICS morphology and its spectrum. 
The morphology of the ICS component mostly tracks the convolution of the spatial distributions of cosmic-ray electrons and 
the ISRF. For certain assumptions the far IR-UV has a softer spatial gradient to that of the electrons, but this is not always 
the case.  In Fig.~\ref{fig:DiffComps} and for the same ISM conditions used to produce the Pi0 and Bremss components, 
we present an example of the ICS component.  In producing the ICS maps we use the \texttt{GALPROP} setup that includes 
the Klein-Nishina cross section \cite{1929ZPhy...52..853K} for the cosmic-ray electrons interacting with the photons, which, at high energies, is smaller than the Thomson cross section. We assume that electrons are interacting 
with a bath of isotropically distributed photons (in our frame), which formally is only correct for the CMB. 
Ref.~\cite{Moskalenko:1998gw} has shown that taking into account the non-isotropic distribution of far IR-visible-UV photons 
enhances the ICS intensity by approximately 10$\%$ at GeV energies and for the latitudes of $|b|<30^{\circ}$ that we study 
here. Moreover, in the GeV range and for the inner galaxy, correcting for the non-isotropic distribution of target photons 
 depends weakly on their energy and even less on their location within the $60^{\circ} \times 60^{\circ}$ window 
\cite{Moskalenko:1998gw}. Thus such an effect can not mimic the much more cuspy morphology of the GCE emission component. Given that in the templates analysis we allow for the ICS normalization to be free by significantly 
more than $10\%$ we ignore the anisotropic distribution of photons effect which we believe is absorbed by the fitting procedure.

In addition to the above diffuse galactic emission components, we include the emission from 
the \textit{Fermi} Bubbles \cite{Su:2010qj, Dobler:2009xz, Fermi-LAT:2014sfa}. The nature of  
the  \textit{Fermi} Bubbles is probably leptonic and mostly ICS associated with activity from the 
central supermassive black hole \cite{Guo:2011eg, Mertsch:2011es, Cheng:2011xd,  Guo:2011ip, Yang:2012fy, Lacki:2013zsa, Cholis:2015dea}.
However, the possibility that the Bubbles have a hadronic origin has also been proposed
\cite{Crocker:2010dg, Crocker:2014fla, Carlson:2014cwa, Abramowski:2016mir}. We take the morphology of the Bubbles to be
isotropic, within the region of the sky that they cover, using the template proposed by \cite{Su:2010qj}. The default Bubbles energy spectrum we take
from \cite{Fermi-LAT:2014sfa}, though we allow a free normalization in each energy bin
when we carry out our fits, as described in much more detail below. The energy spectrum 
comes with both statistical errors associated to its fit to the data and systematic ones, associated 
with underlying assumptions on the remaining emission from the Bubbles region. We use the 
combination of statistical and systematic errors from \cite{Fermi-LAT:2014sfa} added in quadrature 
to enforce a statistical penalty in our fits when the normalization deviates from the central values 
of the \textit{Fermi} Bubbles energy spectrum. Our precise prescription is given in Sec.~\ref{sec:method}.
After testing both the possibility that there is an ``edge brightening'' to the Bubbles or that
there is no such brightening (see \cite{Su:2010qj} and \cite{Fermi-LAT:2014sfa} 
for the two original alternative analyses), we decided not to include an edge brightening to the 
Bubbles templates. By including the edge-brightening we found signs of over-subtraction on the residual maps, which we discuss in 
further detail in our Sec.~\ref{sec:results}.

\subsection{Template Modeling: Including Astrophysical Uncertainties}
\label{sec:TemplateModels}

We describe here the range of multi-messenger astrophysical uncertainties for the inner galaxy that we probe through 
the array of gamma-ray models we produce. The reader interested just in the results of that selection and their impact on fits to 
the total gamma-ray emission can skip to Sec.~\ref{sec:method}. 

Cosmic rays propagate in the Milky Way's ISM through diffusion and convection as described in 
Eqs.~\ref{eq:Diff} and~\ref{eq:Conv}. That propagation continues up to a distance of $z_{L}$ away from 
the galactic disk. We take models that include values for $z_{L}$ in the range of $3-10$ kpc. Larger values 
of $z_{L}$ may be used to describe the propagation of cosmic rays far away from the galactic disk, but 
are not relevant for the galactic center region. We focus on choices that can describe the physical 
conditions within a few kpc from the galactic center, and note that a gradient on propagation assumptions
may exist as we move away from the galactic disk. Such a gradient would allow faster propagation far 
from the disk. 
The diffusion index $\delta$ in Eq.~\ref{eq:Diff} that defines the rigidity/energy dependence of cosmic-rays 
is set to take values between 0.33 and 0.5. 
When fitting only the cosmic-ray spectra in Sec.~\ref{sec:cosmicrays}, we find that cosmic ray 
measurements prefer values closer to $\delta \simeq 0.4-0.5$ in agreement with recent works as 
in \cite{Weinrich:2020ftb}; however, as discussed earlier, here 
we model the inner galaxy where propagation conditions may differ from local ISM properties. The 
underlying assumption used here is in agreement with Ref.~\cite{Johannesson:2016rlh}, where it was 
shown that on large scales diffusion is not homogeneous within the Milky Way's ISM (see also~\cite{Yang:2016jda, Cerri:2017joy}).
The timescale that cosmic rays require to propagate away from a region of the Milky Way via diffusion is set 
also by the diffusion coefficient $D_{0}$, which we take for the inner galaxy to be in the range of $(2-40) 
\times 10^{28}$ cm$^{2}$/s, defined at a rigidity of 4 GV (see Eq.~\ref{eq:Diff}). Our prior range is significantly 
wider than the range $D_{0} \simeq (2-8) \times 10^{28}$ cm$^{2}$/s based only on local ISM cosmic-ray 
measurements \cite{Johannesson:2016rlh, Yuan:2018vgk}.

Convection in the inner galaxy as parameterized by Eq.~\ref{eq:Conv} is described by a convection 
gradient $dv_{c}/d|z|$ which may lie anywhere from 0 (no convection) up to 200 km/s/kpc. 
By comparison, the \textit{AMS-02} data gives a local convection gradient$\lsim 10$ km/s (in agreement 
with \cite{Weinrich:2020ftb, Boschini:2019gow, Yuan:2018vgk, Boschini:2017fxq}), although 
some works \cite{Cuoco:2016eej} have given values for the local convection velocity of up to 45 km/s,
suggesting a substantial uncertainty on the convection gradient.

Diffusive reacceleration can increase the energy of low-energy cosmic rays. This is parameterized by 
the Alfv$\acute{\textrm{e}}$n speed, $v_{A}$ for which we take values in the inner galaxy in the range of $0-50$ km/s. 
By comparison, locally measured cosmic rays suggest values in the range of $\simeq 20-30$ 
km/s, as we show in our Table~\ref{tab:ISM_models}. Similarly \cite{Yuan:2018vgk} constrain $v_{A}$ from 
cosmic ray observations to be locally within $\simeq 26-32$ km/s, while \cite{Johannesson:2016rlh} give two
ranges for $v_{A}$, to be between 0 and 22 km/s when using only light nuclei cosmic-ray species and 
between 16 and 44 km/s when using more massive nuclei. Instead, \cite{Weinrich:2020ftb} and \cite{Niu:2018waj} 
get values for $v_{A}$ as high as 67 km/s and 63 km/s respectively. Our assumed range for the inner galaxy 
encompasses all those values. In fact, we have tested up to values of $v_{A} = 150$ km/s, which would
require a great amount of energy in the magnetic fields. Such high values of $v_{A}$ are very strongly
excluded by the \textit{Fermi} data. The Alfv$\acute{\textrm{e}}$n speed can not be much 
larger than 30 km/s for the entire Milky Way cosmic-ray propagation volume, as low energy cosmic-ray 
electrons would also get reaccelerated and in turn produce an unacceptable amount of power via 
synchrotron radiation \cite{Drury:2016ubm}. These limits depend on the exact assumptions 
of the magnetic field amplitudes and their distribution in the Milky Way. Additional limits on Milky Way's 
ISM Alfv$\acute{\textrm{e}}$n speed via synchrotron emission observations have been presented in 
\cite{2013JCAP...03..036D, Orlando:2013ysa, Orlando:2017mvd}. 

\setlength{\tabcolsep}{2pt}
\begin{table*}[!ht]
    \begin{tabular}{ccccccccccccccc}
         \hline
            Name & $z_{L}$ & $D_{0}$ & $\delta$ & $v_{A}$ & $dv_{c}/d|z|$ & $S^{N}/S^{e}$ & $\alpha_{1}^{p}$/$\alpha_{2}^{p}$ 
            & $\alpha_{1}^{e}$/$\alpha_{2}^{e}$ & $N^{p}$/$N^{e}$ & B-field & ISRF & H2 & HI & HII \\
            \hline \hline
            I & 4.0 & 5.00 & 0.33 & 32.7 & 55 & Pul/Pul & 1.35/2.33 & 1.5/2.25 & 4.13/3.33 & 200030050 & 1.36,1.36,1.0 & 9 & 5 & 1 \\   
            II & 6.0 & 7.1 & 0.33 & 50.0 & 0 & Pul/SNR & 1.89/2.30 & 1.40/2.10 & 2.40/2.20 & 050100020 & 1.0,1.0,1.0 & 2$^{*}$ & 1 & 1 \\ 
            III & 5.6 & 4.85 & 0.40 & 40.0 & 0 & Pul/Pul & 1.50/1.90 & 1.5/2.25 & 2.40/1.55 & 200050040 & 1.4,1.4,1.0 & 9 & 4 & 1 \\   
            VI & 6.0 & 2.00 & 0.33 & 0 & 200 & Pul/SNR & 1.60/2.10 & 1.6/2.30 & 2.32/5.70 & 200030050 & 1.4,1.4,1.0 & 9 & 5 & 1 \\   
            X & 10.0 & 8.00 & 0.33 & 32.2 & 50 & Pul/SNR & 1.40/1.80 & 1.4/2.35 & 1.90/3.20 & 200040050 & 1.4,1.4,1.0 & 0 & 5 & 2 \\   
            XV & 6.0 & 7.10 & 0.33 & 50.0 & 0 & Pul/SNR & 1.89/2.30 & 1.40/2.10 & 2.40/2.20 & 050100020 & 1.0,1.0,1.0 & 0 & 5 & 2 \\                         
            \hline 
        \end{tabular}
    \caption{Galactic diffuse model parameters $z_{L}$ is in kpc, $D_{0}$ is in $\times 10^{28}$ cm$^{2}$/s, $v_{A}$ is 
    in km/s, $dv_{c}/d|z|$ is in km/s/kpc. $N^{p}$ and $N^{e}$ are the cosmic-ray proton and electron differential flux 
    $dN/dE$ normalizations at the galactocentric distance of 8.5 kpc. They are defined at 100 and 34.5 GeV for the 
    protons and electrons respectively and are in units of $\times 10^{-9}$ cm$^{-2}$s$^{-1}$sr$^{-1}$MeV$^{-1}$. 
    See text for full details.}
    \label{tab:ModelsShort}
\end{table*}  

The combination of $z_{L}$, $D_{0}$, $\delta$, $v_{A}$ and $dv_{c}/d|z|$ set how cosmic rays released
into the ISM propagate in space, and $v_{A}$ affects the the low-energy  spectrum of cosmic ray nuclei. In Tab.~\ref{tab:ModelsShort}, we provide six diffuse emission models that provide a good fit
to the gamma-ray sky whose predicted gamma-ray emission we shall present and compare in this section. The choices
for $z_{L}$, $D_{0}$, $\delta$, $v_{A}$ and $dv_{c}/d|z|$ are given in that table. The entire ensemble of diffuse
emission models used to fit the {\it Fermi} data is given in Table~\ref{tab:ModelsLong} of App.~\ref{app:Allmodels}.

Specifying the propagation properties of cosmic rays is necessary but not sufficient to predict gamma-ray templates:
the sources that inject the cosmic rays must be specified as well.
Primary cosmic rays are injected typically by supernova remnants (SNRs) and by energetic environments such as pulsar wind 
nebulae (PWNe). 
Primary nuclei are produced in SNRs, while electrons and positrons are also produced close to 
PWNe at the highest energies. There are different tracers for the location of these sources. Here we take these 
tracers to be either the observed Milky Way pulsars \cite{1993ApJS...88..529T}, which roughly coincide with the 
locations of past SNRs, or the fewer but more directly observable recent SNRs \cite{1998ApJ...504..761C}. In 
Table~\ref{tab:ModelsShort} and Table~\ref{tab:ModelsLong}, at the column denoted by ``$S^{N}/S^{e}$'', we include 
different combinations for the choice of the spatial profile distribution assumed for the cosmic-ray nuclei ($S^{N}$) 
and cosmic-ray electrons ($S^{e}$). ``Pul'' refers to the profile of \cite{1993ApJS...88..529T}, while ``SNR'' to that of 
\cite{1998ApJ...504..761C}. We note that these are chosen as tracers of older or more recent SNRs. In addition to 
the spatial distribution assumptions for the primary cosmic-ray sources, we vary the first two injection indices of 
the cosmic-ray nuclei and electrons as described by Eq.~\ref{eq:Inj}. Those are depicted in the columns 
``$\alpha_{1}^{p}$/$\alpha_{2}^{p}$'' for the protons and ``$\alpha_{1}^{e}$/$\alpha_{2}^{e}$'' for the electrons.
Furthermore, since 
we incorporate fit data only up to around 50 GeV, as given in Table~\ref{tab:PSF_vsE}, the choices for
$\alpha_{3}^{p}$ and $\alpha_{3}^{e}$ that describe the cosmic-ray spectra at energies above 200 GeV have only 
a minor impact in our analysis.  

\begin{figure}
\begin{centering}
\includegraphics[width=3.65in,angle=0]{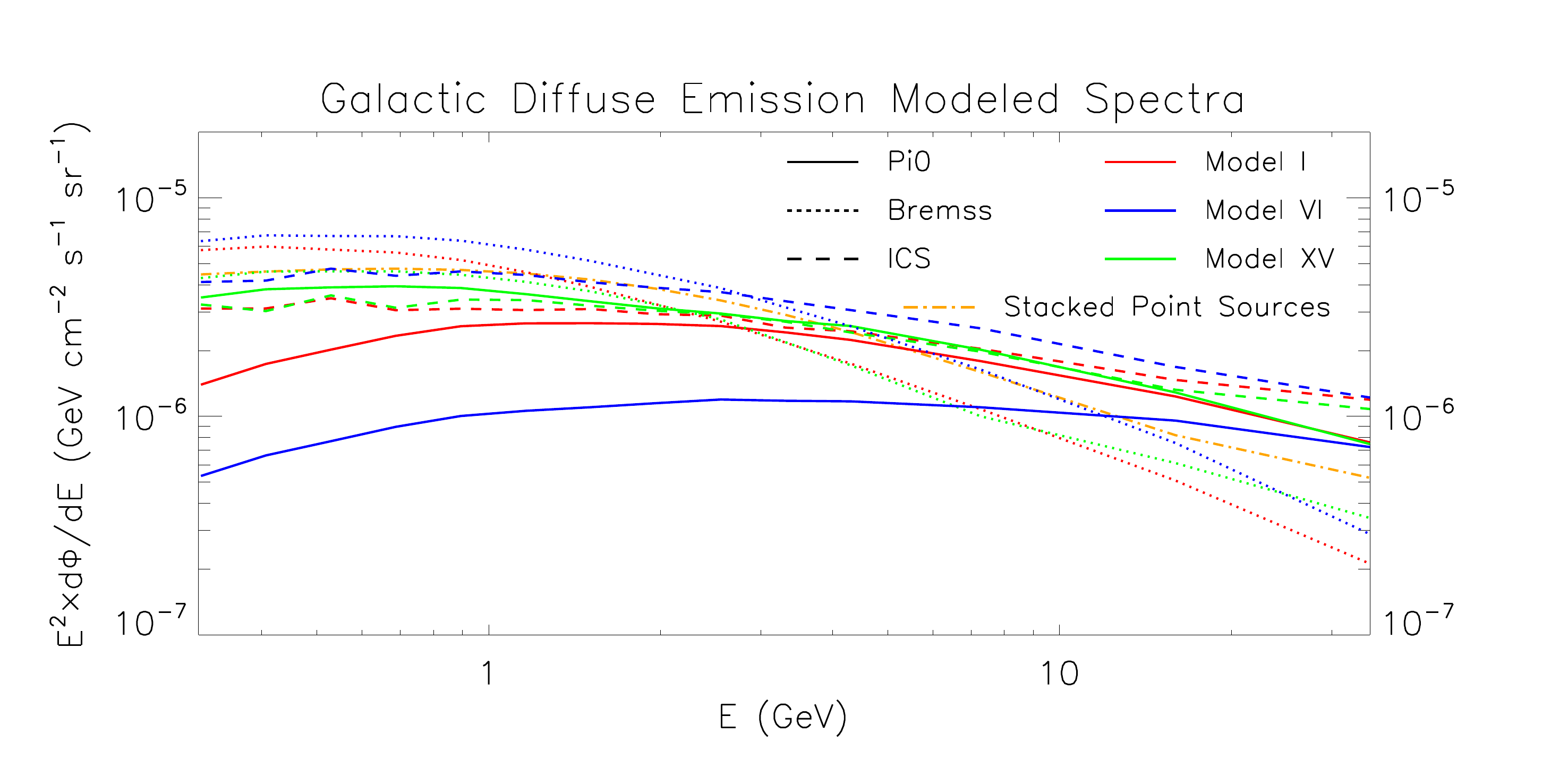}
\end{centering}
\vskip -0.3in
\caption{The predicted averaged diffuse spectra over the $60^{\circ} \times 60^{\circ}$ window excluding 
the galactic disk for three different models; I (red), VI (blue) and XV (green). The solid lines depict the 
predicted Pi0 emission, the dotted lines the Bremss emission and the dashed the ICS emission 
fluxes. The spectra provided include fit normalizations to the data. We also provide for reference the 
stacked spectrum from 4FGL-DR2 point sources in the same window with $|b| > 2^{\circ}$ (orange dashed dotted).}
\label{fig:ModelSpectra}
\end{figure}

Combining these sources of uncertainty, and following a fitting procedure for the inner $60^{\circ} \times 60^{\circ}$
described in detail in Sec.~\ref{sec:method}, we can obtain good fits to the gamma-ray sky that encompass realistic uncertainties while 
maintaining our good description of the cosmic-ray data. In Fig.~\ref{fig:ModelSpectra} we show the
diffuse Pi0, bremsstrahlung and ICS emission spectra for three models.
For comparison, we include the combined emission from the detected point sources of
\textit{Fermi} 4FGL 10-year Source Catalog (4FGL-DR2)~\cite{Ballet:2020hze} with $|b| > 2^{\circ}$, 
even though we mask these point sources throughout the fits described later in this work. 
Even though we allow for a free template normalization between energy bins when fitting the data according to the prescription
given in Sec.~\ref{sec:fit}, different choices for the injection indices will nevertheless affect the allowed diffuse
spectra. This is the case because the Pi0 and bremsstrahlung components share a normalization for reasons discussed
in Sec.~\ref{sec:CRtoGR} above.

The Pi0 and bremsstrahlung emission components have a morphology that is related to the integrated ISM gas 
density at any line of sight. If the cosmic-ray nuclei and electrons were homogeneously distributed, we would only 
need to test a small number of templates that describe the different choices for the gas distribution or what is referred 
to as the column density. Moreover the two templates would have identical morphologies, which as we show in 
Fig.~\ref{fig:DiffComps} is not the case. Given the quality of gamma-ray observations such differences need to 
be accounted for. The main cause for these differences is that electrons and protons have distinct gradients as 
we move from the disk. In Tables~\ref{tab:ModelsShort} and~\ref{tab:ModelsLong}, we focus on varying both the 
underlying maps that give the ISM gas distributions and the cosmic-ray electron and nuclei spatial distribution. 
In the last three columns of those tables, by ``H2'', ``HI'' and ``HII'' we code the spatial distribution assumptions 
for the molecular, atomic and ionized hydrogen. 
\begin{figure*}[t]
\hspace{-0.07in}
\includegraphics[width=2.32in,angle=0]{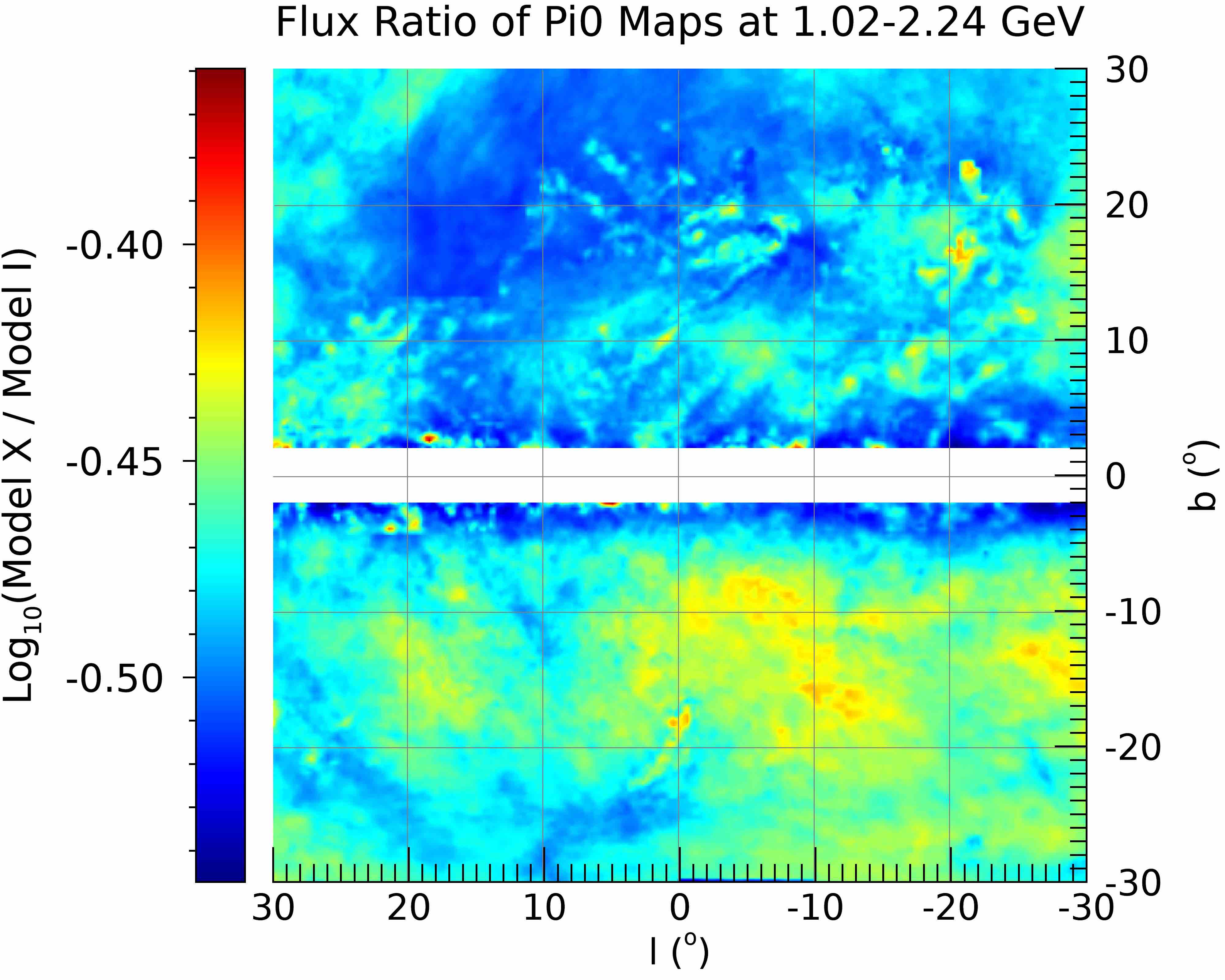} 
\hspace{-0.03in}
\includegraphics[width=2.32in,angle=0]{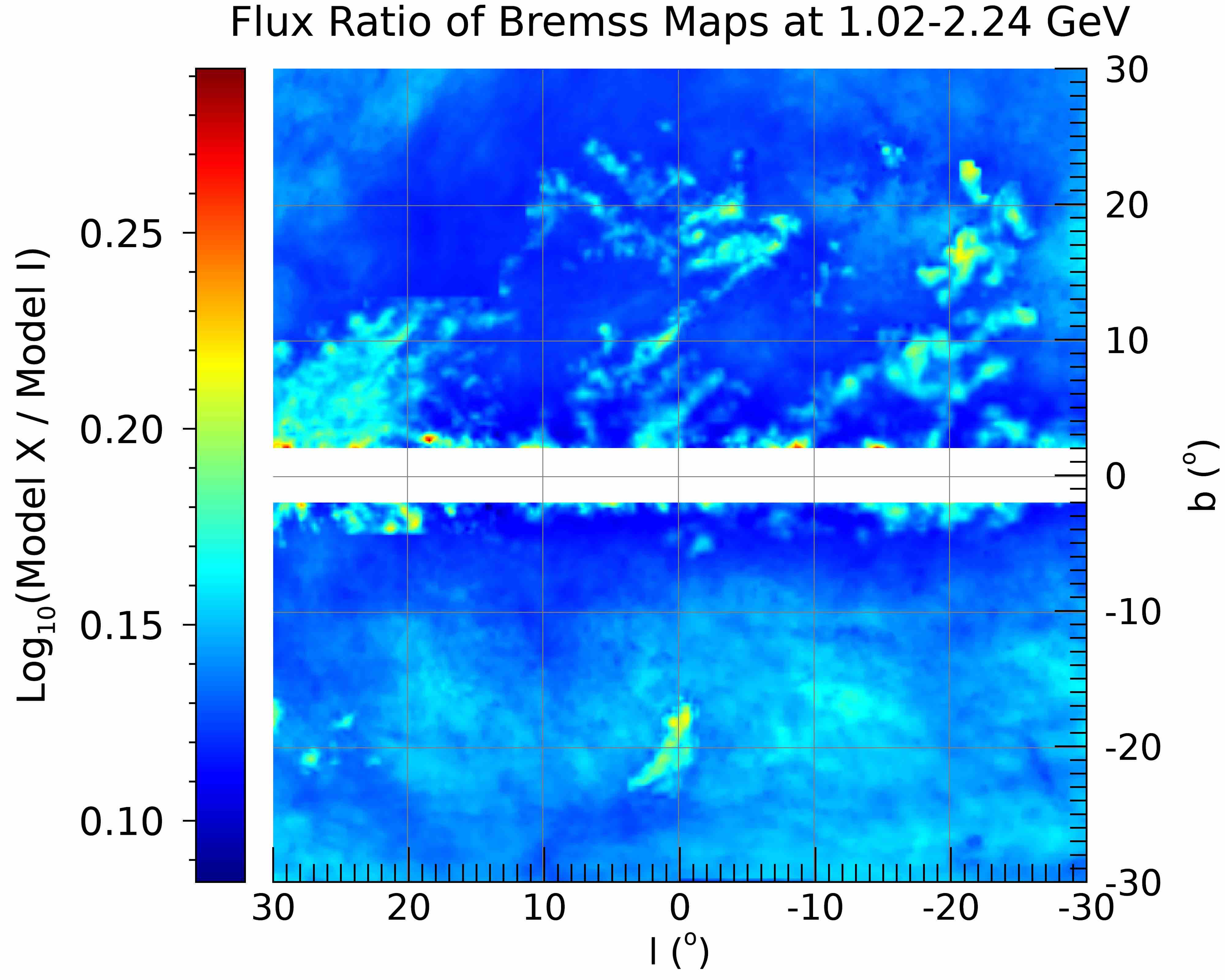}
\hspace{-0.03in}
\includegraphics[width=2.32in,angle=0]{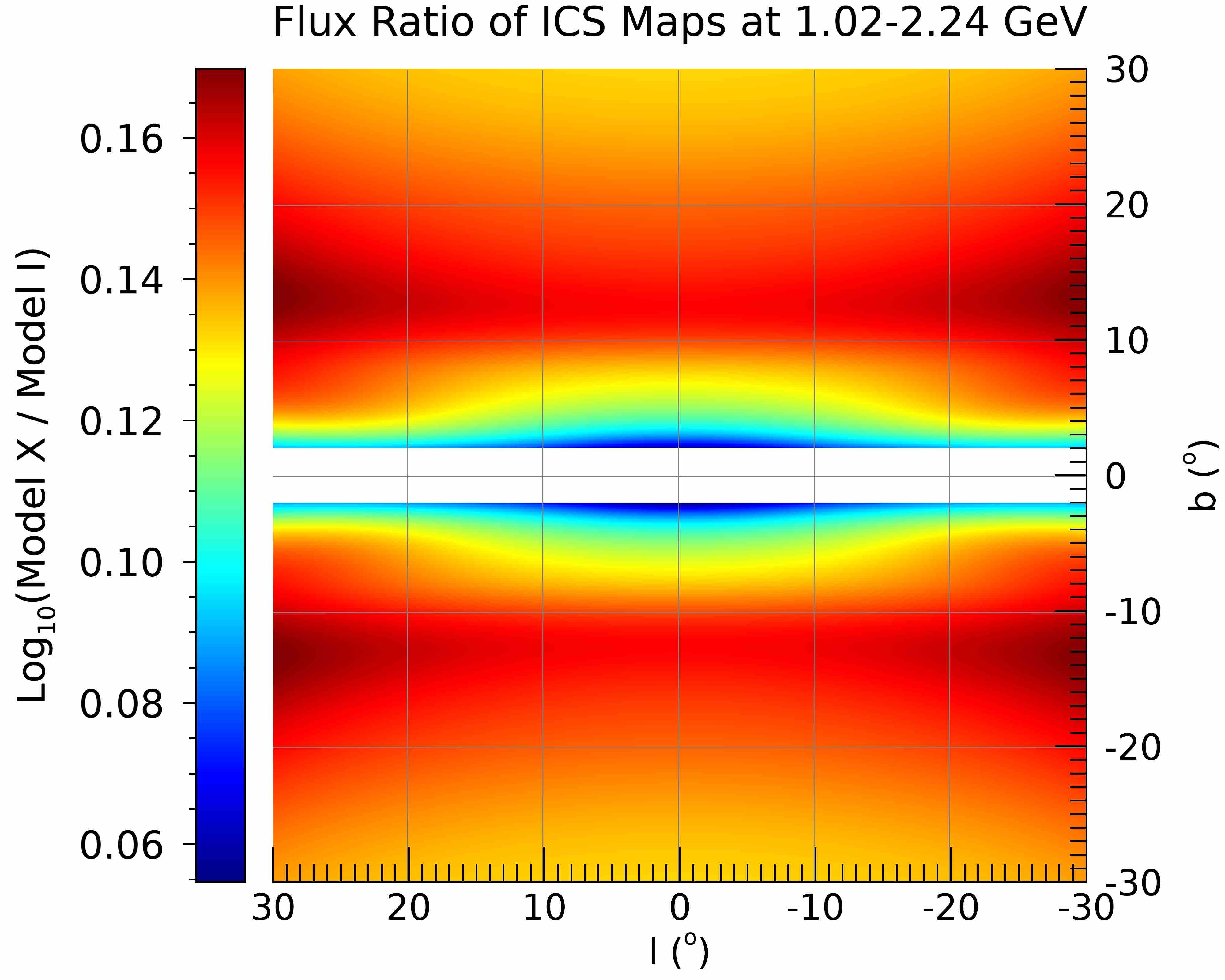} \\
\vspace{0.20in}
\hspace{-0.07in}
\includegraphics[width=2.32in,angle=0]{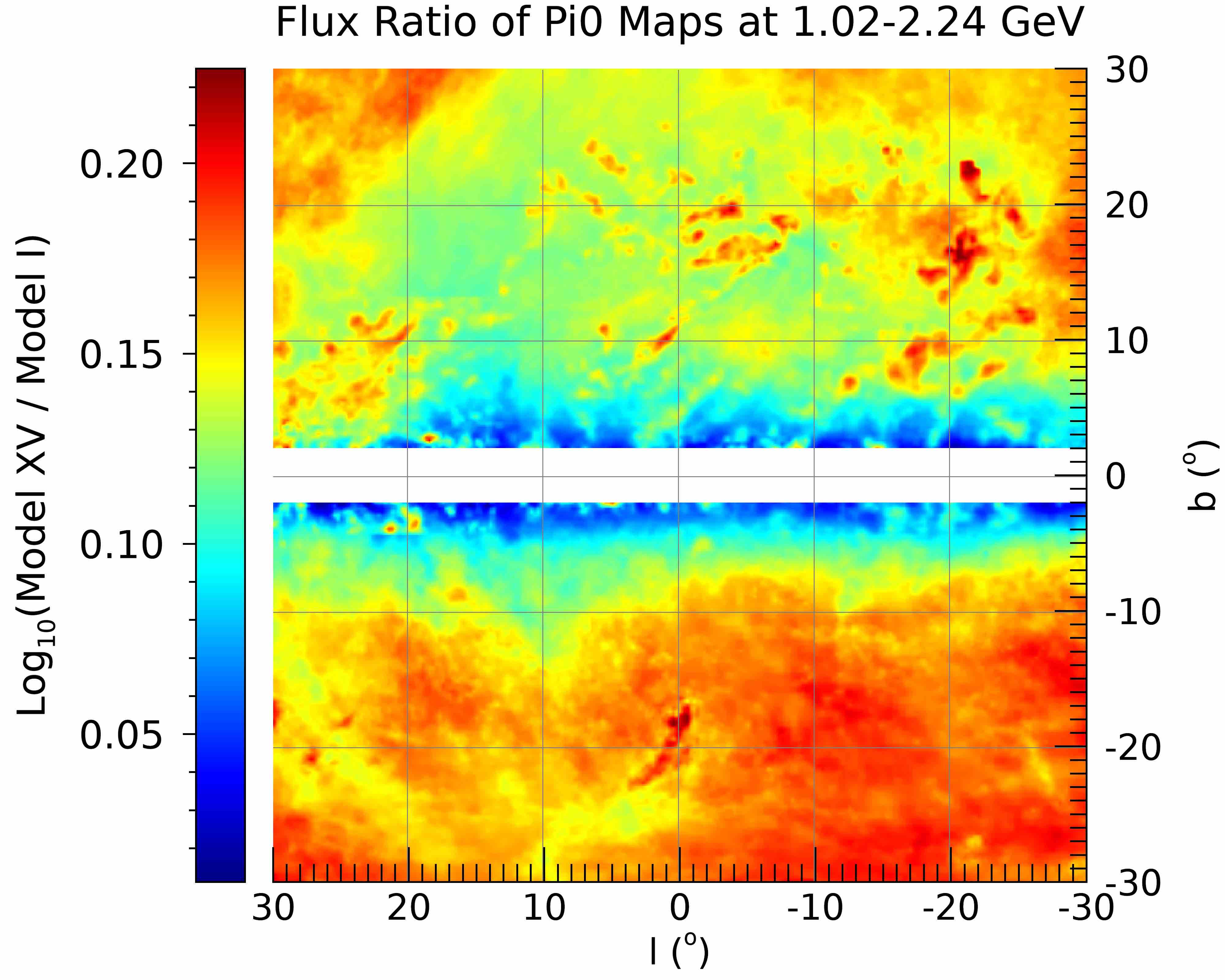} 
\hspace{-0.03in}
\includegraphics[width=2.32in,angle=0]{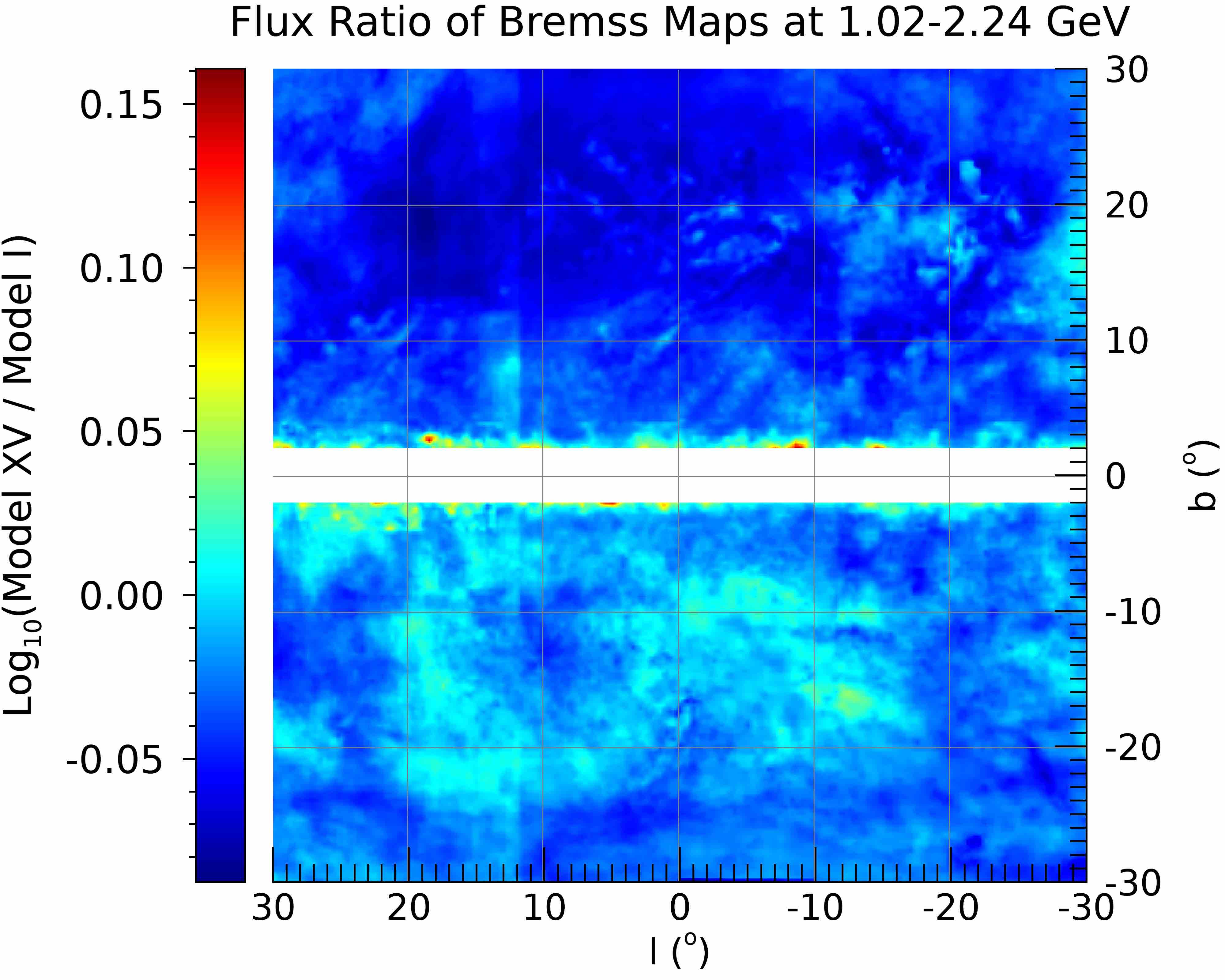}
\hspace{-0.03in}
\includegraphics[width=2.32in,angle=0]{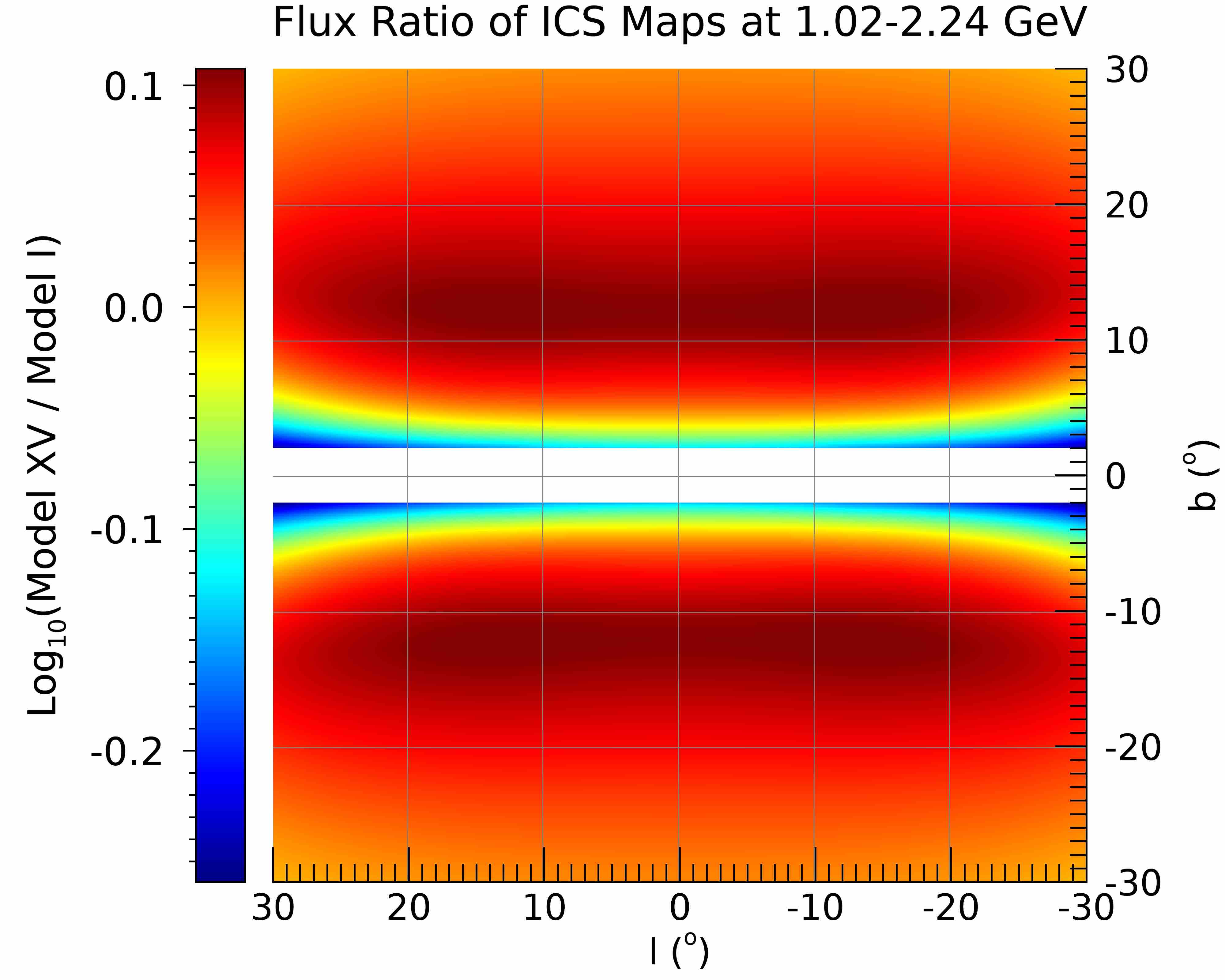}
\vspace{0.00in}
\caption{
The impact of different assumptions on the morphologies of the background diffuse template fluxes. 
On the \textit{top panels} we compare the predictions of Model X to Model I and on the \textit{bottom panels}
the predictions of Model XV to Model I. We present results for the energy range of $1.02-2.24$ GeV, where the 
GCE's significance peaks. We show that the ratio of the predicted Pi0 flux emission on the \textit{left panels}, 
the ratio of the predicted Bremss flux emission on the \textit{central panels}, and the ratio of the predicted ICS 
flux emission on the \textit{right panels}. We have included the equivalent fitting normalizations to the data. 
While the PSF is included in the fits to the data, it is not presented here to exhibit the underlying modeling 
differences.}
\label{fig:RatioOfTemplates}
\end{figure*}  
We give the full details of this notation in App.~\ref{app:Allmodels}, and we also comment that we tested even wider 
variations than the ones presented in this work. The ones not included are either degenerate with the choices
presented or strongly excluded by inner galaxy fits.

In Fig.~\ref{fig:RatioOfTemplates} (first two columns), we compare the prediction for the Pi0 and bremsstrahlung 
emissions from varying assumptions.  We show how the morphology of these emission components changes at the 
energy range of 1-2 GeV. We compare models X and XV to model I. Model I employs different assumptions on the 
molecular and ionized hydrogen gas distribution compared to those of models X and XV, which assume the same ISM 
gas distributions. The gas distribution differences alone can result in certain directions on the sky becoming brighter or 
dimmer depending on the amount of assumed column density. However, if that was the only difference the top and bottom
panels on each column should be identical. The cosmic-ray propagation and the injection assumptions between models X and XV are 
significantly different and result in significantly different morphologies. The ratios that we present in Fig.~\ref{fig:RatioOfTemplates}, 
are calculated \textit{after} the fitting procedure described in Sec.~\ref{sec:method}
is done: the emission for all models shown provides a good  fit to the data. The differences shown between the separate components 
are equivalent to ratios of size $1/3$ to 2, but the overall fit is sufficiently well constrained that when one of
the diffuse emission components is dimmer it is compensated by others increasing in brightness.

The ICS component, which dominates away from the disk and at high energies (see Figs.~\ref{fig:DiffComps} 
and~\ref{fig:ModelSpectra}), depends on the distribution of low-energy photons. Only the CMB is well measured and 
kept fixed in this work, while the density of photons in the IR to the UV is allowed to change. \texttt{GALPROP} 
breaks the ISRF into three components: the ``optical'', the ``IR'' and the ``CMB'', each of which in its conventional
mode is fixed to 1 \cite{GALPROPSite, Porter:2006tb, Porter:2008ve}. We test in our models different normalizations 
for the first two of these components. As the optical and IR photons have distinct spatial distributions, our models test 
different assumptions on the ISRF spectra, spatial morphology, and total energy density within the Galaxy. The impact
on the morphology of the ICS component is shown in the third column of Fig.~\ref{fig:RatioOfTemplates}. We compare the predicted 
ICS emission from model I to models X and XV, at $1-2$ GeV, within the $60^{\circ} \times 60^{\circ}$ window. Again, 
the calculated ratios \textit{include} the normalizations after the template fits described in Sec.~\ref{sec:method}.
Table~\ref{tab:ModelsShort} provides a handful of models that are representative for wide assumptions for the ISM.

\begin{figure*}[!ht]
\hspace{-0.12in}
\includegraphics[width=3.4in,angle=0]{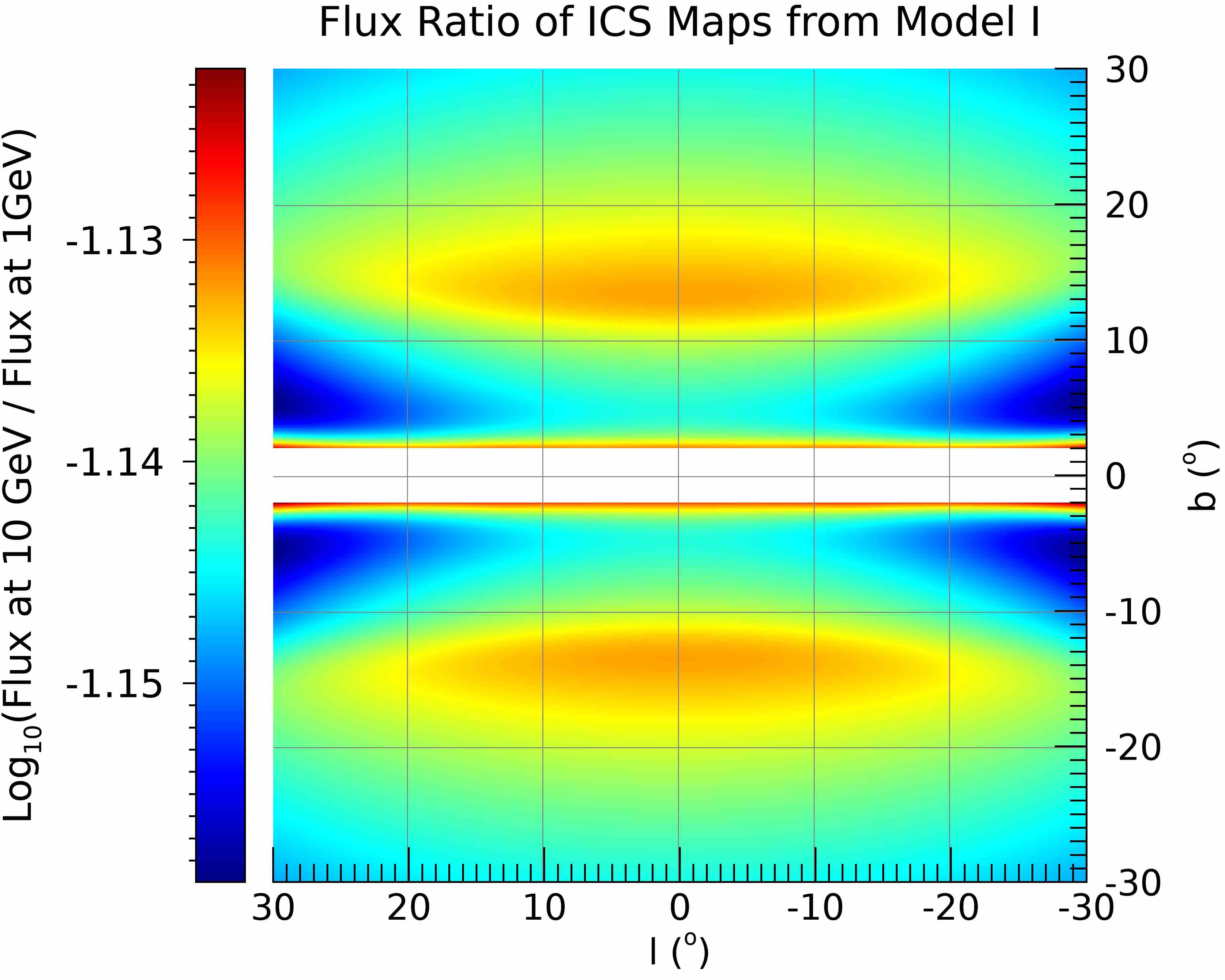} 
\hspace{0.10in}
\includegraphics[width=3.4in,angle=0]{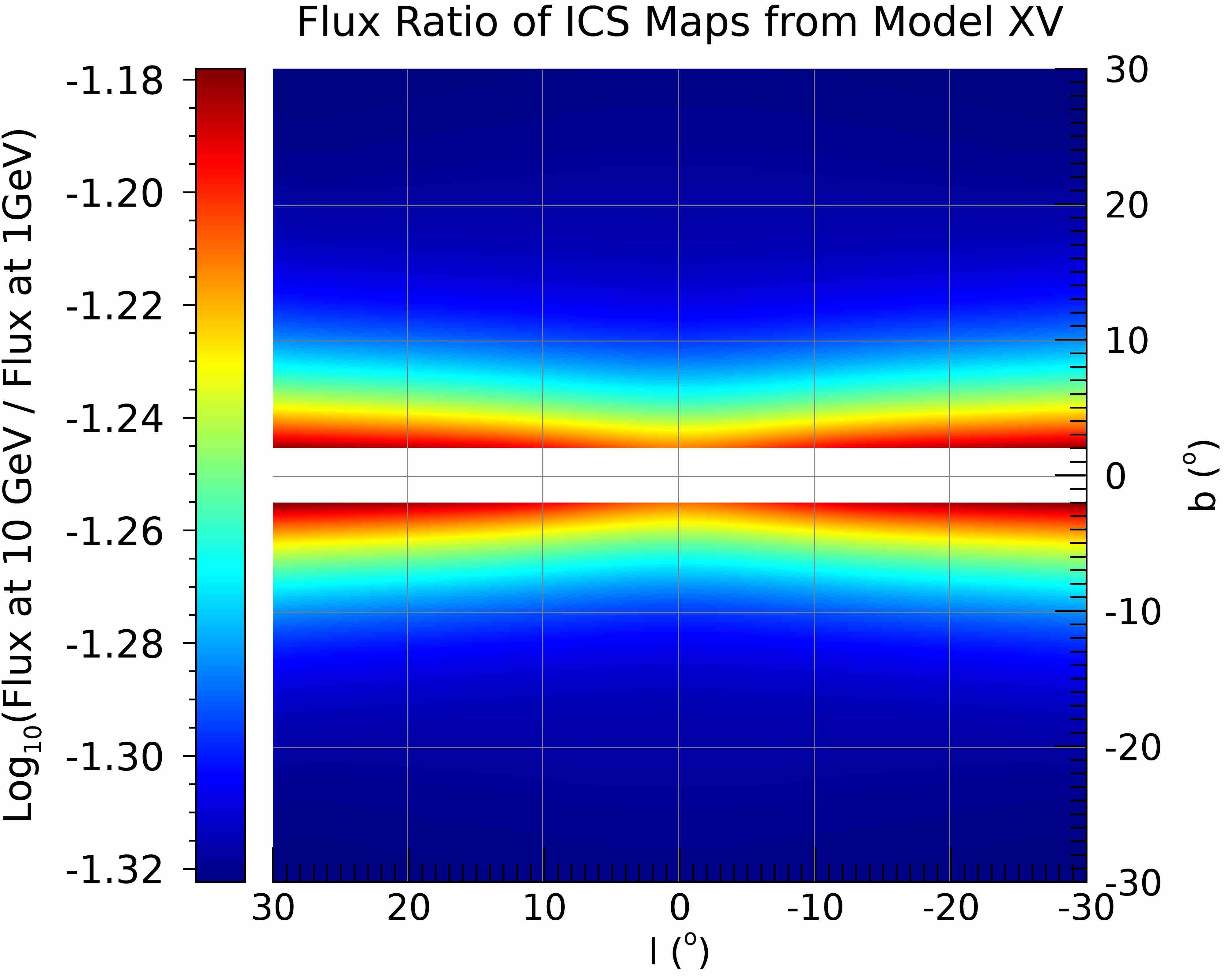}
\caption{Different physical assumptions on the ISM and CR injection conditions predict that 
the diffuse emission template component morphologies can deform with energy in a remarkably 
dissimilar manner. We show for the ICS component the ratio of maps for the predicted flux 
at 10 GeV to the flux at 1 GeV for the $60^{\circ} \times 60^{\circ}$ window. \textit{Left}: Model I,
\textit{Right}: Model XV.  We have included the fitting to the data normalizations. The PSF 
is not presented here in order to emphasize the underlying modeling differences.}
\label{fig:RatioOfTemplateFluxesWithEnergy}
\end{figure*}

As already mentioned, the assumptions on the propagation of cosmic-ray electrons are of great importance in
producing our templates. Electrons, unlike nuclei, suffer from fast energy losses via synchrotron radiation, ICS and 
bremsstrahlung emission. This results in extra modeling degrees of freedom that we need to account for.
Aside from the ISM gas, which affects the bremsstrahlung emission and energy losses, and the ISRF assumptions,
which affect the ICS emission and energy losses, the galactic magnetic field is also responsible for the 
synchrotron energy losses of the cosmic-ray electrons. These energy losses, while they do not result in emission 
at gamma-ray energies, can regulate how much remaining energy electrons have to emit in gamma rays. 
Larger values of the galactic magnetic field (i) suppress the modeled ICS and bremsstrahlung emission
normalizations, (ii) result in softer cosmic-ray electron spectra and subsequent ICS and bremsstrahlung
emission spectra and (iii) make the ICS and Bremss template components have a larger gradient as one moves from
the galactic disk. The subsequent fitting may to some extent absorb the first two of these effects by allowing for
larger values of relevant normalizations, but can not deform the template morphologies. 
We model the galactic magnetic field following the \texttt{GALPROP}  parameterization, 
\begin{equation}
B(r,z)  = B_{0} e^{-r/r_{c}} e^{-|z|/z_{c}}.
\label{eq:Bfield}
\end{equation}
In Tab.~\ref{tab:ModelsShort}, in the B-field column, the first three characters represent the value of $B_{0} \times 10$ 
in $\mu$G. The next three characters the value of $r_{c} \times 10$ in kpc and the last three characters the value of 
$z_{c} \times 10$ in kpc. As an example, Model I assumes $B_{0} = 20$ $\mu$G, $r_{c} = 3.0$ kpc and $z_{c} = 5.0$ kpc.
We allow for the $B_{0}$ normalization values from 2.5 to 20 $\mu$G: a factor of eight in this normalization translates
to an amplitude of synchrotron energy losses of a factor of 64. Moreover, we allow for $r_{c}$ to be 
within the range of $3-10$ kpc and $z_{c}$ of $2-5$ kpc. Together with the different choices for the ISRF and the
ISM gas, this allows us to test a great width of amplitudes and spatial profiles for cosmic-ray electrons energy 
losses. This affects the morphology of the ICS templates at any given energy, as is shown in Fig.~\ref{fig:RatioOfTemplates} 
(right panels), but also how the ICS templates morphology evolves with energy, as shown in 
Fig.~\ref{fig:RatioOfTemplateFluxesWithEnergy}. In that figure we plot the ratio of the ICS emission at 10 GeV to that 
at 1 GeV. Again, the normalization of these results reflect the fitting procedure described in Sec.~\ref{sec:method}.
As we increase the gamma-ray energy, the morphology of the ICS templates deforms from within a few $\%$ within the 
Bubbles regions (left panel of Fig.~\ref{fig:RatioOfTemplateFluxesWithEnergy} for model I), to more than 40$\%$ and 
along the galactic disk (right panel of Fig.~\ref{fig:RatioOfTemplateFluxesWithEnergy} for model XV).

\begin{figure*}[!ht]
\begin{centering}
\includegraphics[width=3.4in,angle=0]{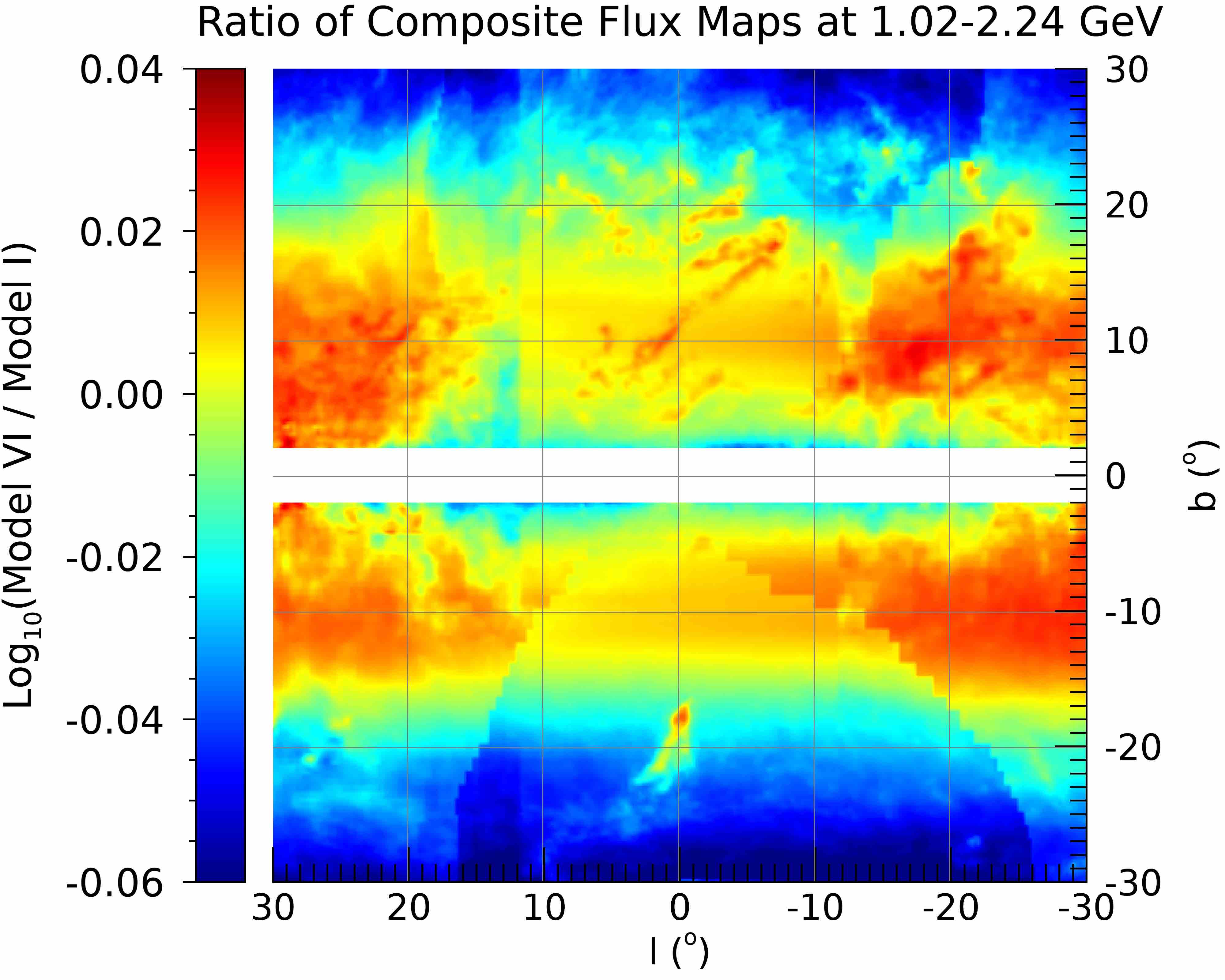} 
\hspace{0.1in}
\includegraphics[width=3.4in,angle=0]{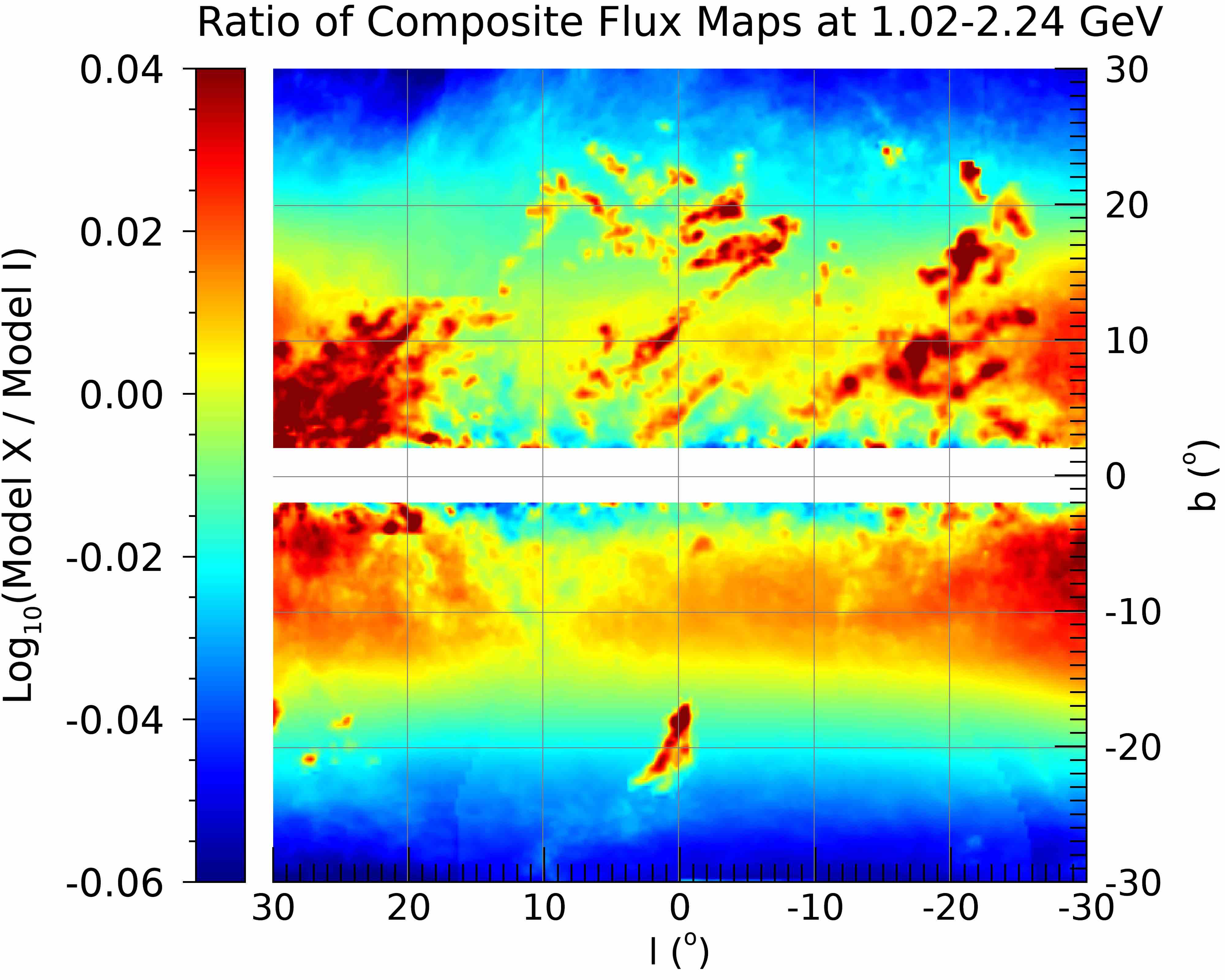}\\
\vspace{0.2in}
\includegraphics[width=3.4in,angle=0]{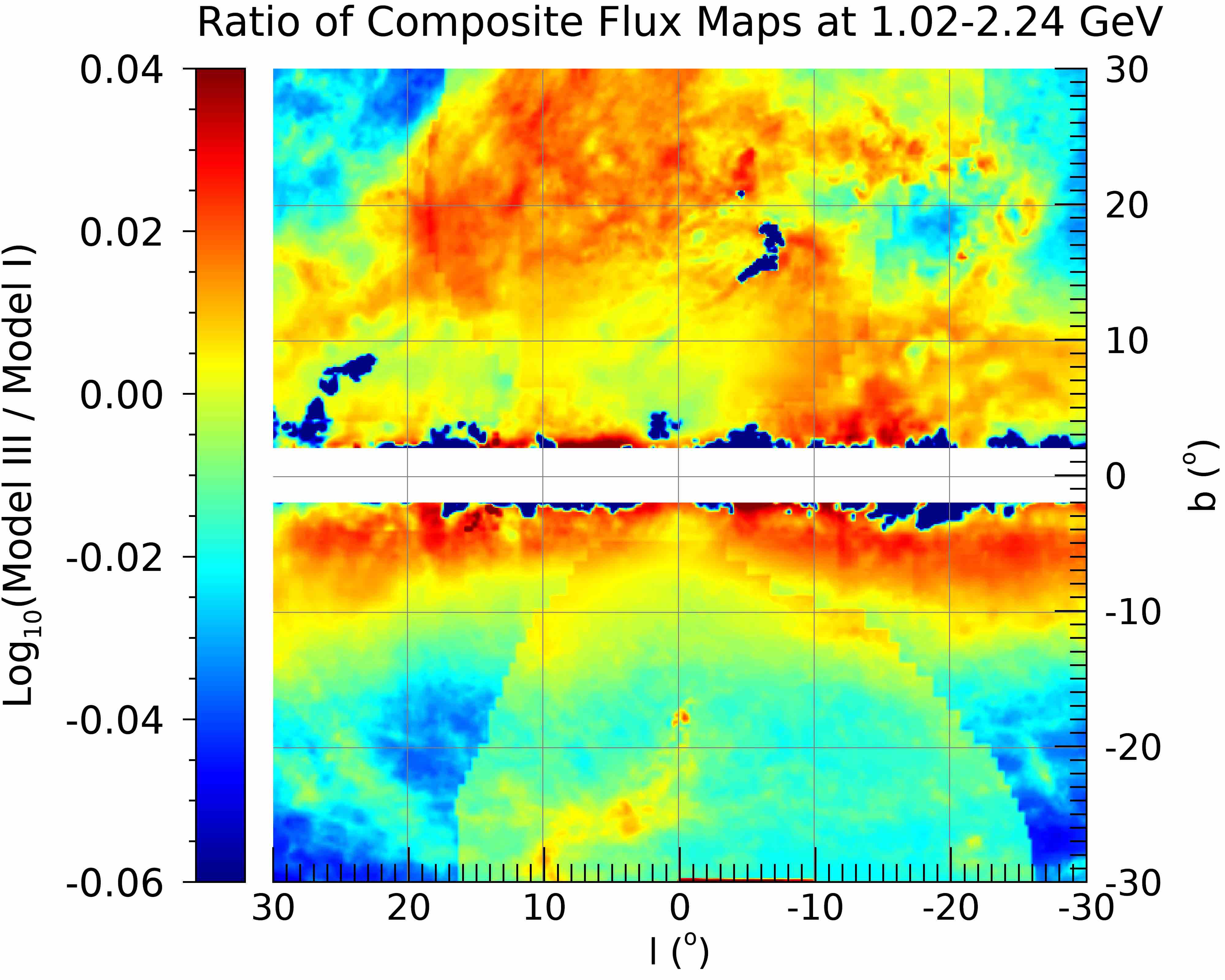} 
\hspace{0.1in}
\includegraphics[width=3.4in,angle=0]{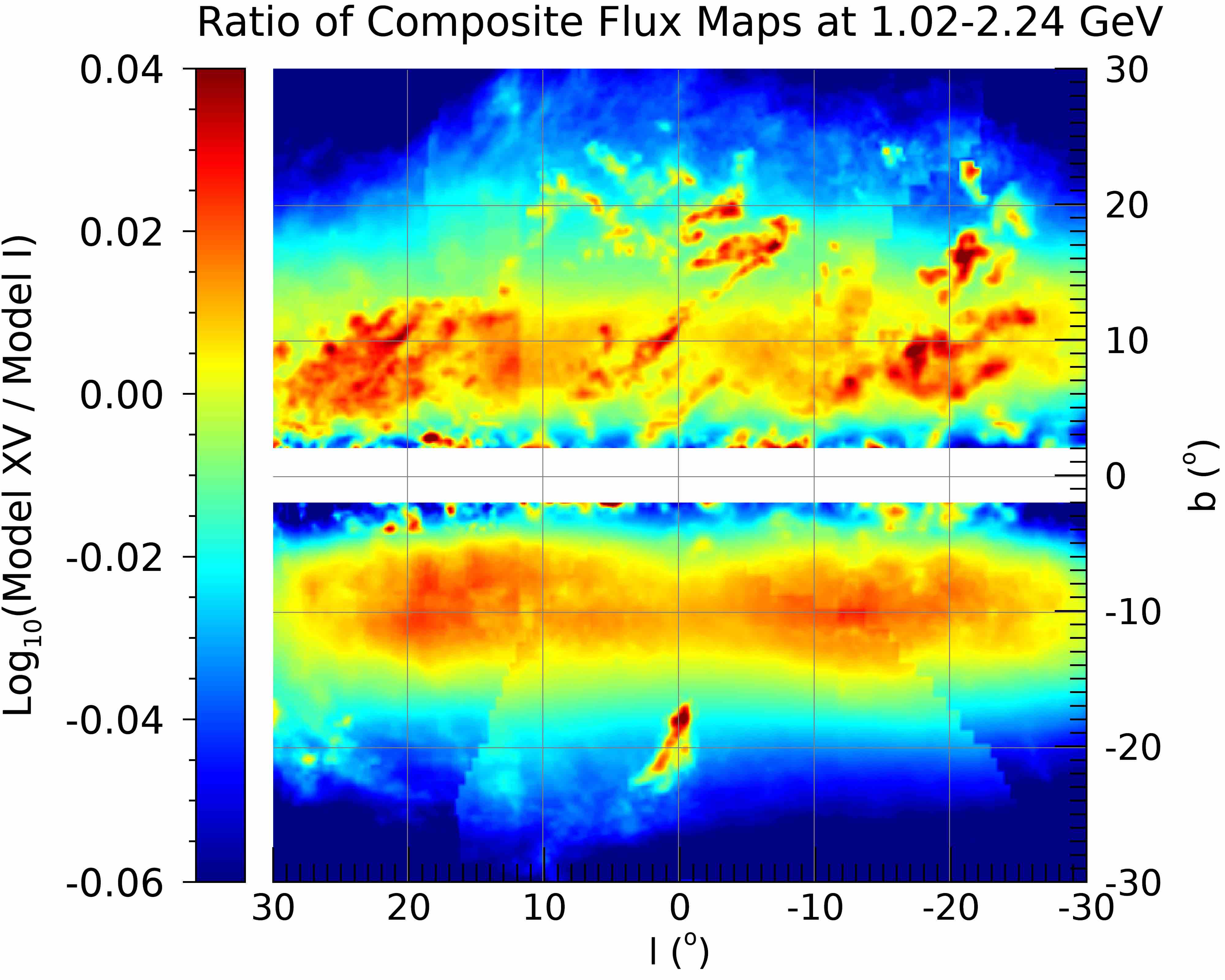} 
\end{centering}
\vspace{0.05in}
\caption{After performing the template fit to the \textit{Fermi} data, we can still observe
$\sim \mathcal O(10\%)$ differences in the predicted composite count maps. We show the ratio of such composite count 
maps produced from different underlying diffuse background models for the energy range of 
$1.02-2.24$ GeV. \textit{Top left}: Model VI to Model I, \textit{top right}: Model X to Model I, 
\textit{bottom left}: Model III to Model I, and \textit{bottom right}: Model XV to Model I.}
\label{fig:RatioOfFittedModels}
\end{figure*}  

Even after the fitting procedure, the template models can differ at the $\sim \mathcal O(10\%)$ level.
We depict representative variation between models in Fig.~\ref{fig:RatioOfFittedModels} 

We provide our galactic diffuse emission maps and the \texttt{GALPROP} input files used to generate them through \texttt{https://zenodo.org/record/6423495\#.YogWjS-ZPGI}

\subsection{Emission from the GCE}
\label{sec:GCE}

The morphology of the GCE is central to the question of its origin. As  our initial assumption, we 
take the GCE morphology to be spherical (see e.g. \cite{Abazajian:2014fta, Daylan:2014rsa, Calore:2014xka, Zhou:2014lva, TheFermi-LAT:2015kwa}).
We also take the GCE to have the profile expected from prompt
gamma-rays emitted by annihilating dark matter particles distributed in space with a profile following a
contracted NFW profile \cite{Navarro:1995iw},
\begin{equation}
\rho(r) = \frac{\rho_{0}}{  \left(r/r_c\right)^{\gamma}\left(1 + {r}/{r_{c}}\right)^{3-\gamma}},
\label{eq:DM}
\end{equation}
such that the GCE morphology scales like the line-of-sight integral of $\rho(r)^{2}$. We fix $r_{c} = 20$ kpc.
We take as a central choice $\gamma = 1.2$, though in Sec.~\ref{sec:results} we explore different choices for the
``cuspiness'' of the NFW profile. The normalization for $\rho_{0}$ is always taken such that at $r = 8.5$ kpc
the density $\rho(r=8.5 \textrm{kpc}) = 0.4 $ GeV/cm$^3$, in agreement with the estimates of the local dark
matter density \cite{Catena:2009mf, Salucci:2010qr}. Recent measurements suggest that the galactocentric distance
is closer to 8.3 kpc \cite{2021arXiv210112098G}, but we retain the choice of 8.5 kpc as this is the assumption
used in \texttt{GALPROP} \cite{GALPROPSite} to produce the galactic diffuse emission maps. Changing the
galactocentric distance to its updated smaller value would result in both the GCE and the diffuse templates
increasing in brightness by $\sim 5\%$, which is smaller than the uncertainty on the local dark matter density
\cite{Gardner:2021ntg}. This change would also be absorbed when fitting to the gamma-ray observations,
as all templates have free normalizations.

For our core analysis, we take the NFW profile to be centered at the exact location of the galactic center. 
We test whether the GCE is indeed spherically symmetric or if instead mild prolateness
(extension along the plane of the galactic disk) or oblateness (extension perpendicular to the plane of the 
galactic disk) is favored. 
We will do this by deforming the opening angle from the galactic center $\psi$ of the NFW template by a factor $\epsilon$,
such that cos($\psi$)=cos($b$)cos($\ell/\epsilon$). We get the regular spherically symmetric NFW for $\epsilon = 1$. A value of 
$\epsilon > 1$ gives elongation along the galactic disk, while $\epsilon < 1$ perpendicular to the disk.
In Sec.~\ref{sec:Uncertainties}, to estimate the size of systematic astrophysical uncertainties along the galactic disk, 
we allow for that additional template to be translated in longitude from the center of the galaxy. For the main part of this
work however, the GCE is centered at $\ell=0^{\circ}, b=0^{\circ}$ and we refer to it as the GCE only in the analysis for which
it is located at the galactic center. 

\subsection{Isotropic Emission}
\label{sec:Iso}

In addition to the previously mentioned diffuse emission templates, we include the isotropic gamma-ray 
emission template. This is a flat, homogeneous flux that accounts for the combined emission 
from: unresolved extragalactic point sources, like distant galaxies or gamma-rays originating from ultra-high 
energy cosmic rays; a high-latitude component of galactic point sources that are below the detection 
threshold; and misidentified cosmic rays. The isotropic gamma-ray spectrum has been measured 
in \cite{Ackermann:2014usa}. Like with the Bubbles spectrum, we take the central value of the isotropic 
spectrum as an input, and we add a statistical penalty in our fits for deviations from the central value in each 
energy bin. The size of the penalty is based on the errors reported by \cite{Ackermann:2014usa}. We add the reported 
statistical and systematic errors of \cite{Ackermann:2014usa} in quadrature.

\subsection{Total Gamma-Ray Diffuse Emission Models}
\label{sec:TotalGR}

Every particular galactic diffuse emission model determines the Pi0, Bremss and ICS templates in space and energy. We add the Pi0, Bremss, ICS, Bubbles, GCE and Isotropic emission templates 
to create a composite map at a given energy. In each energy bin, the coefficients will be simultaneously
fit to the \textit{Fermi} data as described in Sec.~\ref{sec:method}, but each energy bin will be fit
independently. As described above, the Pi0, Bremss and ICS each have an energy-dependent morphology.
These morphologies change as the propagation properties of cosmic rays change with energy.
This affects their spatial distribution and in turn the rate of cosmic-ray 
interactions with the ISM modeled in creating the templates. The Bubbles, GCE and Isotropic emission 
templates are energy-independent. As described in Sec.~\ref{sec:CRtoGR}, the Pi0 and Bremss components are added with the 
same normalization. The Pi0 + Bremss component and the ICS component are allowed to float freely. Thus, for a given choice of
diffuse emission model we have the following definition of a composite diffuse flux map,
\begin{eqnarray}
\Phi_{\textrm{Tot}}^{\textrm{Diff}}(\ell, b, E)  &=& c_{\textrm{gas}} (E) \left[\Phi_{\textrm{Pi0}}(\ell, b, E) 
+  \Phi_{\textrm{Bremss}}(\ell, b, E)\right] \nonumber \\
&+& c_{\textrm{ICS}} (E) \Phi_{\textrm{ICS}}(\ell, b, E) \nonumber \\
&+& c_{\textrm{Bub}} (E) \Phi_{\textrm{Bubbles}}(\ell, b, E) \nonumber \\
&+& c_{\textrm{Iso}} (E) \Phi_{\textrm{Iso}}(\ell, b, E) \nonumber \\
&+& c_{\textrm{GCE}} (E) \Phi_{\textrm{GCE}}(\ell, b, E) .
\label{eq:CompositeMap}
\end{eqnarray}
The gamma-ray flux at a particular sky location and in a given energy bin is thus given by specifying
the five normalization parameters $c_i$. We note that by $\Phi_{\textrm{Tot}}^{\textrm{Diff}}(\ell, b, E)$ we mean the integral of the differential diffuse flux across an energy bin: $\Phi_{\textrm{Tot}}^{\textrm{Diff}}(\ell, b, E) = \int dE \, d \Phi_{\textrm{Tot}}^{\textrm{Diff}}(\ell, b, E)/dE$.

\section{Data, Masks, and Methods}
\label{sec:method}

\begin{figure}[!t]
\begin{centering}
\hspace{-0.1in}
\includegraphics[width=3.4in,angle=0]{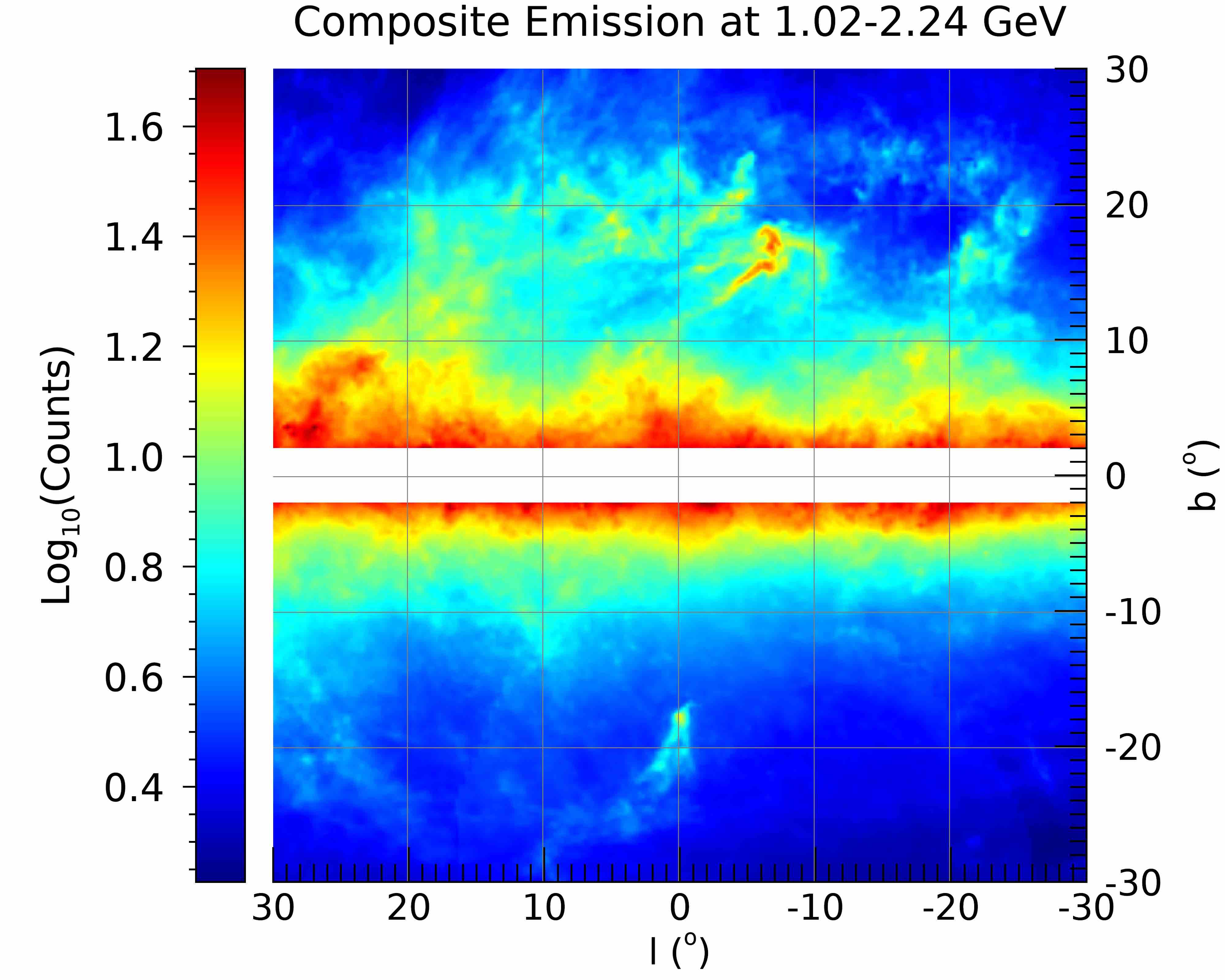}
\end{centering}
\vspace{0.00in}
\caption{A composite counts map after summing up all diffuse emission template components and 
before including the instrumental PSF. This map includes the exposure of the \textit{Fermi}-LAT 
instrument. The emission from point sources is not included as those are masked in the fitting
process.}
\label{fig:HydrogenSpect}
\end{figure}  

\subsection{Gamma-Ray Data}
\label{sec:data}

We use Fermi Pass 8 data, version P8R3, recorded from Aug 4 2008 to April 14 2021, corresponding to 
weeks $9-670$ of \textit{Fermi}-LAT observations\footnote{\url{https://fermi.gsfc.nasa.gov/ssc/data/access/}}.
We use \textit{Fermi} {\tt ScienceTools P8v27h5b5c8} for selection cuts and in order to calculate the relevant exposure-cube files 
and exposure maps\footnote{\url{https://fermi.gsfc.nasa.gov/ssc/data/analysis/}}, which allow us to pass from
fluxes to expected counts. We show an example of an expected counts map in Fig.~\ref{fig:HydrogenSpect}, where we have included 
the instrumental exposure after 12.5 years of observations. The exposure is calculated for each 
pixel using the \textit{Fermi} {\tt ScienceTools P8v27h5b5c8}.

We keep only FRONT-converted {\tt CLEAN} data. In addition, we set the following filters: {\tt zmax = $100^{\circ}$}, 
{\tt DATA$\_$QUAL==1}, {\tt LAT$\_$CONFIG==1}, and {\tt ABS(ROCK$\_$ANGLE) < 52}. Our 
data maps are centered at the galactic center and cover a square window of $60^{\circ}$ per side 
in galactic coordinates in Cartesian pixels of size $0.1^{\circ} \times 0.1^{\circ}$. 
Unlike a {\tt HEALPix} pixelization, our pixels do not have equal area, but we account for this in our fits.

We bin the gamma-ray data in 14 energy bins spanning energies from 0.275 to 51.9 GeV, given in the first column of Tab.~\ref{tab:PSF_vsE}.
The first eleven energy bins have a constant log width; because the gamma-ray flux drops at higher energy, the final three energy bins are wider so that each bin has a similar statistical impact in our fits.


\setlength{\tabcolsep}{6pt}
\begin{table}
    \begin{tabular}{crrc}
    \hline 
            $E_{\textrm{min}}-E_{\textrm{max}} {\rm\,[GeV]}$ &  $\theta_{s} [^{\circ}]$ &  $\theta_{l} [^{\circ}]$ & Masked fraction\\
            \hline
            \hline 
             $0.275-0.357$ & 1.125 & 3.75 & 71.8\%\\
             $0.357-0.464$ &  0.975 & 3.25 & 62.9\%\\
             $0.464-0.603$ & 0.788 & 2.63 & 52.2\%\\
             $0.603-0.784$ & 0.600 & 2.00 & 38.5\%\\
             $0.784-1.02$ &  0.450 & 1.50 & 29.2\%\\
             $1.02-1.32$ & 0.375 & 1.25 & 23.4\%\\
             $1.32-1.72$ & 0.300 & 1.00 & 19.0\%\\
             $1.72-2.24$ & 0.225 & 0.750 & 16.3\%\\
             $2.24-2.91$ & 0.188 & 0.625 & 13.0\%\\
             $2.91-3.78$ & 0.162 & 0.540 & 12.9\%\\
             $3.78-4.91$ & 0.125 & 0.417 & 11.6\%\\
             $4.91-10.8$ & 0.100 & 0.333 & 11.5\%\\
             $10.8-23.7$ & 0.060 & 0.200 & 10.3\%\\
             $23.7-51.9$ & 0.053 & 0.175 & 10.3\%\\
             \hline 
        \end{tabular}
       \caption{The energy bins and the energy-dependence of the ``small'' ($\theta_{s}$) 
       and ``large''  ($\theta_{l}$) radii used to mask known point sources. The last column shows the fraction of pixels masked in our standard mask (4FGLDR2+disk) with respect to the total number of pixels in the inner $40^\circ \times 40^\circ$ Galactic center region ($1.6\times 10^5$ pixels).}
    \label{tab:PSF_vsE}
\end{table}

\subsection{Masks and PSF}
\label{sec:masks}

\begin{figure*}[!ht]
\includegraphics[width=2.18in,angle=0]{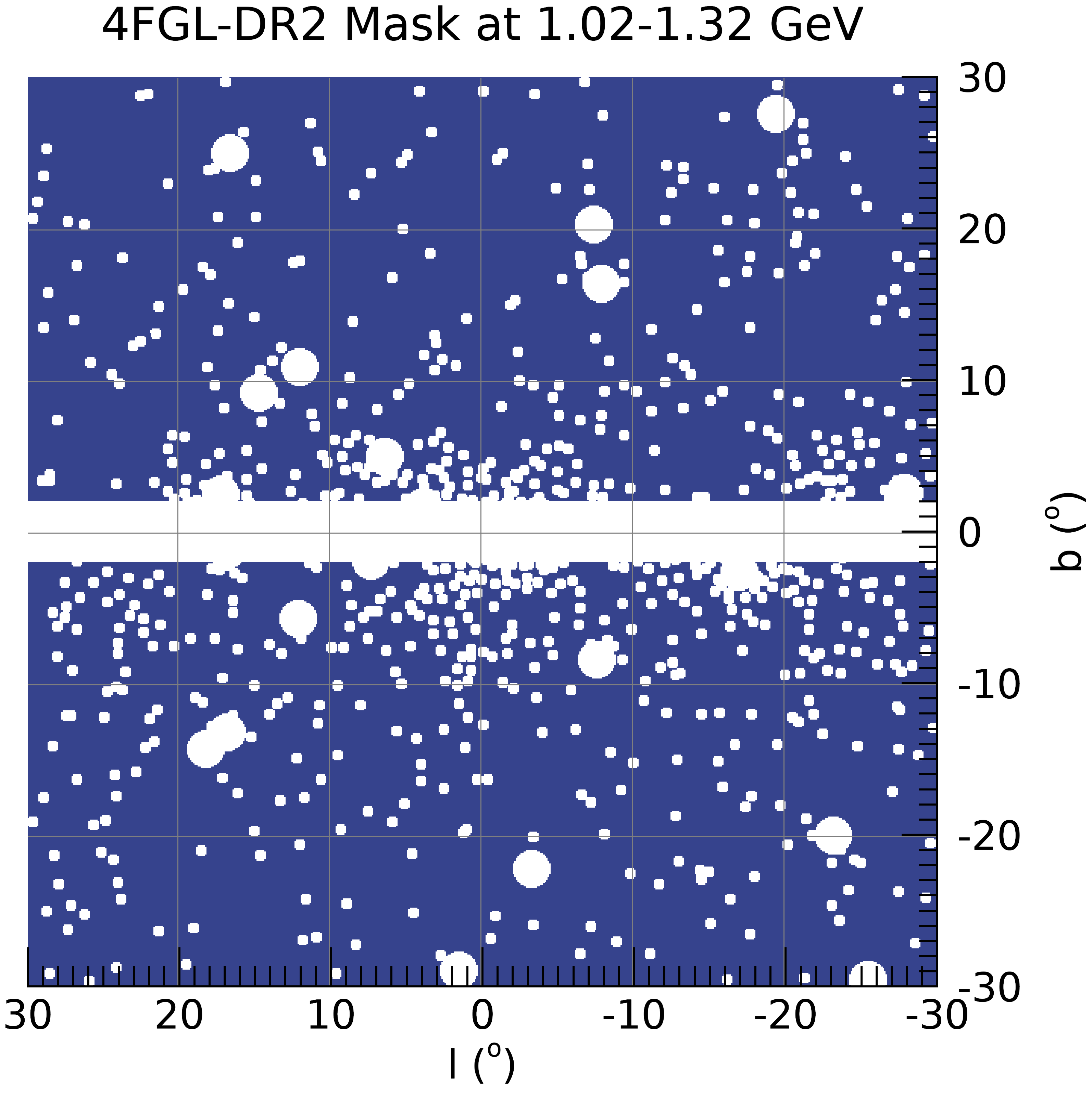} 
\hspace{0.08in}
\includegraphics[width=2.18in,angle=0]{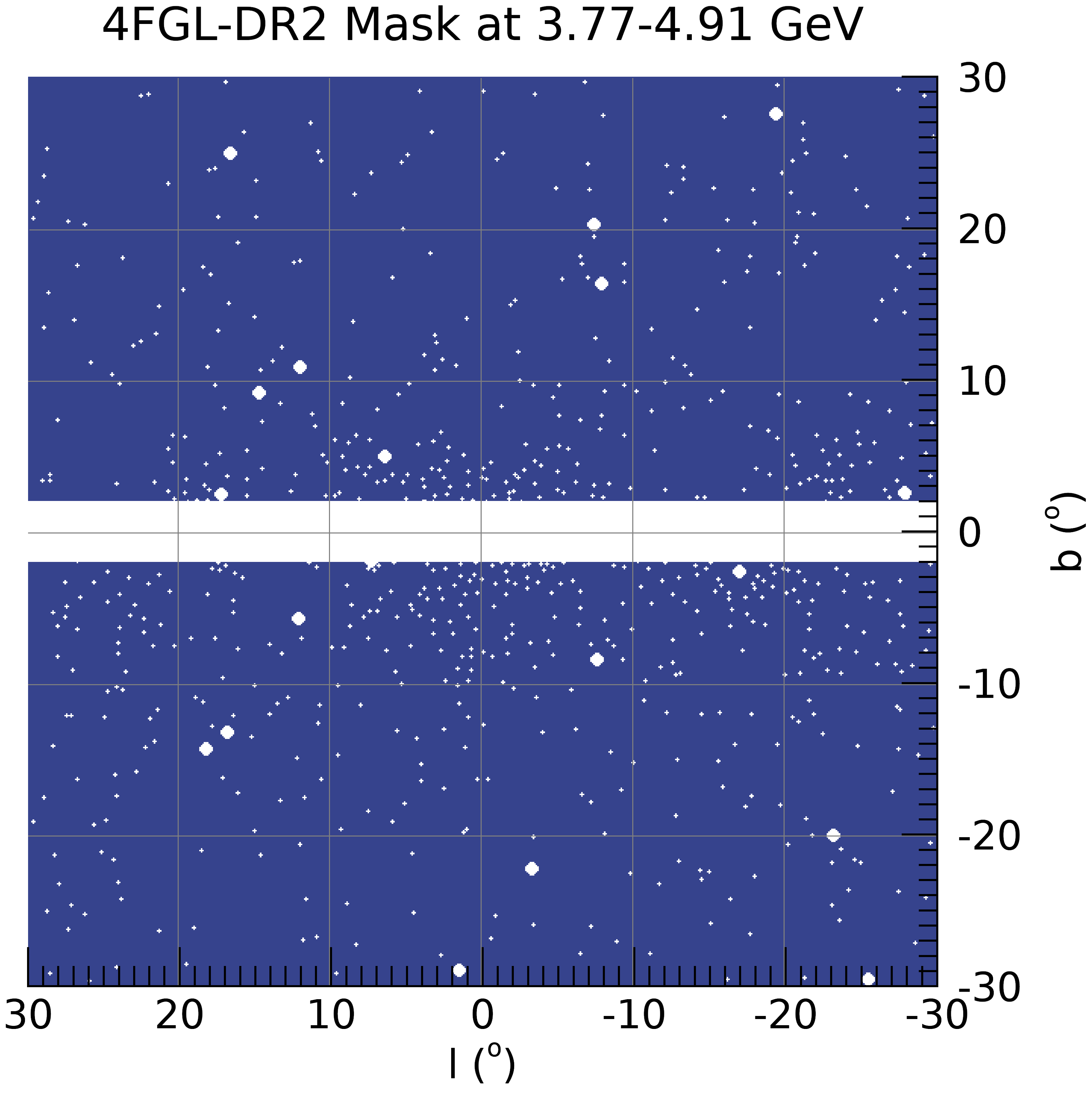}
\hspace{0.08in}
\includegraphics[width=2.18in,angle=0]{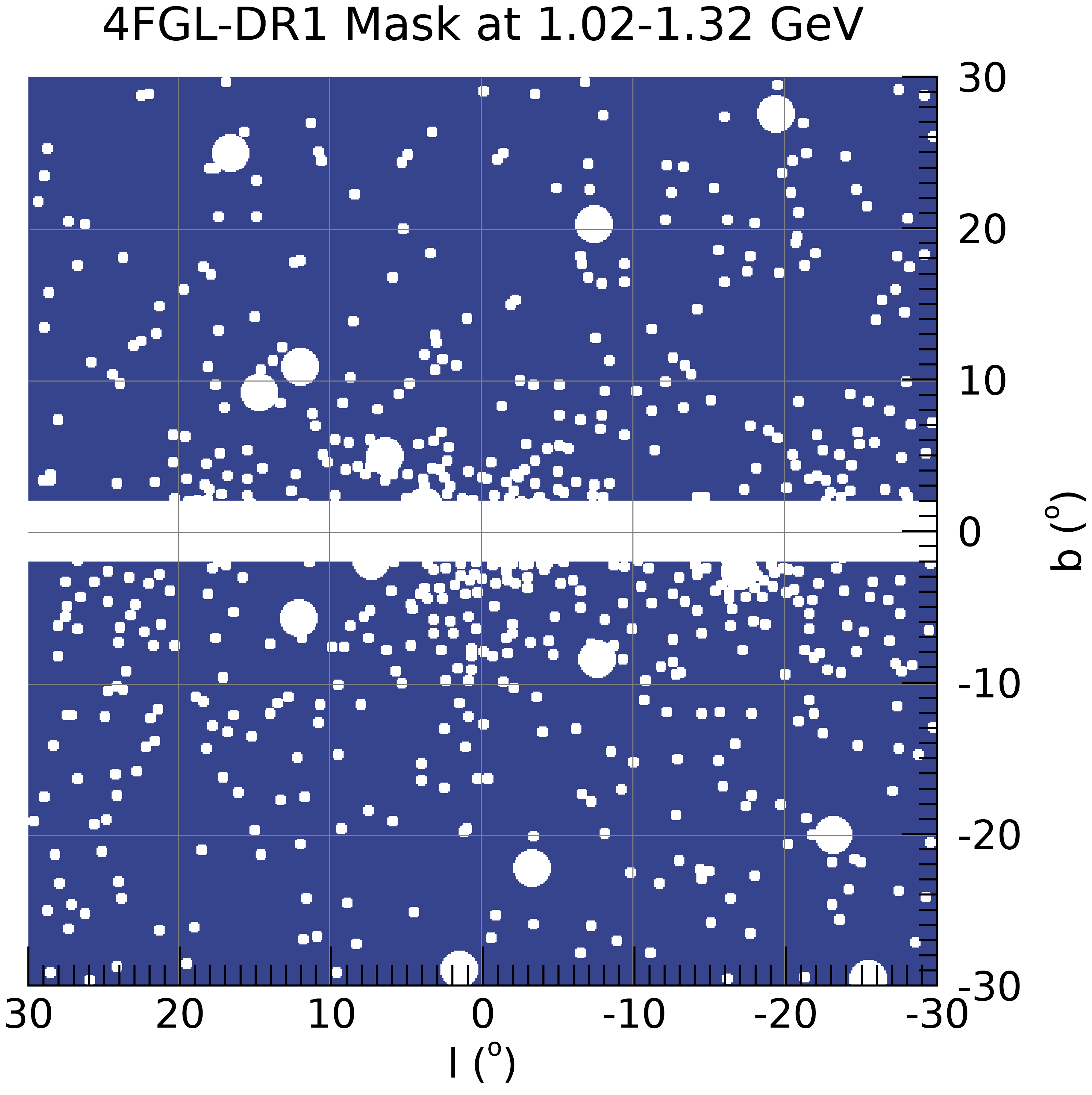} 
\vspace{0.00in}
\caption{The 4FGL-DR2 masks for two energy bins (\textit{left} and \textit{center}) and (\textit{right}) the  4FGL-DR1 mask for the lower 
energy bin at $1.02-1.32$ GeV. White pixels are masked out and not included on the fits.}
\label{fig:Masks}
\end{figure*} 

For our default analysis, we mask the inner $2^{\circ}$ of latitude, i.e.,~$|b|<2^{\circ}$, at all longitudes
and all energies. This excludes the very bright galactic disk emission. It also masks undetected
point sources in that part of the sky. Although the GCE emission plausibly peaks within this latitude range, masking the 
galactic disk allows us to perform a search in a part of the sky where the backgrounds are better controlled 
and the GCE signal to noise is better determined. We note that the detection threshold of point sources increases at lower 
latitudes $|b|$, so relatively bright undetected point sources may potentially bias fits that include this region. 
We explore the impact of variations of this mask in App.~\ref{app:Ellipticity}.

We additionally mask all point sources in the 4FGL-DR2 catalog~\cite{Ballet:2020hze},
as described here. Within the window of $60^{\circ} \times 60^{\circ}$, including the galactic disk,
there are 918 point sources and 26 extended sources in the 4FGL-DR2 catalog. Of these 944 sources,
25 have a detection significance as measured by a \textit{Fermi}-LAT test statistic (TS) of 49 or more,
which makes them very bright. At lower energies the \textit{Fermi}-LAT instrument has a poorer angular resolution,
so the emission from ``point'' sources in fact will appear in several pixels in an energy-dependent way.
To account for these effects, the angle at which we mask a given point source is energy- and significance-dependent:
it is larger at lower energies and for brighter sources, as we show in Fig.~\ref{fig:Masks} for two different energies.
We choose to have a ``small'' and a ``large'' mask radius for sources with TS$~\leq 49$ or TS$~>49$, respectively, labeled $\theta_s$ and $\theta_l$.
The values of $\theta_s$ and $\theta_l$ are given at all energies in Tab.~\ref{tab:PSF_vsE}.
For the point spread function (PSF), we use the Fermi-LAT double-Moffat-function profile with parameters specified in~\cite{psf-vals}.
Since we are interested in the GCE  and its possible point source origin, we also test with the older 4FGL point source catalog that was derived after 8 years of observations \cite{Fermi-LAT:2019yla}, which we name ``4FGL-DR1''. The  4FGL-DR1 catalog 
excludes fewer pixels for a given energy bin than those of the same energy bin from the 4FGL-DR2 catalog, as shown in the right panel of Fig.~\ref{fig:Masks}. We explore the impact of this and other variations of the point source masking procedure in App.~\ref{app:Ellipticity}, leaving a full exploration of this topic for future work.

\subsection{Template Fitting}
\label{sec:fit}

For each of our 14 energy bins, we start by constructing the composite diffuse flux map $\Phi_{\textrm{Tot}}^{\textrm{Diff}}(\ell, b, E|\{c\})$ 
of Eq.~\ref{eq:CompositeMap} in the $60^{\circ} \times 60^{\circ}$ inner galaxy region, where $\{c\}$ refer to the template normalizations.
We then multiply by the energy-dependent exposure map, referred to as $\mathcal E (\ell, b, E)$, as described in Sec.~\ref{sec:data}.
This provides the map of expected counts due to all emission components,
\begin{eqnarray}
\mathcal C^0 (\ell, b, E|\{c\})  = \Phi_{\textrm{Tot}}^{\textrm{Diff}}(\ell, b, E|\{c\}) \cdot \mathcal E (\ell, b, E).
\label{eq:CountsNoPSF}
\end{eqnarray}
We then multiply the expected counts by the fraction of FRONT-converted {\tt CLEAN} events 
and convolve with the energy-dependent PSF $\mathcal P (E)$ for FRONT-converted {\tt CLEAN} 
events, following \cite{psf-vals} and \cite{Zhong:2019ycb}. This gives,
\begin{equation}
\mathcal C^{\mathcal P}(\ell, b, E|\{c\}) = \mathcal C^0(\ell, b, E|\{c\}) \ast \mathcal P(E) ,
\label{eq:CountsWithPSF}
\end{equation}
where $\ast$ indicates a convolution at every point.
Finally, we  multiply the convolved maps by the mask, referred to as $M(\ell, b, E)$, which accounts for the fact that
we do not model all sky locations. In this case, the expectation for the counts at a given pixel and energy is,
\begin{equation} \label{eq:ExpectedCounts}
\mathcal C(\ell, b, E|\{c\}) = \mathcal C^{\mathcal P}(\ell, b, E|\{c\}) \cdot M(\ell, b, E).    
\end{equation}
$\mathcal C(\ell,b,E)$ in Eq.~\ref{eq:ExpectedCounts} is the quantity that we will compare against the masked data map $\mathcal D(\ell,b,E)$.

All maps are pixelized in $0.1^{\circ} \times 0.1^{\circ}$ pixels. Considering the mask along the Milky Way disk,
and the fact that additional pixels are subject to point source masking, there are at most $3.36 \times 10^{5}$ 
pixels per energy bin. To reduce the computational expense but maintaining our sensitivity to the sky
region of greatest interest, we further restrict to a $40^\circ \times 40^\circ$ 
inner region, which reduces the maximum number of pixels for each energy bin to $1.44\times 10^5$.
In the fit, we will have discrete labels for each pixel,
\begin{align} \label{eq:ExpectedCountsPixel}
    \mathcal C (\ell, b, E|\{c\}) &\to \mathcal C_{j,p}(\{c\})
    \\ \mathcal D (\ell, b, E | \{c\}) &\to \mathcal D_{j,p}(\{c\}),
\end{align} 
where $j$ refers to the energy bin and $p$ to the pixel number. We use \texttt{emcee}~\cite{ForemanMackey:2012ig}, a Markov chain Monte Carlo (MCMC) program, to fit the coefficient parameters $c_{\textrm{gas},j} $
, $c_{\textrm{ICS},j} $
, $c_{\textrm{Bub},j} $
, $c_{\textrm{Iso},j} $
, and $c_{\textrm{GCE},j} $ 
of Eq.~\ref{eq:CompositeMap}, which are inherited by $\mathcal C_{j,p}$.
Because the data $\mathcal D_{j,p}$ are given by counting a discrete number of events,
the likelihood of a given model is a Poisson likelihood, and the best fit model will maximize this
quantity. Using {\tt emcee}, we in fact seek the parameters that minimize the negative log-likelihood,
\begin{align}
- 2 \ln(\mathcal{L}|\{c\})_j ={}& 
2 \sum_{p} \left[\mathcal C_{j,p} + \ln(\mathcal D_{j,p}!) - \mathcal D_{j,p} \ln \mathcal C_{j,p}  \right]
\nonumber  \\
{}& + \chi_{\textrm{Bubbles},j}^{2} + \chi_{\textrm{Iso},j}^{2},
\label{eq:Likelihood}
\end{align}
where the index $p$ goes over unmasked pixels. The ``external $\chi^2$'' functions
$\chi^2_{\textrm{Bubbles},j}$ and $\chi^2_{\textrm{Iso},j}$ are constraints that act as 
penalties when the $c_{\textrm{Bub},j}$ and $c_{\textrm{Iso},j}$ values in the fit deviate too much from their 
spectra measured at high latitudes \cite{Fermi-LAT:2014sfa,Ackermann:2014usa}. For our \texttt{emcee}
runs, we use the \texttt{EnsembleSampler} with 100 walkers and 1000 steps to assess convergence.
After getting the MCMC chains, we discard the first 300 steps and draw posterior probabilities
for the parameters using \texttt{ChainConsumer}~\cite{Hinton2016}.
For many of our fits we allow the normalization of the GCE, $c_{\textrm{GCE},j}$, to be negative, but we always require the total modeled masked counts, $\mathcal C_{j,p}(\{c\})$, to be non-negative for all pixels.

The total log-likelihood for a model is given by a sum over all the energy bins, 
\begin{equation}
    \ln(\mathcal{L}|\{c\}) = \sum_j \ln(\mathcal{L}|\{c\})_j.
    \label{eq:LikelihoodSum}
\end{equation}
We stress that we fit each energy bin independently, but we use the total log-likelihood in Eq.~\ref{eq:LikelihoodSum}
to assess the best fit parameters for a particular model and to compare competing models.
 
\section{The GCE after the Template Fits}
\label{sec:results}

\begin{figure*}[t]
\hspace{-0.2in}
\includegraphics[width=2.44in,angle=0]{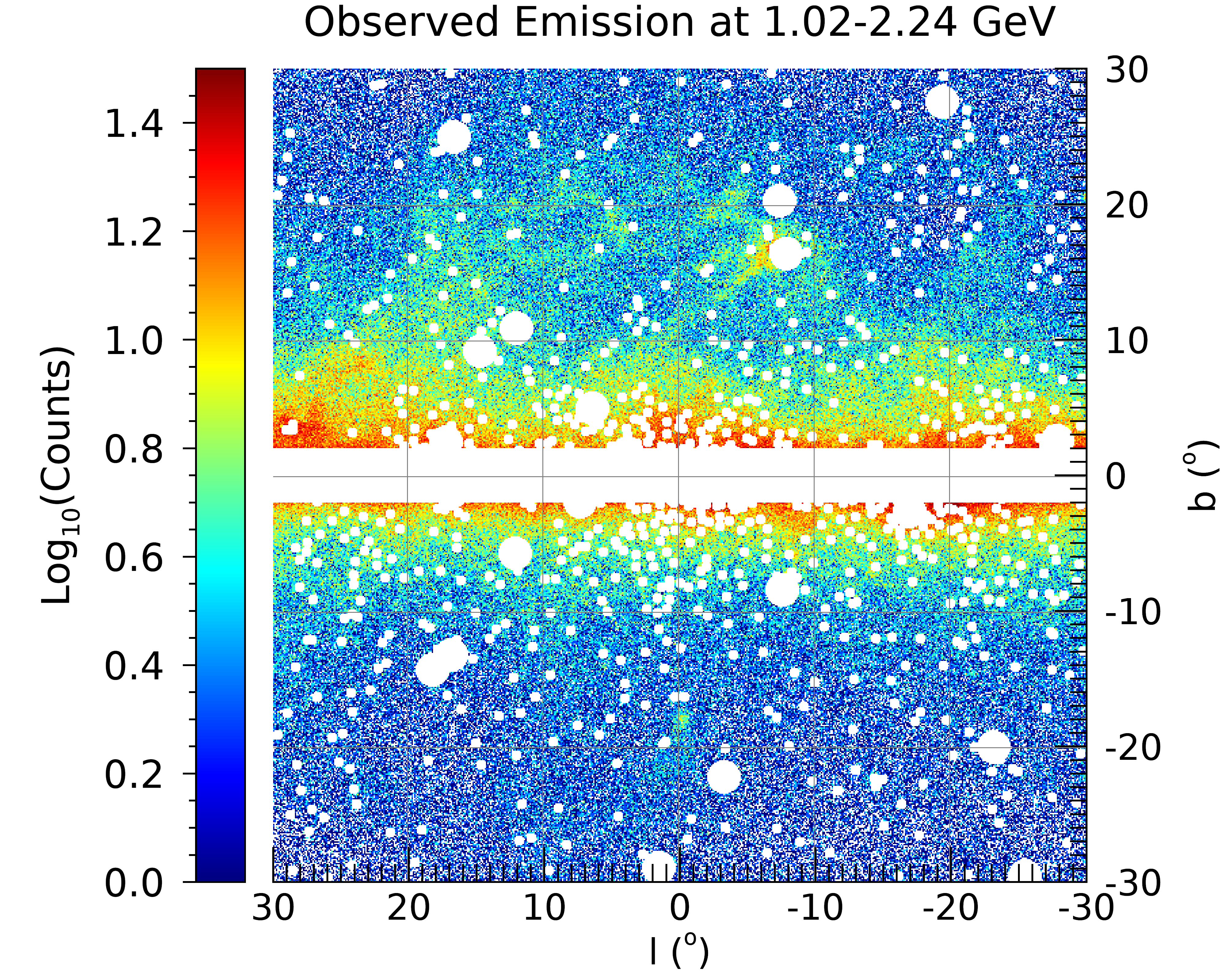} 
\hspace{-0.2in}
\includegraphics[width=2.44in,angle=0]{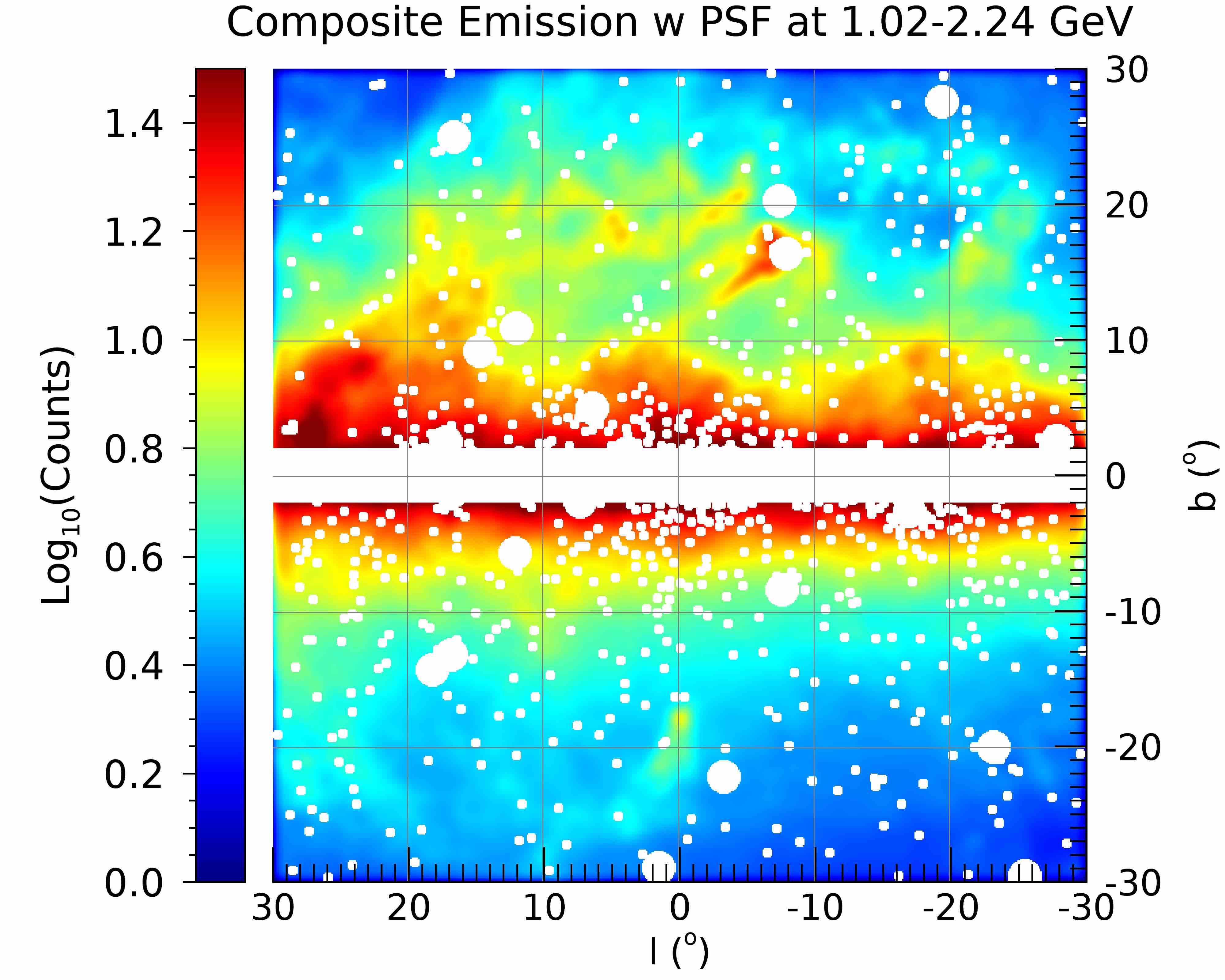} 
\hspace{-0.2in}
\includegraphics[width=2.44in,angle=0]{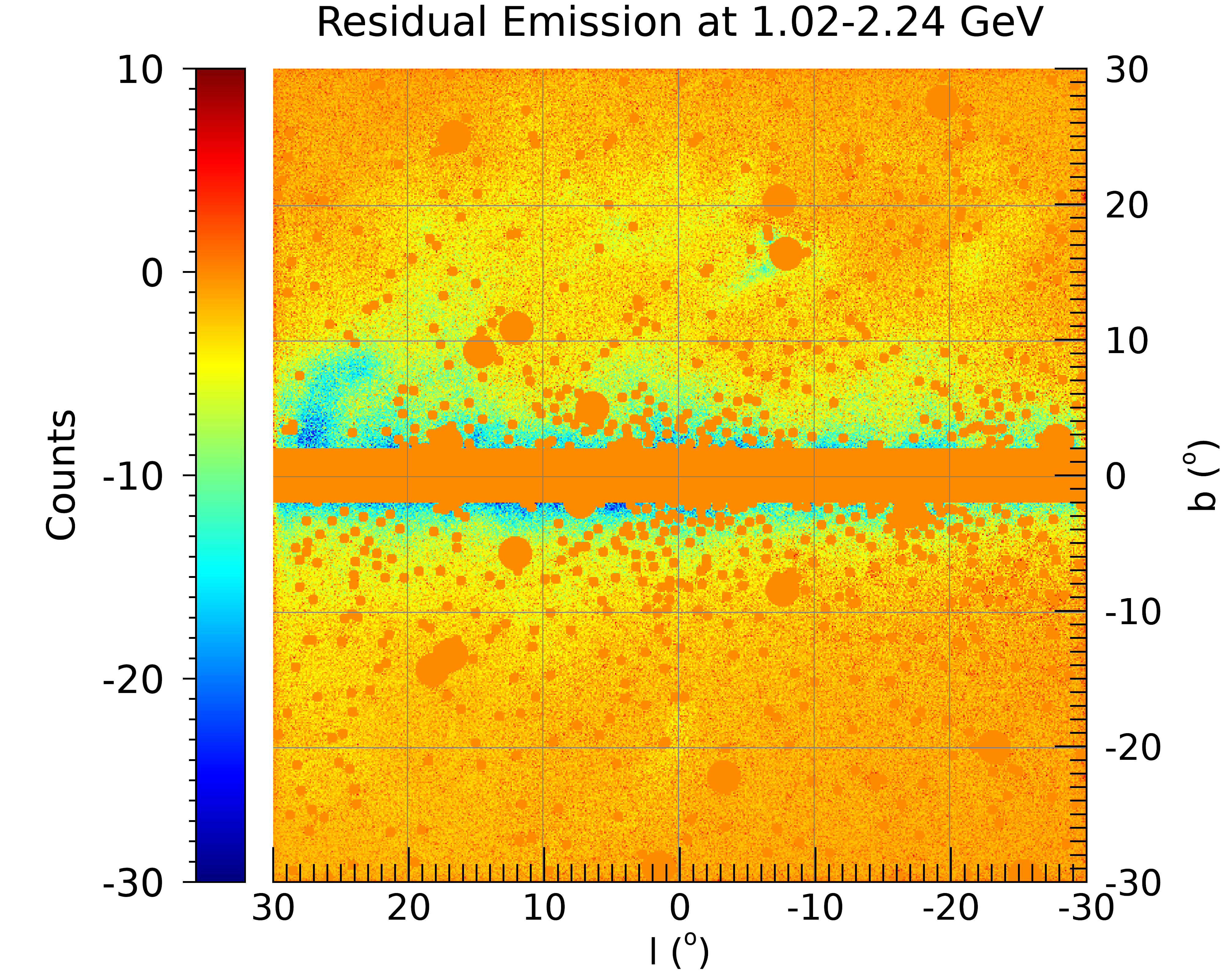}\\
\vspace{0.2in}
\hspace{-0.2in}
\includegraphics[width=2.44in,angle=0]{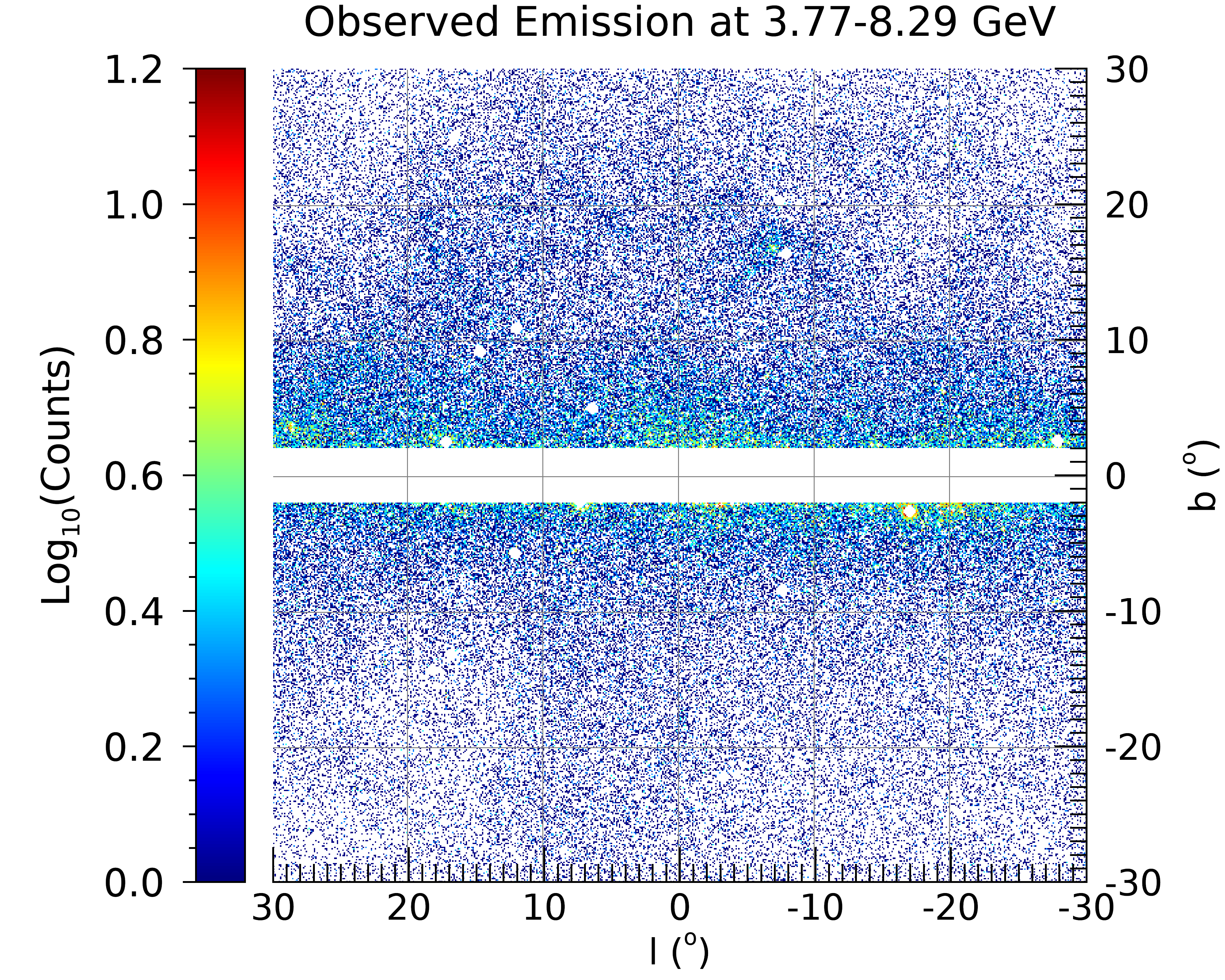} 
\hspace{-0.2in}
\includegraphics[width=2.44in,angle=0]{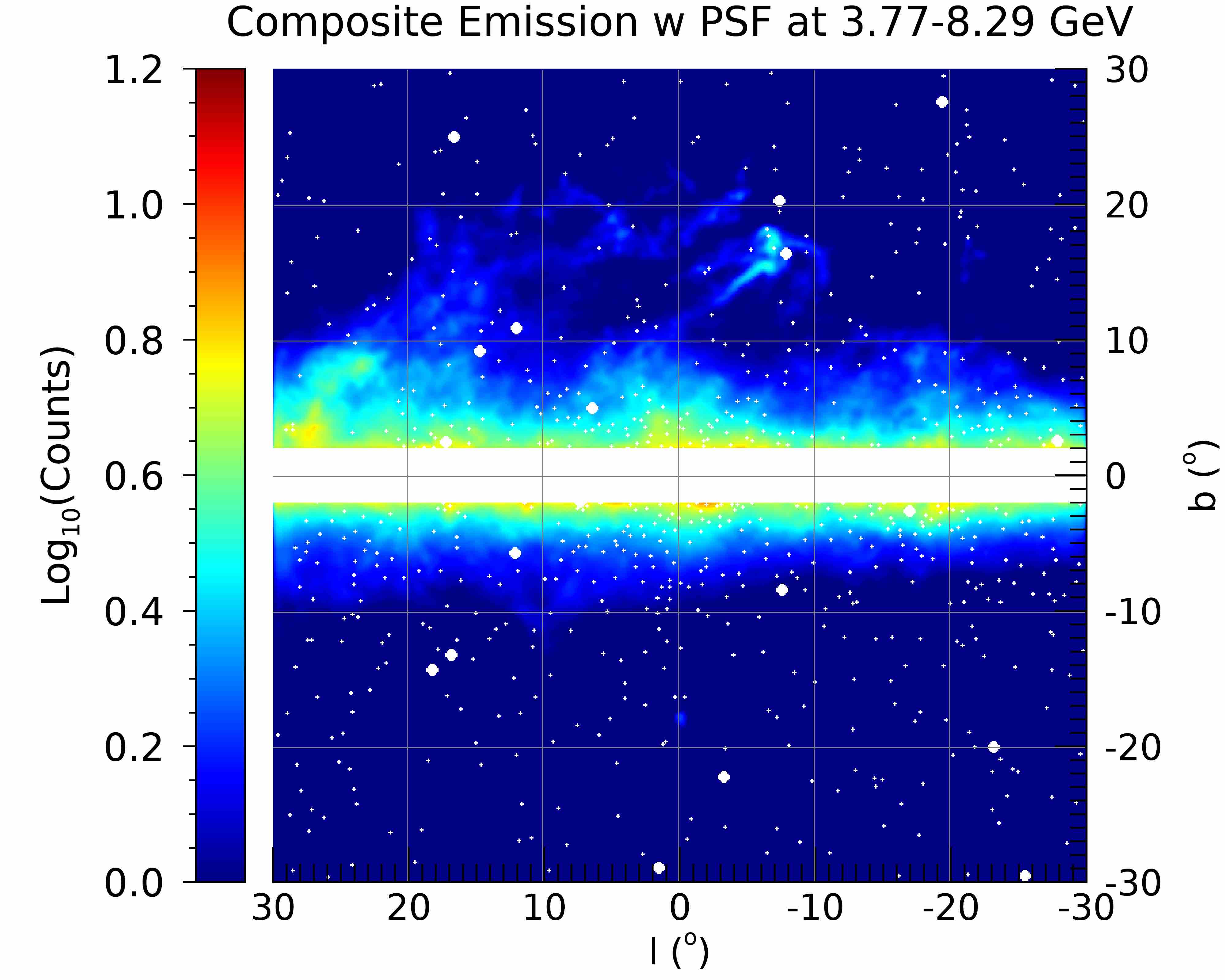} 
\hspace{-0.2in}
\includegraphics[width=2.44in,angle=0]{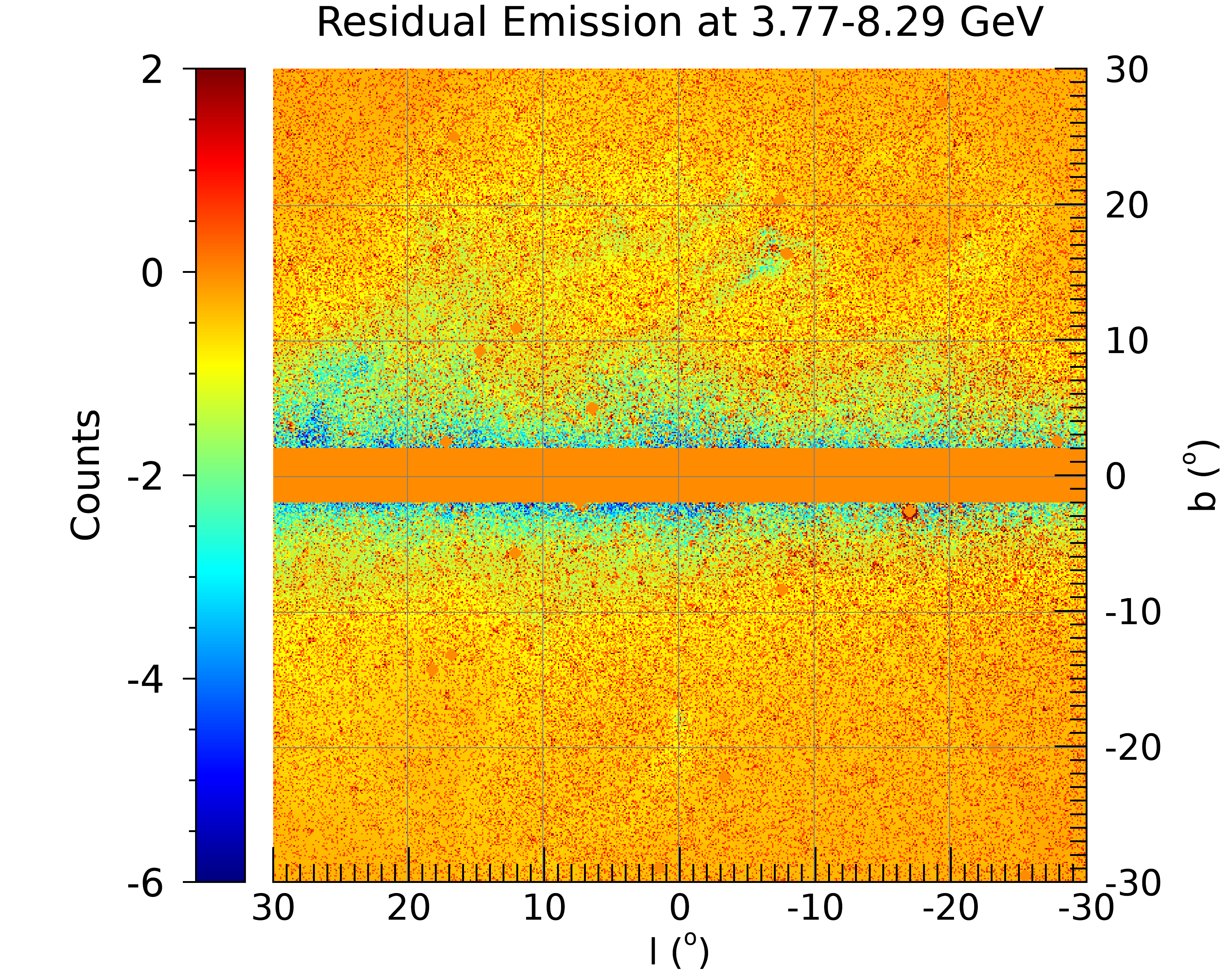}\\
\vspace{0.2in}
\hspace{-0.17in}
\includegraphics[width=2.44in,angle=0]{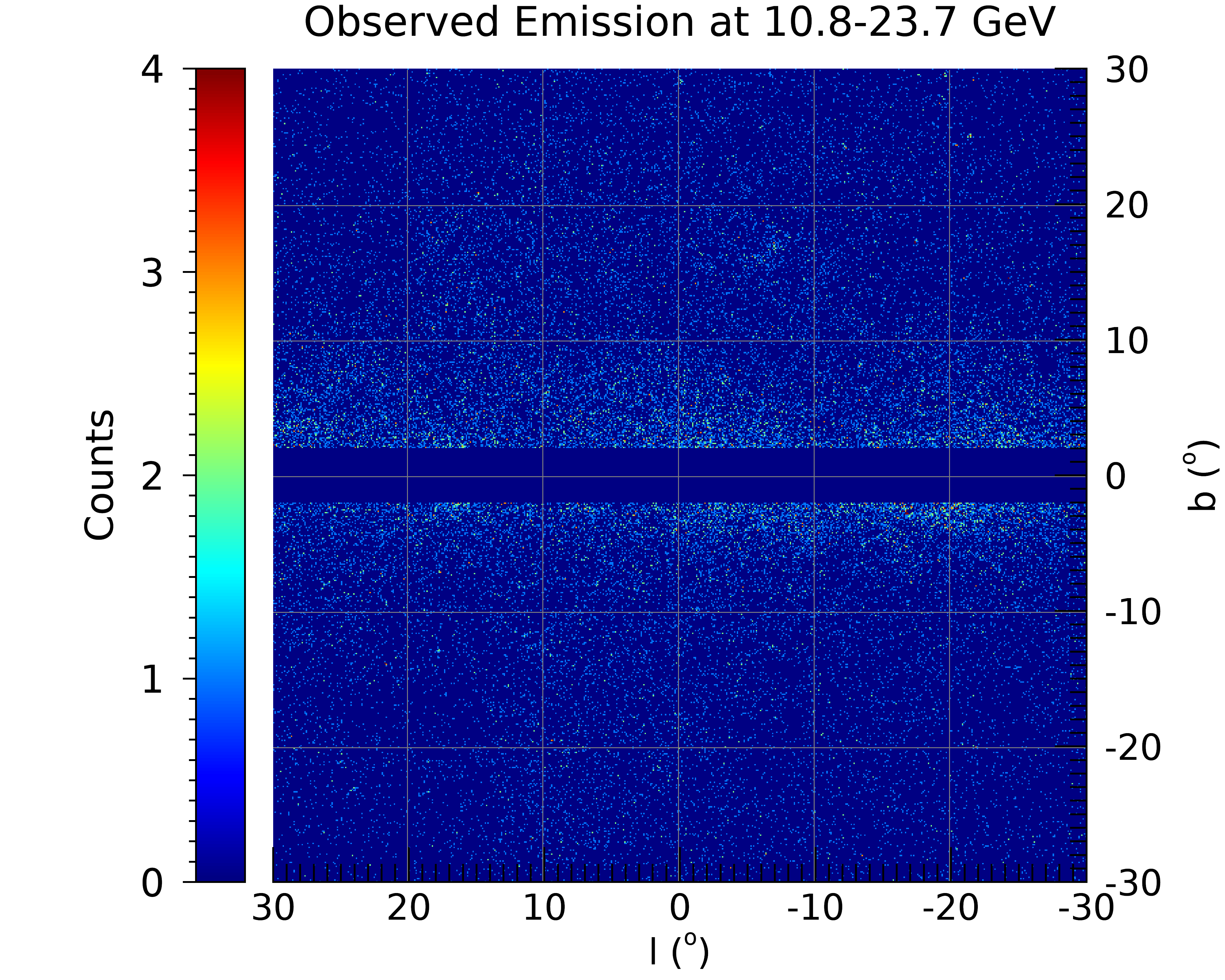} 
\hspace{-0.2in}
\includegraphics[width=2.44in,angle=0]{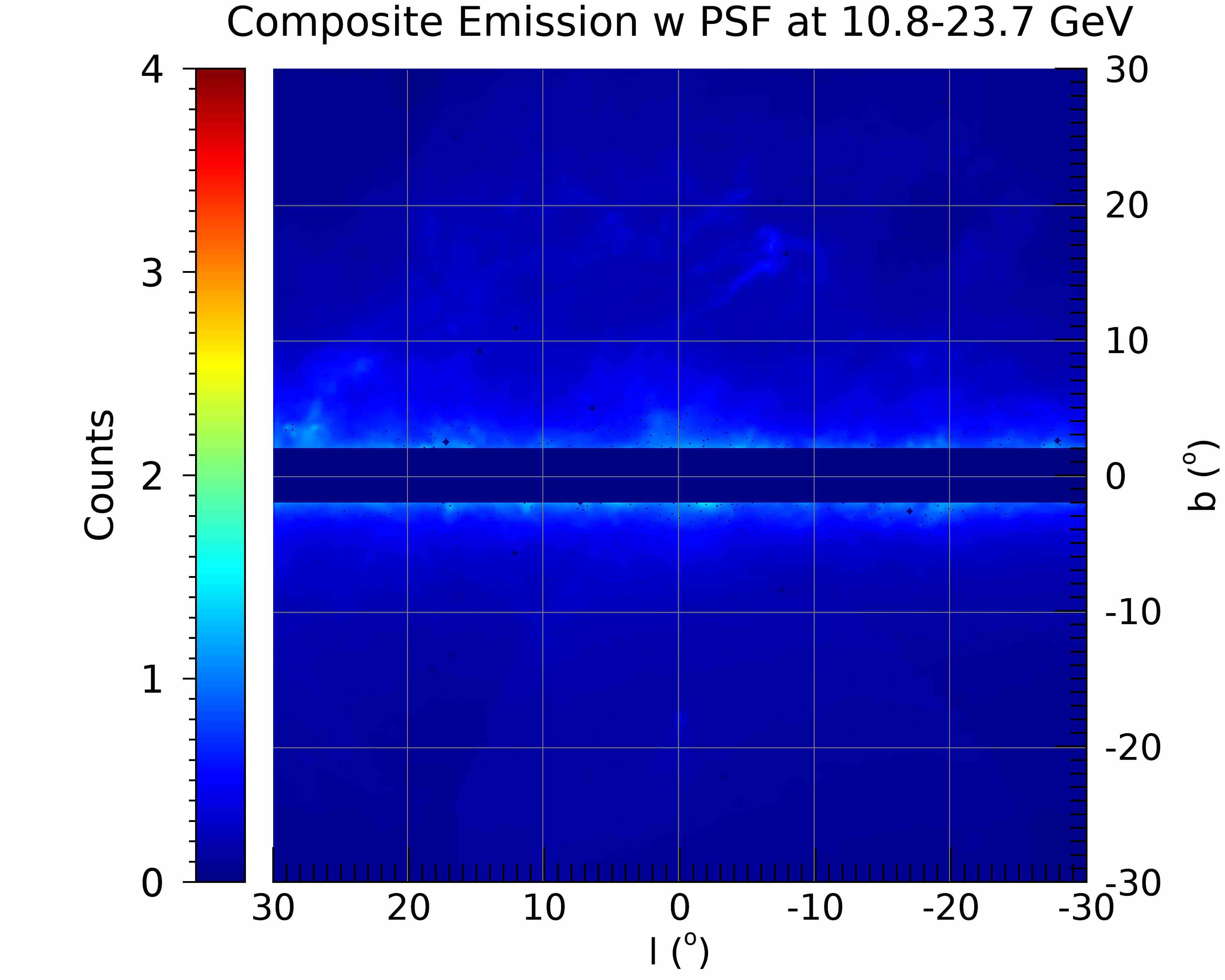} 
\hspace{-0.2in}
\includegraphics[width=2.44in,angle=0]{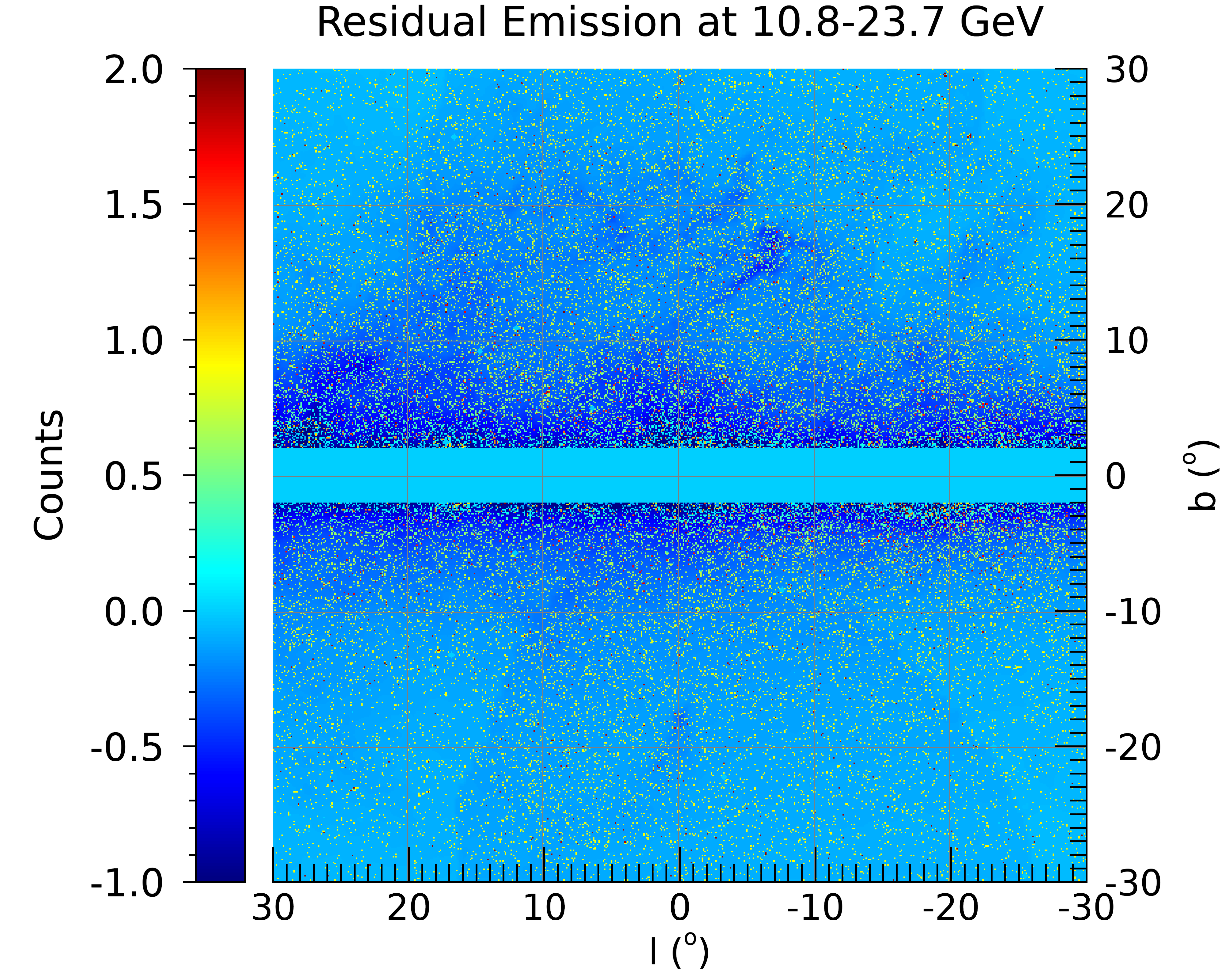}
\vspace{0.05in}
\caption{Comparison between the predictions for one composite diffuse model (using background 
Model I) and the observed data for three different energy ranges. In the \textit{top panels} we show 
results for the energy range of $1.02-2.24$ GeV, in the \textit{middle panels} for the energy range of 
$3.77-8.29$ GeV, and in the \textit{bottom panels} the range of $10.8-23.7$ GeV. \textit{Left panels} show 
the observed counts maps to which we fit, including the mask. \textit{Middle panels} show the 
composite best fit model with the PSF included. The \textit{right panels} show the difference $\equiv$ 
Data-Model, i.e., ``Residual Emission''. Including the GCE component in the fit of the inner galaxy's 
emission does not cause any over-subtraction. The most evident residuals appear at 
positive longitudes (see top right panel), near a couple of bright point sources (see e,g. $\ell= 
-17^{\circ}$, $b=-3^{\circ}$), but far from the GCE.}
\label{fig:Fit}
\end{figure*}  

In Fig.~\ref{fig:Fit}, we present as an example Model I as evaluated at its best fit parameter values and we
compare its performance with respect to the observed gamma-ray data. We present results for three
different energy ranges: the $1.02-2.24$ GeV bin in which the energy output of the GCE emission is
approximately peaked; the $3.77-8.29$ GeV bin in which the GCE brightness is nearly constant, and 
for which it achieves its highest relative brightness; and the $10.8-23.7$ GeV bin where the
GCE is still detected at high confidence, but the emission has decreased by a factor $\sim 2$. While 
our fits are performed only in the inner $40^{\circ} \times 40^{\circ}$ window, we show the
larger region of the $60^{\circ} \times 60^{\circ}$ window. In the left panels of Fig.~\ref{fig:Fit}, 
we show the observed \textit{Fermi}-LAT count maps. In the middle panels we show the composite
diffuse model count maps, including all components of Eq.~\ref{eq:CompositeMap} with best-fit 
normalizations and after accounting for the PSF. In the right panels we show the residual count maps. Our highest positive 
residuals are the result of point source leakage, while the most negative residuals are from gas 
emission far away from the inner galaxy and in fact in sky regions that are not fitted. Once subtracting the 
GCE, there is no visible positive or negative residual within the inner galaxy. We tested both the 
case of the Fermi Bubbles with their edge as given by \cite{Su:2010qj} and the case where the 
outer two degrees of the Bubbles are brightened by a factor of 2. We find that the edge-brightened 
Bubbles over-predict the observed fluxes in the relevant regions.

\begin{figure}
\begin{centering}
\includegraphics[width=3.65in,angle=0]{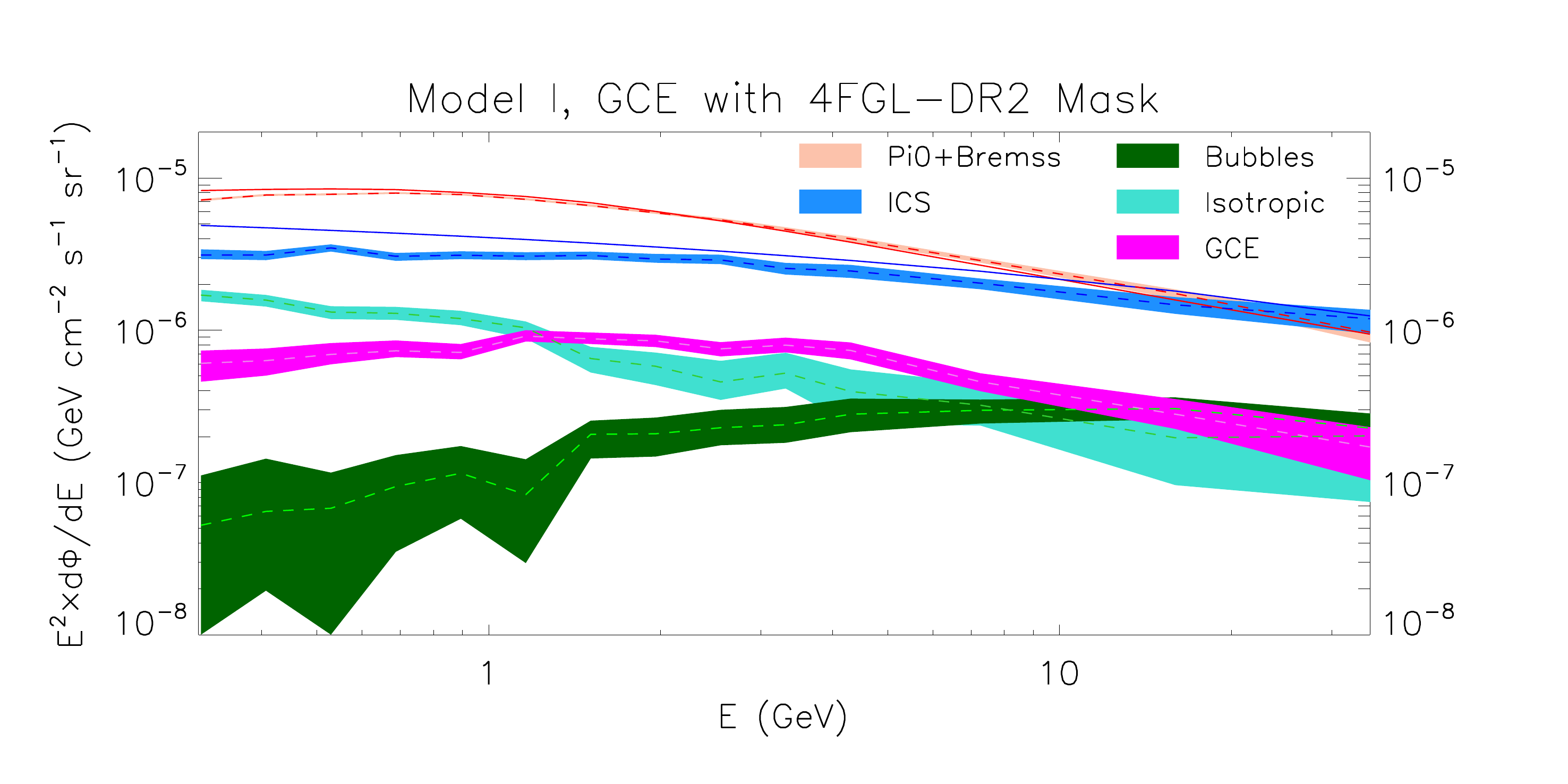}
\end{centering}
\vskip -0.3in
\caption{The modeled diffuse emission components in units of differential flux times energy squared, 
$E^{2} \times d\Phi/dE$. The dashed lines give the relevant Pi0+Bremss, ICS, Bubbles, Isotropic and 
GCE components after fitting to the data. The bands show the 2$\sigma$ ranges derived from the fit. 
We have used the background Model I for the Pi0+Bremss and ICS components; the solid lines show their pre-fit fluxes.  The GCE is a prominent emission between $\simeq$ 1 and 
10 GeV, and in that energy range is more luminous 
than either the isotropic emission or that from the Bubbles. 
}
\label{fig:Bands}
\end{figure}

In Fig.~\ref{fig:Bands}, we give the modeled diffuse emission components using Model I again as a reference.
We show fluxes in units of $E^{2} \times d\Phi/dE$, averaged over the $40^{\circ} \times 40^{\circ}$ region 
and excluding the galactic disk, i.e. $|b|\leq 2^{\circ}$. While in the fit 4FGL-DR2 point sources with $|b|>2^{\circ}$ are masked, we include the predicted diffuse background 
model emission from these masked regions for simplicity. 
We use Model I, because for $E > 0.5$ GeV the best-fit fluxes for the dominant Pi0+Bremss component of the
galactic diffuse emission is always within $\sim 10\%$ of the original template model assumption, and the ICS
component is largely within this same realm of accuracy. Only at the lowest energies, for which its contribution is subdominant, is the ICS roughly $50\%$ of the initial prediction. 
We show this by presenting the pre-fit predictions for the Pi0+Bremss and the ICS components
as solid lines, and showing the fluxes after the fit as dashed lines with shaded 2$\sigma$ error bands. The results
shown here are robust against the choice of point-source mask and are not specific to the 4FGL-DR2 mask which
is our standard choice. We point out that even after averaging over the $40^{\circ} \times 40^{\circ}$ window and including the mask within 2 degrees of the Galactic disk, the
GCE is a prominent emission between $\simeq$ 1 GeV and 10 GeV: it is more luminous than either the isotropic 
emission or that from the Bubbles. This is mostly because in the inner $5^{\circ}$ of the Galaxy the GCE 
is responsible for $\sim 10\%$ of the total emission. 

\begin{figure*}
\hspace{0.3in}
\includegraphics[width=7.4in,angle=0]{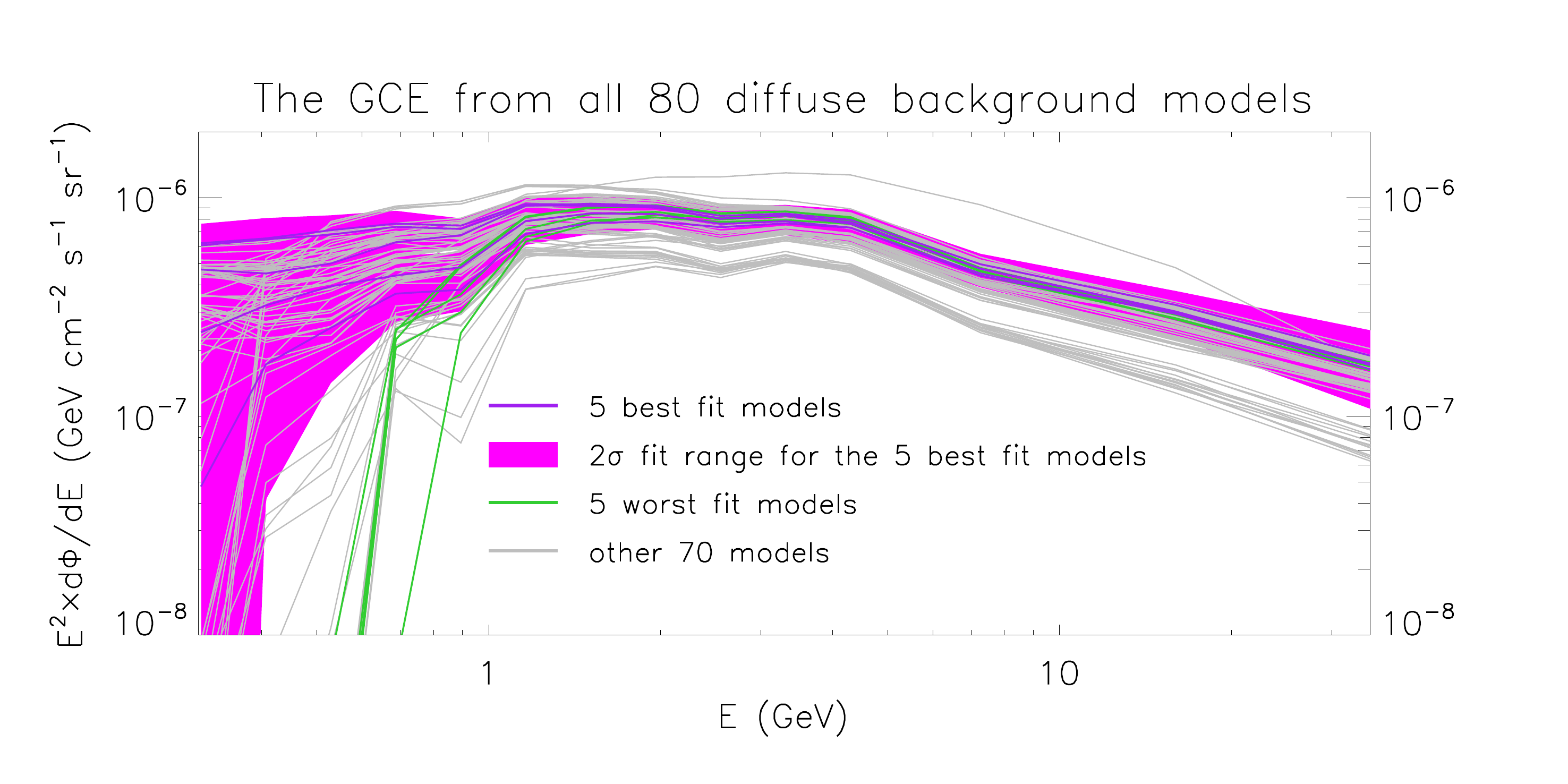}
\vskip -0.3in
\caption{The GCE derived in conjunction with all 80 diffuse galactic emission background models from our fits in the 
$40^{\circ} \times 40^{\circ}$ window. The purple lines 
show the GCE derived using the five galactic emission background models (Models X, XV, XLVIII, XLIX and LIII) 
that give composite models with the statistically best performance. Those lines heavily overlap above 
0.5 GeV. The magenta band shows their combined relevant 2$\sigma$ ranges. The green lines give 
the equivalent GCE from the five statistically worse performing models (based on models II, LXIV, 
LXIX, LXX and LXXI). The gray lines show the GCE from the remaining 70 galactic emission background 
models. The GCE above 0.7 GeV and up to 50 GeV is always present irrespective of the galactic emission 
background assumptions. 
}
\label{fig:GCE_emission_ALL_models}
\end{figure*}

In Fig.~\ref{fig:GCE_emission_ALL_models}, we show for all 80 galactic diffuse emission models 
the best-fit GCE emission (see Sec.~\ref{sec:method} for details on the fitting procedure). In this
figure, we assume a spherical GCE centered at the origin with  $\gamma=1.2$. 
We omit the galactic diffuse emission components, the Bubbles, and the isotropic emission for clarity. 
The GCE fluxes are shown at the central values of each of the energy bins represented in Tab.~\ref{tab:PSF_vsE}.
The GCE flux shown is obtained from a joint fit as described above, and the five best diffuse emission models, considered
to be those models that minimize the total negative log-likelihood, are highlighted in magenta. These are
Models X, XV, XLVIII, XLIX and LIII. We show the combined outer envelope of their 2$\sigma$ statistical uncertainties 
as the shaded magenta band. Likewise, the five worst models, those with the largest negative log-likelihood, 
are highlighted in green. The other 70 models are intermediate in fit quality between the best five and worst five 
fit models; we show them in gray. 
For these main results, in which we seek to determine the qualities of a putative excess signal near the galactic center,
we require the normalization of all fit components, including the GCE, to be positive. Only at the lowest energies and
with the worst fit models to the $40^\circ \times 40^\circ$ region do we find a preference for $c_{\textrm{GCE},j}$ to be zero or
negative.

We find that in every fit to the models generated in our work, the GCE is non-zero from 0.7 GeV and up to the highest energies that we model and test. The underlying 
background uncertainties can absorb the GCE only below 0.7 GeV. Moreover, we find that the GCE has a fairly 
robust spectrum. The ratio of the maximal to the minimal flux fit value among these 80 models is never
more than a factor of 4 at energies above 1 GeV. The differences between the GCE flux fits become even smaller 
once we focus on the best fit models (given in the magenta lines). For these five best models, the differences 
in their predicted GCE flux is of order $10 \%$ at energies above 1 GeV. Within the energy range we test, the GCE is
detected at greater than 2$\sigma$ significance for all energies above 0.4 GeV for these five best-fit models.
We note that even for the models shown by green lines (the five statistically less preferred galactic diffuse emission 
model assumptions), the predicted GCE fluxes are non-zero. Those give GCE spectra suppressed at sub-GeV energies,
but they still agree above 1 GeV with the models that provide superior fits to the overall emission. The ISM 
gas distribution assumptions have the largest impact at low energies. Any morphological choice that is 
in tension with the low-energy \textit{Fermi} data can in principle be compensated in the fit by a sufficiently negative GCE component.

\subsection{The Morphology of the GCE}
\label{sec:MorphologyResults}

\begin{figure*}[t]
\begin{centering}
\hspace{-0.16in}
\includegraphics[width=2.47in,angle=0]{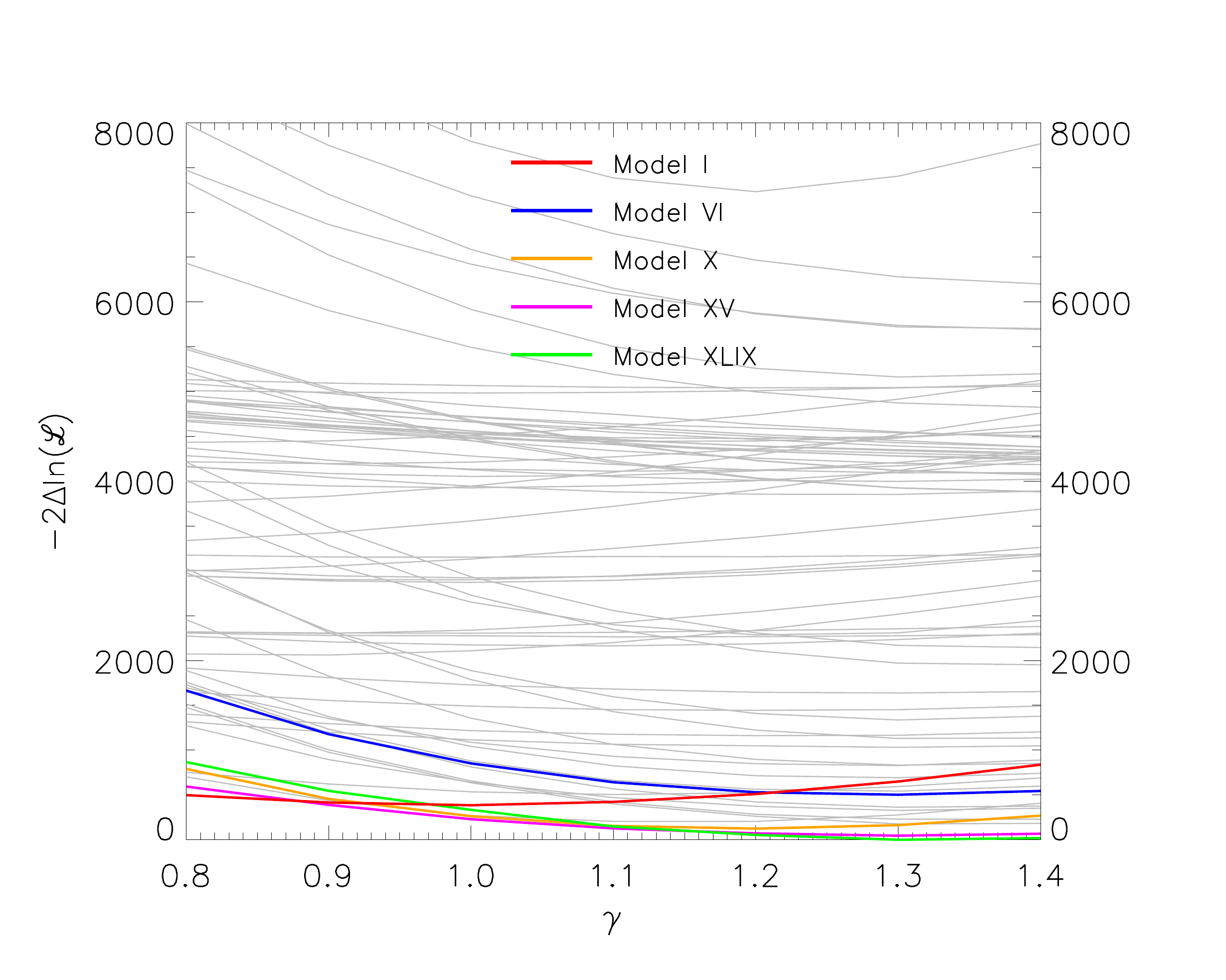} 
\hspace{-0.21in}
\includegraphics[width=2.47in,angle=0]{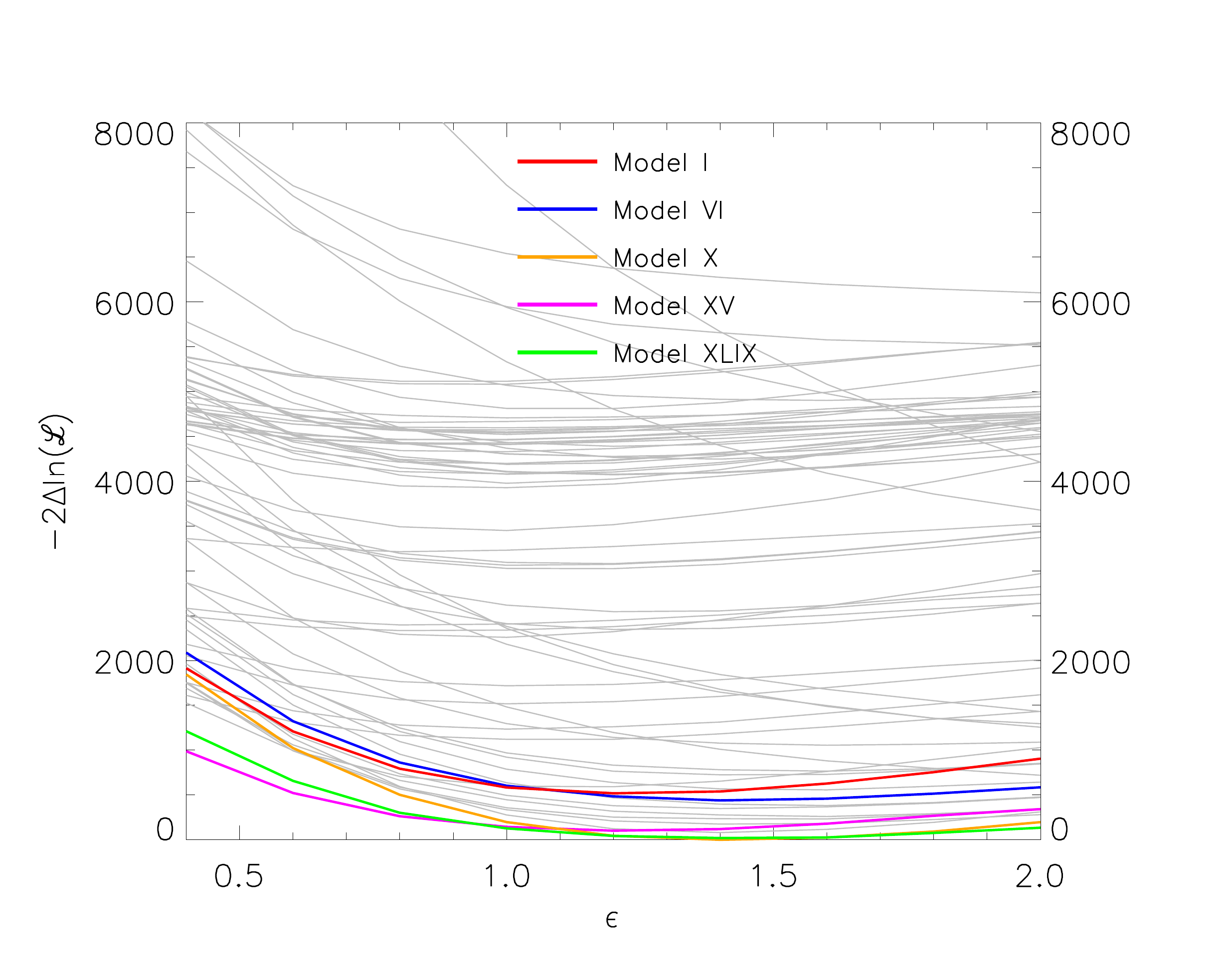} 
\hspace{-0.21in}
\includegraphics[width=2.47in,angle=0]{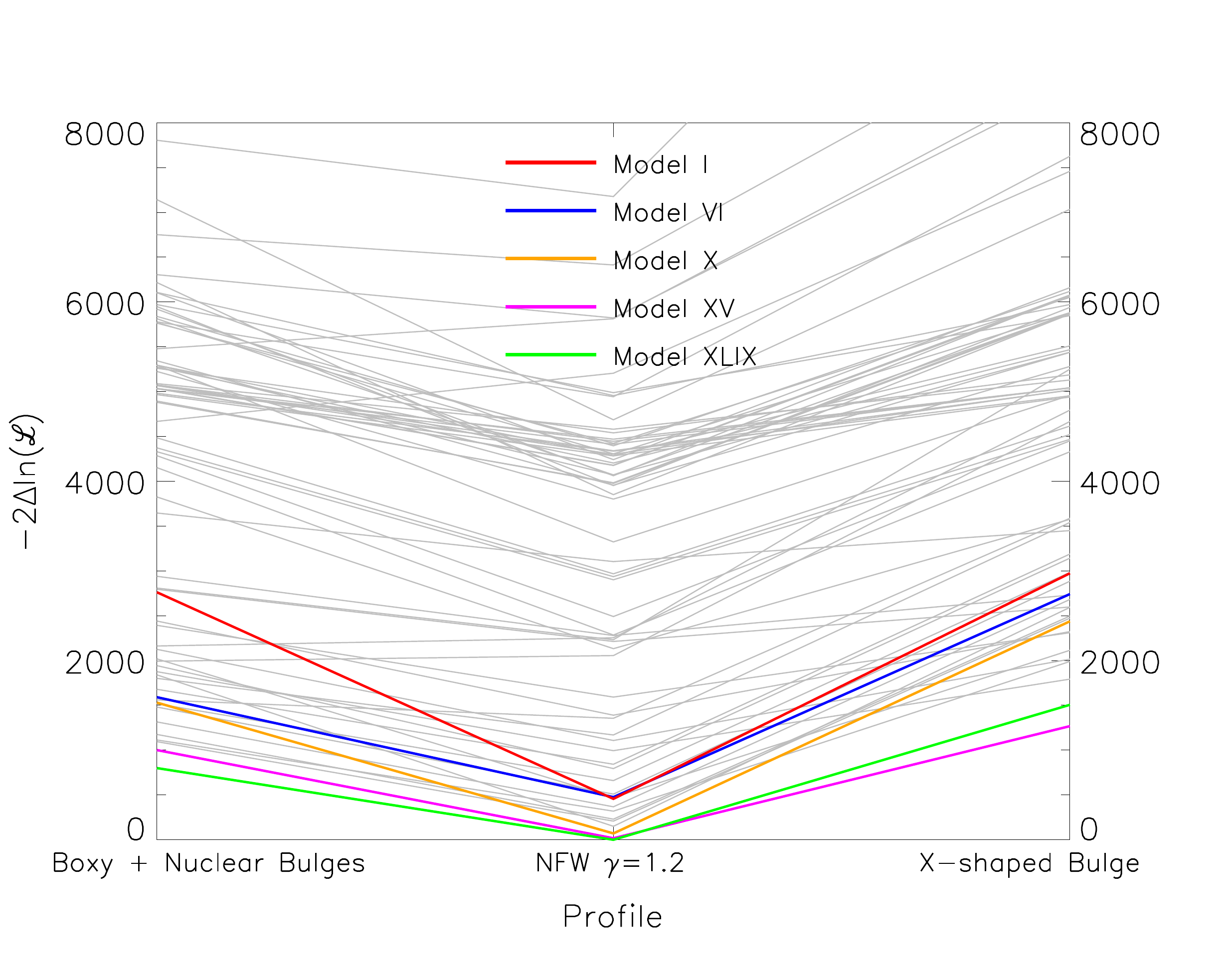}
\end{centering}
\vskip -0.1in
\caption{Testing the morphology of the GCE for all 80 diffuse galactic emission background models. The 
$y$-axis gives difference in the quality of the fit between choices for the GCE morphology, defined as $-2 \ln(\mathcal{L})$ for the alternative model minus $-2 \ln(\mathcal{L})$ of the best fit GCE from the entire sample.
Colored lines show the resulting GCE using five specific galactic emission 
background models, while the gray lines show the GCE under every single background model. \textit{Left}: 
assuming spherical symmetry we change the NFW cuspiness parameter  $\gamma$ of Eq.~\ref{eq:DM}, 
within the range of 0.8 and 1.4. \textit{Center}: using $\gamma=1.2$ we change the ellipticity parameter 
$\epsilon$ (see text for details). We find some preference for a GCE being 
elongated either along the galactic disk, which we explore further in App.~\ref{app:Ellipticity}. \textit{Right}: 
We compare the GCE with $\gamma=1.2$ and $\epsilon=1.0$ to the Boxy Bulge and the X-shaped Bulge 
that have been suggested in \cite{Bartels:2017vsx, Macias:2016nev} as alternative morphologies for the GCE. 
The NFW with $\gamma=1.2$ is systematically preferred in our fits.}
\label{fig:Morphology}
\end{figure*}

In identifying the possible physical mechanisms behind the GCE, its exact morphology is essential. Multiple alternatives to the spherical NFW emission morphology have been suggested \cite{Bartels:2017vsx, Macias:2016nev}. 
We study the GCE morphology by testing a sequence of alternative GCE templates including those of 
\cite{Bartels:2017vsx, Macias:2016nev} with each of our 80 galactic diffuse emission models. We show 
the results of these tests in Fig.~\ref{fig:Morphology}. 

In the left panel of Fig.~\ref{fig:Morphology}, maintaining the assumption of spherical symmetry
for the GCE, we test the ``cuspiness'' of the profile, i.e., how strongly it peaks as a function of
galactocentric radius. That is modeled by the parameter $\gamma$ in Eq.~\ref{eq:DM}, 
where $\gamma=1.0$ is the regular NFW dark matter profile \cite{Navarro:1995iw}, and $\gamma=1.2-1.3$
is favored by past results \cite{Daylan:2014rsa, Calore:2014xka, TheFermi-LAT:2015kwa} and even more recently by 
\cite{DiMauro:2021raz} that analyzed the GCE region in quadrants, using an alternative set of background models.
Every combination of GCE morphology and galactic diffuse emission model is provided, each of which is
compared to the global best fit result obtained across all models that maintain spherical symmetry.
We compare the difference in the log-likelihood $-2 \Delta \ln(\mathcal{L})$ to the best fit from  
across these combinations (the factor of two appears because the test statistic $-2 \Delta \ln(\mathcal{L})$
should follow a $\chi^2$ distribution). This provides a way to determine the statistically preferred
morphology in a global sense. Colored lines highlight five specific galactic emission background models.
These are among the models that give the best log-likelihoods, and they also showcase that the $-2 \Delta \ln(\mathcal{L})$
value can be minimized for different values of $\gamma$ depending on different diffuse model assumptions.
The gray lines show the results for the remaining background model assumptions that fall within the range
$-2 \Delta \ln(\mathcal{L}) < 8 \times 10^{3}$. We test $0.8 \leq \gamma \leq 1.4$ and find that the range
$1.2 \leq \gamma \leq 1.3$ is still generally preferred, though in some cases, such as for Model I, the
value $\gamma=1.0$ can be a relatively better fit. Galactic diffuse models which are statistically
much worse fits overall may even obtain their best fit for $\gamma < 1.0$; such values, which are strongly excluded by our work,
could in principle have been expected if the GCE traced the morphology of an underlying stellar population. Values
of $\gamma$ larger than 1.0 are expected if the dark matter halo has adiabatically contracted \cite{2012ApJ...752..163S}.
Values of $\gamma$ larger than 1.5 lead to profiles that are sharply peaked at very small galactocentric distances, 
but due to the mask on $|b|\leq 2^\circ$ such profiles are not anticipated to be 
easy to differentiate from the results
obtained for $\gamma=1.4$. 

In the central panel of Fig.~\ref{fig:Morphology}, we maintain $\gamma=1.2$ and vary the ellipticity parameter $\epsilon$,
as defined in Sec.~\ref{sec:GCE}. A value of $\epsilon > 1$ corresponds to elongation along the galactic disk (a prolate 
profile), while $\epsilon < 1$ corresponds to elongation perpendicular to the disk (an oblate profile). We find 
a mild preference for the GCE being elongated along the galactic disk, compatible with a recent analysis by 
\cite{DiMauro:2021raz}. The difference in $-2 \Delta \ln(\mathcal{L})$ 
between an $\epsilon$ of 1.0 vs 1.4 for the best fit models is $\lsim 100$. Such a difference in the likelihood is small:
for comparison, the best fit model with a GCE is preferred to the best fit model without a GCE by  $2.8\times 10^{3}$.
Only background models that provide very poor global fits to the observations show a strong preference for GCE emission
that is elongated along the disk. Moreover, a GCE elongated along the galactic disk 
is degenerate with the mask that we use. We explore the impact of variations of the mask in App.~\ref{app:Ellipticity}.
There we find that if, rather than masking $|b|<2^\circ$ at all longitudes ($|\ell|<20^\circ$), we instead mask $|b|<2^\circ$ only for
a limited longitude range such as $|\ell|<5^\circ$ or $|\ell|<8^\circ$, the preference 
for $\epsilon \simeq 1$ returns. By changing the mask, a subsequent absolute change in the $-2 \Delta \ln(\mathcal{L})$ 
by $\mathcal O(10^2)$ may be expected as alternative masks remove $\mathcal O(10^2)$ pixels in the inner galaxy. 
We note the possibility that mask assumptions may bias the extracted GCE ellipticity. Alternative masking procedures
will be interesting avenues for future work.

Finally, in the right panel of Fig.~\ref{fig:Morphology}, we compare the GCE fits obtained with our baseline assumptions 
of $\gamma=1.2$ and $\epsilon=1.0$ to fits assuming two other morphologies that have been suggested to correlate with 
the GCE. Thus, instead of using the 
morphology described by the square of the NFW profile from Eq.~\ref{eq:DM} defined by $\gamma$ and $\epsilon$, we repeat the procedure of Sec.~\ref{sec:method} 
using the Boxy Bulge with its Nuclear Bulge component and the X-shaped Bulge templates favored in \cite{Bartels:2017vsx, Macias:2016nev}, 
respectively, as alternative morphologies for the GCE. 
Some of the new diffuse models produced in this work, which differ from the diffuse models used in \cite{Bartels:2017vsx, Macias:2016nev}, lead to a preference for these alternative templates, but these combinations of diffuse models and alternative morphologies provide significantly poorer overall fits to the data, and the spherical NFW profile with $\gamma=1.2$ is globally 
preferred in our fits at the level of $-2 \Delta \ln(\mathcal{L}) \gtrsim 1000$.
We conclude that these alternative morphologies cannot accommodate the observations in the $2^\circ \leq |b| \leq 20^\circ$, $|\ell| \leq 20^\circ$ region of interest with as good of a fit as the moderately contracted
spherically symmetric NFW profile when used with the diffuse templates generated in this work. We also explore further the possible contribution of the Stellar Bulge to the GCE in the $2^\circ \leq |b| \leq 20^\circ$, $|\ell| \leq 20^\circ$ region in App.~\ref{app:BulgesTests}.

\subsection{The GCE in the North and the South}
\label{sec:resultsgceNS}

\begin{figure*}[!ht]
\hspace{0.3in}
\includegraphics[width=7.4in,angle=0]{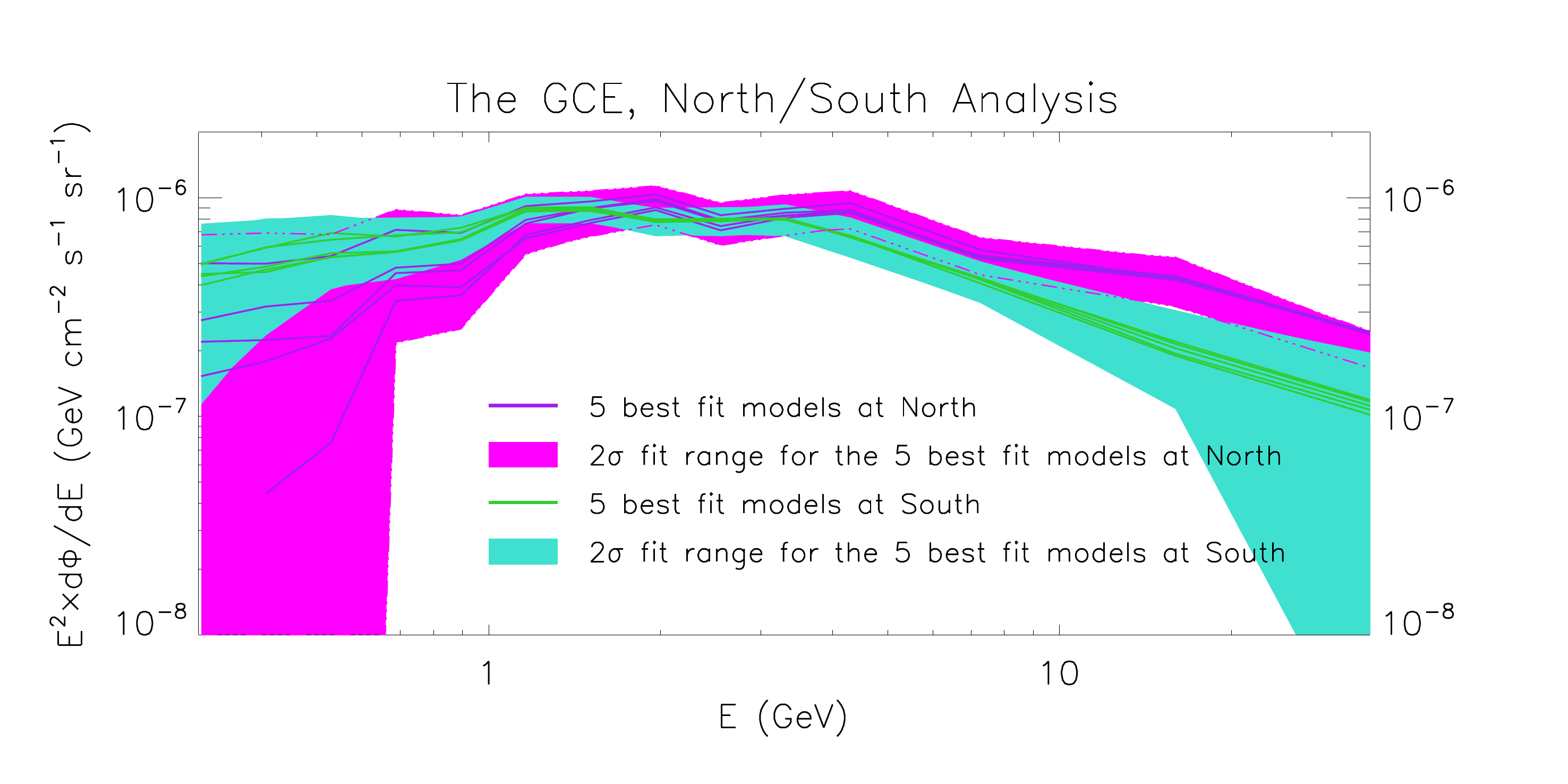}
\vskip -0.3in
\caption{Fit values of the GCE including data from $|\ell|<20^\circ$ and $0<b<20^\circ$ (``North'') or 
$-20^\circ<b<0$ (``South''). We allow the GCE normalization to vary independently in each region and each energy bin. 
The magenta lines show the GCE for the five best fits in the northern hemisphere, derived using Models XV, XLI, XLIX, 
L and LIII, with the magenta band giving their combined 2$\sigma$ ranges. The cyan lines give the GCE for the five 
best fits in the southern hemisphere, derived using Models X, XLVII, XLVIII, LII and LIII, with the cyan band showing 
their combined 2$\sigma$ ranges.}
\label{fig:GCE_emission_N_vs_S}
\end{figure*} 

\setlength{\tabcolsep}{3pt}
\begin{table*}[!tb]
    \begin{tabular}{ccccccccccccccc}
    \hline 
        ROI & $\Phi_1$ & $\Phi_2$ & $\Phi_3$ & $\Phi_4$ & $\Phi_5$ & $\Phi_6$ & $\Phi_7$ 
        & $\Phi_8$ & $\Phi_9$ & $\Phi_{10}$ & $\Phi_{11}$ & $\Phi_{12}$ & $\Phi_{13}$ & $\Phi_{14}$ \\
        \hline\hline 
        $40^{\circ} \times 40^{\circ}$ & $2.4^{+0.8}_{-0.6}$ & $3.2^{+0.8}_{-0.5}$ & $3.9^{+0.5}_{-0.6}$ & $4.4^{+0.4}_{-0.5}$ & $4.8^{+0.4}_{-0.4}$ & $7.8^{+0.4}_{-0.3}$ & $8.5^{+0.4}_{-0.3}$ & $8.4^{+0.4}_{-0.4}$ & $7.6^{+0.4}_{-0.4}$ & $7.8^{+0.5}_{-0.3}$ & $7.3^{+0.6}_{-0.4}$ & $4.4^{+0.2}_{-0.4}$ & $2.9^{+0.3}_{-0.3}$ & $1.7^{+0.3}_{-0.3}$\\
        South & $4.4^{+1.2}_{-1.4}$ & $4.8^{+1.2}_{-1.0}$ & $5.6^{+0.8}_{-0.9}$ & $5.7^{+0.7}_{-0.7}$ & $6.5^{+0.5}_{-0.6}$ & $8.9^{+0.6}_{-0.4}$ & $8.9^{+0.5}_{-0.5}$ & $7.9^{+0.5}_{-0.5}$ & $8.0^{+0.7}_{-0.4}$ & $7.9^{+0.5}_{-0.7}$ & $6.6^{+0.5}_{-0.7}$ & $4.3^{+0.4}_{-0.4}$ & $2.2^{+0.5}_{-0.4}$ & $1.2^{+0.5}_{-0.4}$ \\
        North & $0.0^{+0.8}_{-1.0}$ & $0.4^{+0.9}_{-0.9}$ & $0.8^{+0.8}_{-0.9}$ & $3.4^{+0.6}_{-0.8}$ & $3.6^{+0.5}_{-0.7}$ & $6.8^{+0.6}_{-0.5}$ & $7.9^{+0.6}_{-0.5}$ & $9.0^{+0.5}_{-0.6}$ & $7.4^{+0.5}_{-0.7}$ & $8.3^{+0.7}_{-0.6}$ & $8.7^{+0.8}_{-0.6}$ & $5.3^{+0.4}_{-0.5}$ & $4.2^{+0.4}_{-0.5}$ & $2.4^{+0.3}_{-0.0}$\\
        \hline
    \end{tabular}
    \caption{Fluxes of the GCE emission, as in Figs.~\ref{fig:GCE_emission_ALL_models} and \ref{fig:GCE_emission_N_vs_S}. 
    The central values are from the respective best fit models and the attached error bars are {\it statistical} error bars only, 
    in units of $10^{-7}\,{\rm GeV \, cm^{-2}\,  s^{-1} \, sr^{-1}}$.}
    \label{tab:gcevals}
\end{table*}

In Fig.~\ref{fig:GCE_emission_N_vs_S}, we repeat the same analysis for all diffuse models listed in 
Tab.~\ref{tab:ModelsLong}, assuming a spherical GCE centered at the origin with inner slope 
$\gamma=1.2$, but now restricting the fit only to either the northern or southern hemisphere. As in 
Fig.~\ref{fig:GCE_emission_ALL_models}, we show only the GCE component of the fit, 
omitting the galactic diffuse emission components, the Bubbles, and the isotropic emission. 
For clarity, we restrict to the five best models in each hemisphere, although for completeness we 
again tested all 80 models for the north and the south independently and found the GCE to be present 
in both hemispheres for all models. The bands give the combined 2$\sigma$ fit ranges. As described above, we allow the normalization of the GCE, 
$c_{\textrm{GCE},j}$, to be negative, although we require that the total counts, $\mathcal C_{j,p}$ as defined 
in Eq.~\ref{eq:ExpectedCountsPixel}, is non-negative in each pixel.

The magenta lines in Fig.~\ref{fig:GCE_emission_N_vs_S} show the results for the GCE normalization 
in the northern sky only, using the five best models for that region of the sky. The five best-fit models 
in the north are Models XV, XLI, XLIX, L and LIII from Tab.~\ref{tab:ModelsLong}. The absolute best fit 
of these models prefers a negative value for the GCE in the first bin, which we graphically represent by 
dropping that data point. As before, the fit strongly prefers positive values of the GCE at all energies 
above 1 GeV.

The cyan lines give the equivalent information from the southern sky only, using the five best models 
for that region of the sky. The five best fit models in the south are Models X, XLVII, XLVIII, LII and LIII. 
The fit strongly prefers positive values of the GCE up to the final energy bin in our analysis, at which 
point the fit is compatible with zero GCE at the 2$\sigma$ level.

For the bulk of the energy range we have tested, the results from Fig.~\ref{fig:GCE_emission_N_vs_S} 
are compatible with one another and also with the results for the full $40^\circ \times 40^\circ$ region shown in 
Fig.~\ref{fig:GCE_emission_ALL_models}. Below 1 GeV and above 10 GeV, however, we begin to 
observe some differences in the fits: at low (high) energies the preferred value of the GCE is roughly 
half (twice) as bright in the north as it is in the south. At high energies, the results from the north and 
the south even appear to be in tension at the 2$\sigma$ level in our penultimate energy bin. However, 
even at high energies, they both remain compatible with the analysis of the $40^{\circ} \times 40^{\circ}$
shown in Fig.~\ref{fig:GCE_emission_ALL_models}. Across the whole energy range, these differences do become 
meaningful -- the GCE has a softer spectrum in the southern sky than in the north, which leads to 
evidently different outcomes in the fits to different underlying models, which we discuss in Sec.~\ref{sec:Interpretations}.

For ease of reference, in Tab.~\ref{tab:gcevals} we provide the central values for the best fit model for 
the $40^{\circ} \times 40^{\circ}$ window analysis, the northern-hemisphere fit, and the southern-hemisphere 
fit as well as their attached error bars. These are {\it statistical} error bars, meaning they come from the 
statistical error on evaluating the GCE normalization in the single best fit model. For the $40^{\circ} \times 40^{\circ}$ window the GCE 
errors are evaluated using background model XLIX. For the northern hemisphere they are evaluated using
background model XV and for the southern hemisphere using background model XLVIII. We discuss the 
systematic error bars in the next section.

\section{Systematic Uncertainties}
\label{sec:Uncertainties}

\begin{figure*}[!p]
    \centering
    \includegraphics[height=0.95\textheight]{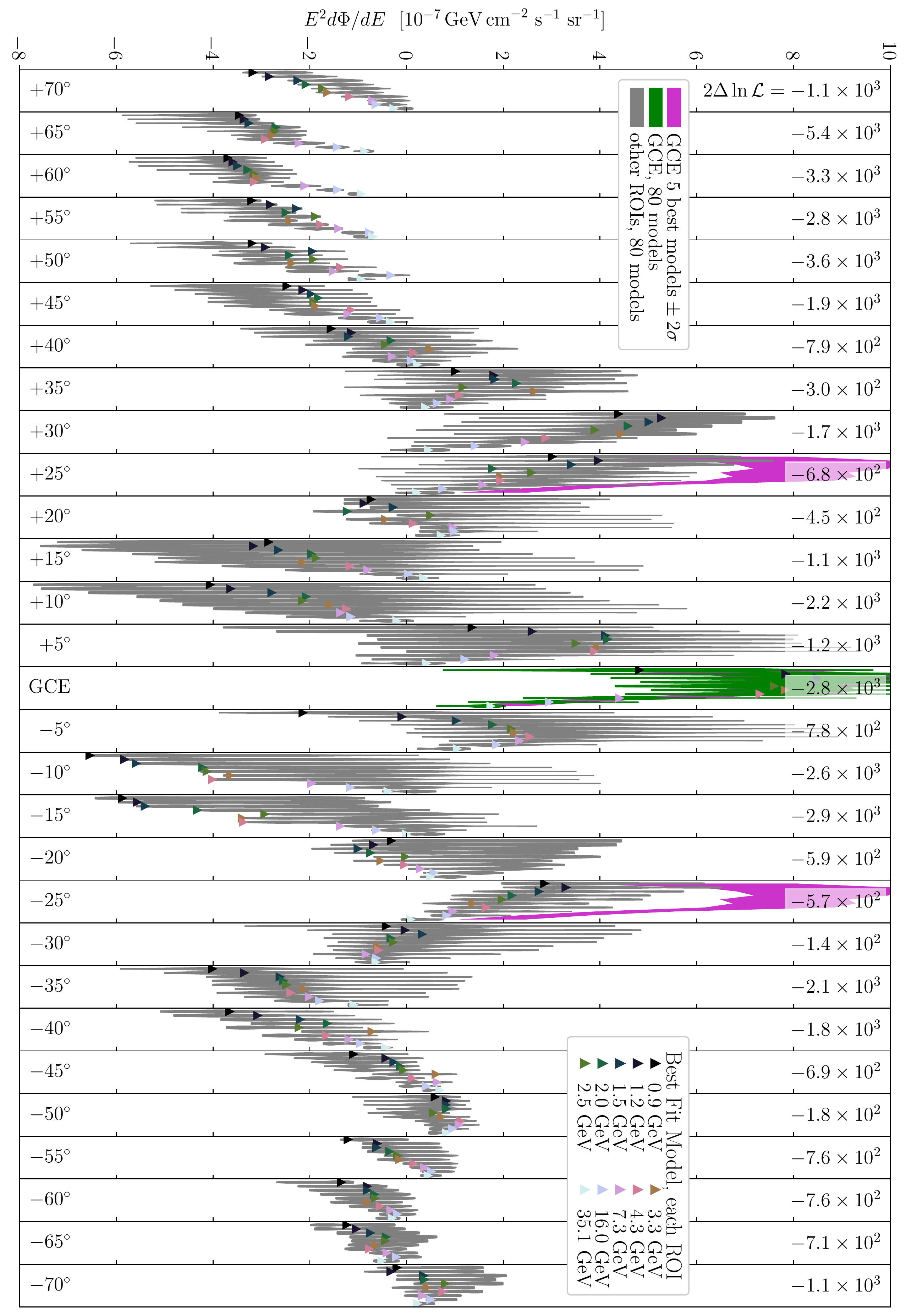}
    \caption{Fits to the translated GCE template in 28 different ROIs, as well as at the galactic center. The triangular symbols give the results with the best fit model in each ROI. The gray regions are violin plots showcasing the distribution of allowed normalizations across all models; the green shows the same for the galactic center only. The magenta region is the same as in Fig.~\ref{fig:GCE_emission_ALL_models}, showing the envelope of the $\pm2\sigma$ range of the five best models in the galactic center. The $2\Delta \ln \mathcal L$ values in each ROI compare the best fit model {\it with} the GCE-like template minus the best fit model {\it without}.}
    \label{fig:otherrois}
\end{figure*}

Because the GCE is a subdominant component of the total galactic gamma-ray emission, as shown in Fig.~\ref{fig:Bands}, 
attaching a quantitative significance to a measurement of the GCE spectrum is nontrivial. Moreover, any underlying 
uncertainty in the astrophysical assumptions discussed in Sec.~\ref{sec:GDGRE} has an effect in several energy bins and 
across a significant fraction of the pixels that we use to fit our models to the observations. Thus, 
the significance of the signal will be overestimated if each data point is taken to be uncorrelated with its neighbors. In this section, we attempt to quantify the 
impact these astrophysical systematic and correlated uncertainties have. We remind the reader however that 
the GCE is present under all the 80 galactic diffuse models that we 
use to probe these astrophysical uncertainties. Furthermore, it is present both in the northern and southern hemisphere 
with its spectrum and morphology quite robustly quantified.  
For this reason, a systematic uncertainty exploration is important to characterize the significance, but not the existence, of the excess.

Because the template-based fits that we have produced have incorporated energy and spatial information via the 
underlying multi-messenger-informed galactic diffuse emission maps, this can lead to correlated errors across energy bins and pixels. 
If neglected, this could cause us to overestimate the significance of the detection of the GCE. In addition, two other 
poorly characterized extended emission components are known to exist in lower-background areas of the sky along 
the Galactic disk\footnote{\url{https://fermi.gsfc.nasa.gov/ssc/data/access/lat/Model_details/Pass7_galactic.html}} 
\cite{Calore:2014nla, Balaji:2018rwz}, potentially leading to concerns about the ``look elsewhere'' effect incurred 
by template fitting. 

With the goal of evaluating the correlated systematic error bars needed to assess the significance of our findings, in this 
section we 
estimate the 
impact of these considerations by performing null analyses. We begin by performing the same fitting procedure as 
described in Sec.~\ref{sec:fit}, over 28 non-independent regions of interest (ROIs): before masking, each is 
$40^\circ \times 40^\circ$ in size and spans from $-20^\circ<b<20^\circ$ in latitude. They differ by the longitude at 
which they are centered, spanning from $5^{\circ}\leq |\ell| \leq 70^{\circ}$ in $5^\circ$ steps. As before, we mask the Galactic disk and
all 4FGL-DR2 point sources and test all 80 galactic diffuse emission models described 
in Sec.~\ref{sec:GDGRE} and Table~\ref{tab:ModelsLong} for the appropriate region.  Also, we fit all 14 
energy bins independently. The GCE is the only template that is translated along the disk; it is always taken at its
``standard'' choice of $\gamma=1.2$ and $\epsilon=1.0$. Each ROI therefore contains a GCE-like fit component 
for each of the 14 energy bins in Tab.~\ref{tab:PSF_vsE}. We allow the normalization of the GCE-like emission, 
$c_{\textrm{GCE},j} $, to be negative while requiring the total counts, $\mathcal C_{j, p}$ as defined in 
Eq.~\ref{eq:ExpectedCountsPixel}, to be non-negative for each pixel. 

We note that the spherically symmetric and translated GCE may not model exactly 
the morphology of any possible excess emission in regions further away from the galactic center. That is to be 
anticipated, as any excess emission would be emission not fully accounted by the galactic diffuse emission 
models. Such an emission may have its own morphology, due to the details in the spatial distribution of cosmic-ray 
sources located at the spiral arms, or regions with dense ISM gases, or higher ISRF, or 
stronger local magnetic fields. However, we expect that at first order the translated GCE will absorb 
any mismodeling, either by being positive (to account for genuine excesses) or negative (to account for regions of higher-than-expected gas densities, for example). More importantly in ``looking elsewhere,'' an excess emission that has a very different morphology 
to the one we have detected at the galactic center would not qualify as a similar excess emission. 

We show the results of our fits for all ROIs in Fig.~\ref{fig:otherrois}. In each frame of this figure, we show the 
results for ten energy bins, dropping the four lowest energy bins for clarity. We represent the central 
values from all 80 models as violin plots, where each gray band is a probability density for a given energy bin, giving 
equal weight to each model. The colorful triangles in each energy bin show the central values for the spectrum 
of the best fit model in that region, where the best fit is considered to be the model that minimizes the 
sum of the negative log likelihoods across all independent energy bins, as before. The best fit flux values 
vary between $\approx-8$ and $\approx+8$ in units of $10^{-7}{\rm GeV\, cm^{-2}\, s^{-1}\, sr^{-1}}$. 
The brightest (dimmest) best-fit flux values tend to occur at the lowest (highest) energy levels, 
which is to be expected as the overall flux decreases at higher energy, although the sign of the 
bright low-energy emission is negative in some portions of the sky and positive in others. 
The $2\Delta \ln \mathcal L$ values in each ROI give, for all 14 energy bins in that ROI, double the negative log-likelihood 
of the best-fit model {\it with} the GCE-like emission minus the negative log-likelihood of the best-fit model {\it without} the GCE-like emission. 
More negative values correspond to a more significant improvement of the fit with a GCE-like emission.

We show the GCE for reference in the central panel of Fig.~\ref{fig:otherrois}. We show the best fit spectrum 
with the same symbols as for the other ROIs, we depict the $\pm2\sigma$ envelope of the five best models 
in magenta (the same as the magenta region of Fig.~\ref{fig:GCE_emission_ALL_models}), and we show 
in green violin plots the distribution of all 80 models. We reproduce the magenta region in two other ROIs 
whose preferred GCE-like spectra is relatively bright. It is clear from this figure that the GCE is unique in 
two respects: it is brighter by more than a factor of two at all pertinent energies, and it is substantially 
harder than the spectrum preferred in any other ROI. 
The GCE at $\ell=0^\circ$ is not the most favored as measured by the $2\Delta \ln \mathcal L$ values, being surpassed by values at $\ell=+65^\circ, +60^\circ,$ and $ +50^\circ$ and equalled or nearly so at $\ell=+55^\circ$ and $-10^\circ$. However, all of these significantly detected emissions have a strong preference to be negative, and reflect over-modeling errors rather than detections of novel emission components. The GCE at $\ell=0^\circ$ is indeed uniquely favored when restricting to positive excesses only.

Another point that becomes evident from Fig.~\ref{fig:otherrois} is 
the extent in latitude of the emission that we refer to as the GCE. 
For the windows centered at $\ell = \pm 5^{\circ}$, the translated GCE-like emission template spectra are a factor of $2-4$ smaller than at $\ell =0^{\circ}$,
and at $\ell = \pm 10^{\circ}$ it becomes negative and the spectra are spectrally very soft. If the emission 
centered at the galactic center that we refer to as the GCE was significantly extended across galactic 
longitudes, the translated analysis here would have given similar spectra across the $\ell = [-10^{\circ}, +10^{\circ}]$ 
region. 

This procedure both serves as additional proof of the unique properties of the excess found in the GC and, 
as describe now in more detail, provides a way to evaluate its significance. 

\begin{figure}[!t]
\includegraphics[width=8cm]{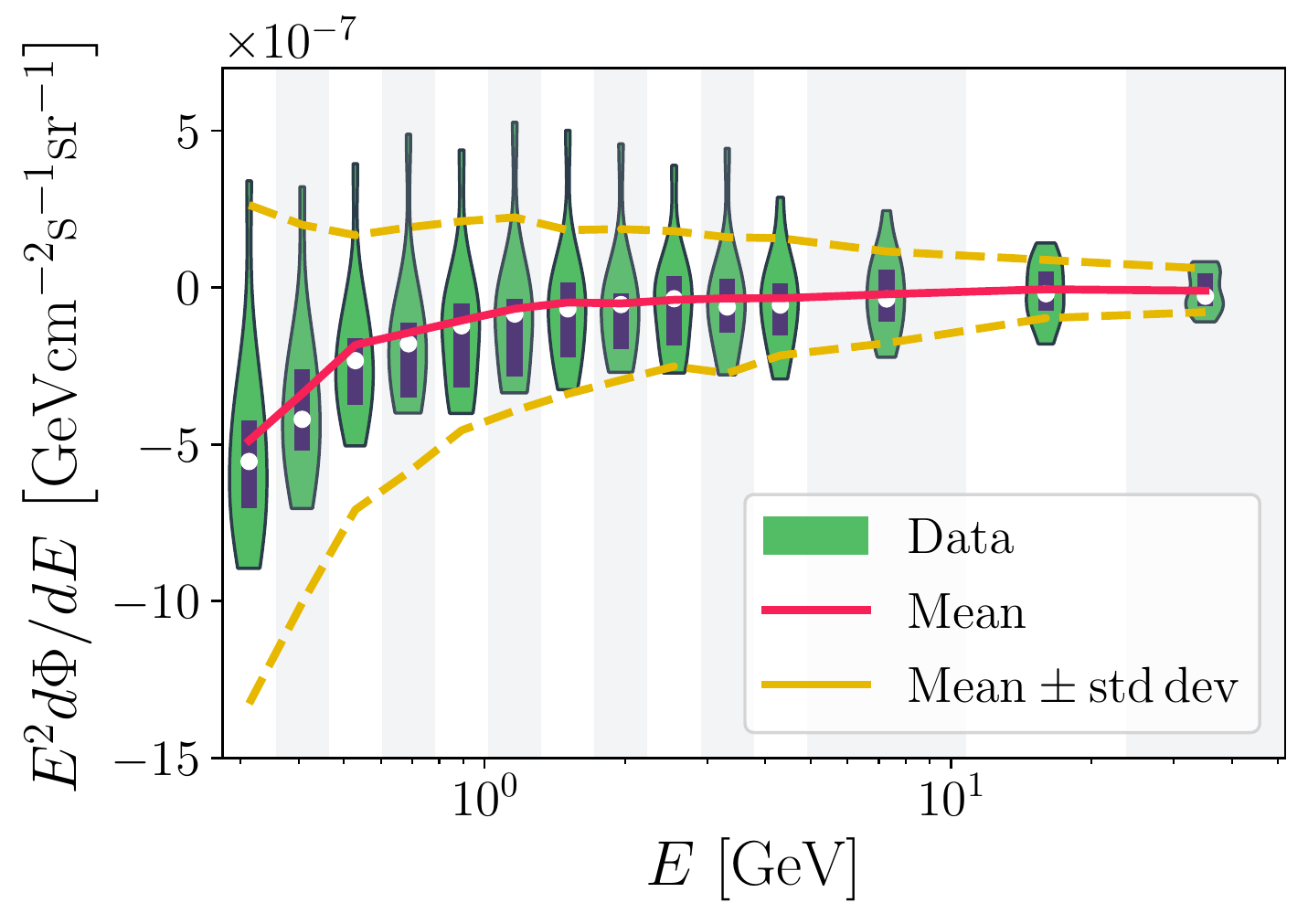}
\centering 
\caption{Violin plots (green)
representing the probability density distributions of GCE-like residual fluxes from the 22 ROIs, whose centers 
are translated on the galactic disk. The white dots inside the violins represent the median values and the 
purple bars represent the interval between the 1st and 3rd quartile. The red line represents the mean. 
In yellow, we present $\pm1\sigma$ standard deviation.}
\label{fig:violin_22ROIs_14Ebins}
\end{figure}

In Fig.~\ref{fig:violin_22ROIs_14Ebins}, we present a related perspective on the distribution of fits to GCE-like 
emission in other regions of the sky. To more closely reproduce the procedure followed to generate the results 
in Fig.~\ref{fig:GCE_emission_ALL_models}, and thus to ascertain the possibility of systematic errors in those 
results, this figure depicts the distribution of the best fits only. Thus, for every region centered at values
of latitude from $\ell = -70^{\circ}$ to $\ell = +70^{\circ}$ in $5^\circ$ increments, 
we only use the spectrum of the translated GCE-like emission that is derived with the galactic diffuse background model that 
best fits the observations. In order to limit contamination from the GCE whose 
significance we seek to evaluate, we omit the GCE itself as well as the fits from the six regions centered at 
$\ell = \pm5^\circ,\pm10^\circ,\pm15^\circ$. Thus, the results in Fig.~\ref{fig:violin_22ROIs_14Ebins} provide 
the distributions of best-fit values with respect to the remaining 22 ROIs centered at $20^\circ \leq |\ell | \leq 70^\circ$. 
We show these distributions across all 14 energy bins. For each energy bin the violin plot
is evaluated using the 22 flux values from the 22 best fit GCE-like emission spectra of the 22 translated ROIs. 
As 
anticipated from Fig.~\ref{fig:otherrois}, 
the average favored GCE-like emission is negative, which is especially clear
at the lowest energies.

\subsection{Covariance Matrix 
from Different ROIs}
\label{sec:covmatx}

The fits across the different ROIs show in Fig.~\ref{fig:violin_22ROIs_14Ebins} enable the construction of a data covariance matrix. 
This allows us to assess the impact of systematic astrophysical uncertainties and bias in recovering GCE-like features along the 
galactic disk, and therefore provides an estimate of the systematic error budget incurred by our template-fit procedure. 

The data covariance matrix we construct is based on the GCE-like fits performed in different regions. 
As for the distributions shown in Fig.~\ref{fig:violin_22ROIs_14Ebins}, we omit the GC and the regions centered 
at $\ell = \pm5^\circ,\pm10^\circ,\pm15^\circ$ and we restrict to best-fit fluxes only. Thus, explicitly, 
we define the covariance matrix as
\begin{equation}\label{eq:cov_model}
    \Sigma_{ij,{\rm mod}}=\left< E^{4}\frac{d \Phi}{dE_{i}} \frac{d \Phi}{dE_{j}} \right> - \left< E^{2}\frac{d \Phi}{dE_{i}}\right> \left<  E^{2}\frac{d \Phi}{dE_{j}} \right>,
\end{equation}
where the notation $\langle \cdot \rangle$ represents an average with respect to the 22 different ROIs centered 
at $20^\circ \leq |\ell | \leq 70^\circ$ in steps of $5^\circ$ and $\frac{d \Phi}{dE_{i}}$ is the best-fit GCE-like flux 
from the $i$th energy bin, for each given ROI. Each entry is computed by taking the difference of the average 
of the product and the product of the averages for the flux of two energy bins. The matrix is by construction 
symmetric and positive semi-definite, as is expected. 
The units for each entry are the square of those of the fluxes from Fig.~\ref{fig:GCE_emission_ALL_models}.

For completeness, we also tested the results using all 80 models instead of the best-fits only, and we find that 
the entries of $\Sigma_{ij,{\rm mod}}$ increase by roughly a factor of 2. Likewise, if we had used the best fits 
for all regions down to $\ell = \pm 5^\circ$, the entries of $\Sigma_{ij,{\rm mod}}$ would increase by roughly 
$30\%$. These approaches will likely overestimate the errors associated to the template-fitting procedure, 
since many of these models can be discarded based on their log-likelihoods.

\subsection{Truncated Covariance Matrix}

Once the covariance matrix is configured, we 
approximate it via a Principal Component Analysis (PCA). 
We do this to remove the small eigenvalues of the covariance matrix, which are likely due to noise, thereby making its inversion more robust.
In fact, we want to ensure that we do not count noise from other regions as a source of systematic errors. In our 
analysis of the GCE spectrum, we include the statistical noise of the 
best fit model in the central $40^{\circ} \times 40^{\circ}$ region.   
After doing the eigendecomposition of the covariance matrix to obtain its eigenvalues $\lambda$ and orthonormal eigenvectors $v$, we define
\begin{equation}
   w_i=\lambda_i/\sum_i \lambda_i , \qquad \qquad
   {\rm PC}_{ij} =  \sqrt{\lambda_i} v_{ij} \label{eq:weights_pca},
\end{equation}
where the indices $i$ and $j$ each go over the 14 energy bins. The  matrix is recovered by $\Sigma_{jk,{\rm mod}} = \sum_{i=1}^{14} {\rm PC}_{ij}^{\sf T} {\rm PC}_{ik}$, where ${\sf T}$ denotes the transpose. By ``the $i^{\rm th}$ principal component'' or ``${\rm PC}_i$'', we will mean the $i^{\rm th}$ 14-entry vector given in Eq.~\ref{eq:weights_pca}. As is evident, the overall sign of each PC is ambiguous, since 
only their product with themselves is known.

\begin{figure}[!t]
\includegraphics[width=8cm]{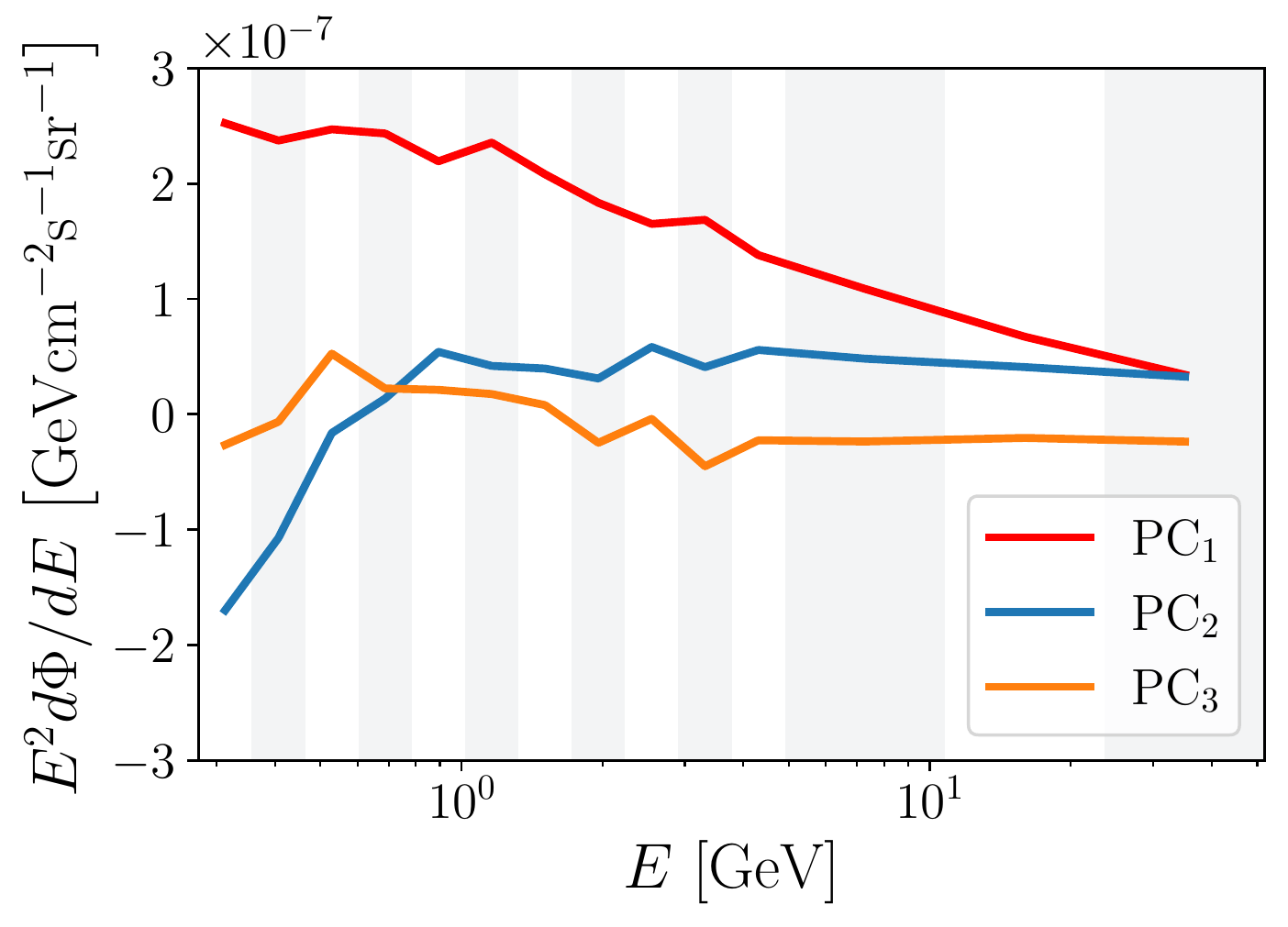}
\centering
\caption{The first three principal components from the singular value decomposition of the covariance matrix. 
}
\label{fig:pca_cov}
\end{figure}

\setlength{\tabcolsep}{3pt}
\begin{table*}[!tb]
    \begin{tabular}{c|cccccccccccccc}
    \hline
        ${\rm PC}_i$ & $\Phi_1$ & $\Phi_2$ & $\Phi_3$ & $\Phi_4$ & $\Phi_5$ & $\Phi_6$ & $\Phi_7$ 
        & $\Phi_8$ & $\Phi_9$ & $\Phi_{10}$ & $\Phi_{11}$ & $\Phi_{12}$ & $\Phi_{13}$ & $\Phi_{14}$ \\
        \hline\hline
        ${\rm PC_1}$ & $2.52$ & $2.37$ & $2.47$ & $2.43$ & $2.19$ & $2.35$ & $2.08$ & $1.83$ & $1.65$ & $1.69$ & $1.38$ & $1.09$ & $0.67$ & $0.34$ \\
        ${\rm PC_2}$ & $-1.70$ & $-1.07$ & $-0.16$ & $0.14$ & $0.54$ & $0.42$ & $0.40$ & $0.31$ & $0.58$ & $0.41$ & $0.56$ & $0.48$ & $0.41$ & $0.33$ \\
        ${\rm PC_3}$ & $0.27$ & $0.06$ & $-0.53$ & $-0.22$ & $-0.21$ & $-0.18$ & $-0.08$ & $0.25$ & $0.04$ & $0.45$ & $0.23$ & $0.24$ & $0.20$ & $0.24$ \\
        ${\rm PC_4}$ & $0.20$ & $-0.15$ & $0.15$ & $-0.14$ & $0.06$ & $-0.04$ & $-0.04$ & $-0.27$ & $0.08$ & $-0.25$ & $0.11$ & $0.25$ & $0.27$ & $0.17$\\
        \hline
    \end{tabular}
    \caption{The first four principal components of the systematic uncertainty contribution to the covariance matrix, defined as in Eq.~\ref{eq:weights_pca}, in units of $10^{-7}\,{\rm GeV \, cm^{-2}\,  s^{-1} \, sr^{-1}}$.}
    \label{tab:pcavals}
\end{table*}

We display the results of the PCA of the systematic error covariance matrix in Fig.~\ref{fig:pca_cov}, where we plot the top 3 PCs in units of residual flux, and we provide the values of first four PC vectors in Tab.~\ref{tab:pcavals}. 
From Fig.~\ref{fig:pca_cov}, it is evident that PC 1 dominates the other two PCs, as should be expected. The values of these first three principal components are very close to those obtained in \cite{Calore:2014nla}. The stability of the data covariance matrix despite increases in the quantity of (and improvements in the quality of) the underlying data, the new 4FGL-DR2 point source catalog, the entirely new set of templates generated in this work, and various different modeling choices adopted over time supports our claim that this procedure captures real systematic limitations to the template fit.

We find that $\sum_{i=1}^{j=3} w_i \simeq 0.99$, so by using the first 3 PCs our truncated matrix is able to represent the original matrix with 99$\%$ accuracy. 
Thus, we take
\begin{equation} \label{eq:systematic_errsq}
    \Sigma_{jk,{\rm mod}} \simeq \Sigma_{jk,{\rm mod}}^{\rm trunc} \equiv \sum_{i=1}^3 {\rm PC}_{ij}^{\sf T} {\rm PC}_{ik}.
\end{equation}
%
%
We note that our fits to the GCE in the following section are robust to including three or more principal components, but the fit quality degrades substantially in all cases if we use two or fewer PCs of the systematic covariance matrix.
For reference, we provide the entries of the first four  components of the PCA in 
Table~\ref{tab:pcavals}. We use the first three of these components, as in Eq.~\ref{eq:systematic_errsq}; we report the fourth to demonstrate that it indeed makes a smaller contribution to the covariance matrix.

\section{Interpretations}
\label{sec:Interpretations}
In this section, we consider possible interpretations of the GCE as characterized in the preceding sections
, emphasizing two possibilities: a population of millisecond 
pulsars (MSPs) and the annihilation of dark matter (DM) particles, as well as the combination of these two 
underlying explanations. There are also other possibilities: for instance, cosmic-ray burst activity from the 
region around the supermassive black hole. However, the MSPs interpretation comes with a relatively 
well measured spectrum from gamma-ray observations toward known galactic MSPs from other regions on the sky. 
Also, the DM prompt emission spectrum can be modeled once the DM mass and annihilation channels are fixed. 
Thus these two interpretations provide spectra that we can easily test. With cosmic-ray bursts from 
the inner galaxy there is no independent observation or theory that can give us a probable gamma-ray spectrum 
that could then be tested in the fit. One would have to perform multiple simulations of bursts that each would 
give its own suggested spectrum and morphology in the inner galaxy which would then be fit to the residual GCE spectra
obtained in Fig.~\ref{fig:GCE_emission_ALL_models} and \ref{fig:GCE_emission_N_vs_S} in this work. 
As a general pattern we expect cosmic-ray bursts to give a gamma-ray emission spectrum described by either 
a simple power-law, or one power-law at low gamma-ray energies that transitions to a softer 
spectrum at higher gamma-ray energies \cite{Carlson:2014cwa, Petrovic:2014uda, Cholis:2015dea}.
These can be 
modeled phenomenologically using a broken power-law or a single power law with an exponential cutoff.

For each model which has a predicted spectrum determined by some free parameters $\theta_k$, we will define a $\chi^2$ test statistic,
\begin{equation} \label{eq:chi2}
    \chi^2  \! = \! \sum_{ij} \! \left( \! {\rm GCE}_i - \! \sum_k f_{ik}( \theta_k)  \! \right) C_{ij}^{-1}  \! \left( \! {\rm GCE}_j - \! \sum_\ell f_{j \ell}(\theta_\ell)  \! \right).
\end{equation}
The values ${\rm GCE}_i$ are the ones depicted in Figs.~\ref{fig:GCE_emission_ALL_models} and \ref{fig:GCE_emission_N_vs_S}, 
which have been given in Tab.~\ref{tab:gcevals} for reference. 
The covariance matrix is $C_{ij} = \sigma_i^2 \delta_{ij} +\Sigma_{ij,{\rm mod}}$, where $\Sigma_{ij,{\rm mod}}$ is 
defined in Eq.~\ref{eq:systematic_errsq}. 
We clarify again, that the covariance matrix is evaluated by studying the systematic galactic diffuse emission modeling 
uncertainties as described in Section~\ref{sec:Uncertainties}. Those are calculated by using the $40^{\circ} \times 40^{\circ}$ 
regions of interest along the galactic disk excluding the central one the as shown in Fig.~\ref{fig:violin_22ROIs_14Ebins}; and by 
using the entirety of our 80 galactic diffuse emission models.
We test the DM, MSP and phenomenological burst-like spectra on the data from the $40^{\circ} \times 40^{\circ}$
region and also from the north or south only regions. We will define $\Sigma_{ij,{\rm mod}}$ in the north (south) to 
be $0.55^2 (0.45)^2$ as 
large as $\Sigma_{ij,{\rm mod}}$ in the full sky, since the north (south) accounts for roughly 55\%(45\%) of the total 
log-likelihood of the $40^{\circ} \times 40^{\circ}$ window. Because we use 14 energy bins to characterize the 
GCE, the indices $i$ and $j$ run from 1 to 14. The indices $k$ and $\ell$ run from 1 to the number 
of free parameters for each model, $N_{\rm fp}$, which ranges from 1 to 4 for the spectra we consider. For the MSP explanation, 
since we fix its spectrum we take only one free parameter, its normalization. For DM we have two parameters, the 
mass and the annihilation channel, which we 
assume is to only a single species of Standard Model particle. 
The 
power-law plus exponential spectrum has a normalization, power-law index, and cutoff energy, 
while the broken power-law spectrum has a normalization, two power-law indices, and the location of a break.

The best fit point for a given model is the one that minimizes the $\chi^2$. We will use $\widehat \chi^2$ to refer to the value of Eq.~\ref{eq:chi2} at this best fit point. We will also use a $p$-value to describe the goodness of the fit at this point. This is
\begin{equation}
    \hat p = 1 - {\rm CDF}_{\chi^2| 14- N_{\rm fp}}(\widehat \chi^2),
\end{equation}
where ${\rm CDF}_{\chi^2| 14- N_{\rm fp}}$ is the cumulative distribution function of the $\chi^2$ distribution with $14-N_{\rm fp}$ degrees of freedom. A $p$-value $\hat p \gtrsim 0.1$ is suggestive of a good fit.

We present results separately for the full $40^\circ \times 40^\circ$ region of interest, the southern sky only, and the northern sky only. 
The $p$-value of every model that we consider is very small when we consider the $40^{\circ} \times 40^{\circ}$ region as the region of interest; 
it is even worse when restricting to the northern hemisphere. However, in some of the scenarios we consider, 
the value of $\hat p$ is larger than 0.1 when considering the southern hemisphere only. The northern hemisphere within the 
$40^{\circ} \times 40^{\circ}$ window is relatively brighter than the southern one. This is mostly due to diffuse emission from dense ISM gases.
Compared to the south, even a small fractional error in the model prediction 
of the diffuse emission in the north can lead to systematic errors on interpretations of GCE emission in 
that region. This has been anticipated by, and 
lends credence to, claims in \cite{Leane:2019xiy, Leane:2020nmi}. 
For these reasons, we suggest that interpretation based on the results from analysis of the southern hemisphere 
on its own is likely valid. The results for all of the models fit to the data in the full sky, the southern hemisphere, 
and the northern hemisphere are collected for easy comparison in Tab.~\ref{tab:Interps}.

\subsection{Millisecond Pulsars}

MSPs have been suggested as an explanation for the GCE previously \cite{Abazajian:2012pn, Abazajian:2014fta, Lee:2014mza, Lee:2015fea, Petrovic:2014xra, Brandt:2015ula, OLeary:2015qpx, Bartels:2017vsx, Ploeg:2020jeh, Gautam:2021wqn}
, although their suitability as an explanation for the entirety of the GCE has been subject to various criticisms \cite{Hooper:2013nhl,  Cholis:2014lta, Zhong:2019ycb}.\footnote{New methods based on photon statistics have been developed, e.g.~\cite{Collin:2021ufc, List:2021aer, Mishra-Sharma:2021oxe}.}
The qualitative reason that MSPs are believed to be a good explanation for the GCE is because their energy spectrum is known to peak near a few GeV \cite{Fermi-LAT:2013svs, Cholis:2014noa} and they may have TeV halos (e.g., \cite{Hooper:2018fih, Guepin:2018jkb, Hooper:2021kyp}). 
Since this would be a new population of sources, its morphology is not well constrained \cite{Abazajian:2012pn, Abazajian:2014fta, Bartels:2017vsx}.
We expect it 
to follow the morphology of dense stellar environments such as the bulge. 
Dense stellar regions are environments that are known to host MSPs and may also be the
places where most MSPs form. Given that in Fig.~\ref{fig:Morphology} we tested the spherically symmetric GCE against
known bulge alternatives as the Boxy Bulge and the X-shaped Bulge and found it to be preferable by more than $2 \Delta \ln(\mathcal{L}) \gtrsim 1000$ 
adds significant tension to the MSP interpretation, purely from morphological arguments.

We may nevertheless utilize the spectral fit of Eq.~\ref{eq:chi2}, which takes the results of Fig.~\ref{fig:GCE_emission_ALL_models} 
and thus is tantamount to assuming that the MSP morphology is spherically symmetric and falls off with galactocentric radius like $r^{-2\times 1.2}$.
We model the MSP spectrum as
\begin{equation} \label{eq:mspspec}
   d \Phi_{\rm MSP}(E_\gamma)/dE_\gamma \propto E_\gamma^{-\alpha_{\rm MSP}} \times \exp(-E_\gamma/E_{\rm cut}).
\end{equation}
We fix $\alpha_{\rm MSP} = 1.57$ and $E_{\rm cut} = 3.78 \GeV$, which were derived in Ref.~\cite{Cholis:2014noa} by stacking the gamma-ray emission 
from known MSPs, and which agree with those obtained in \cite{Ploeg:2020jeh}. The only free parameter when we fit a population of MSPs to the GCE is the overall normalization in front of the proportionality in Eq.~\ref{eq:mspspec}. If an astrophysically novel population of MSPs has a very different energy spectrum compared to Eq.~\ref{eq:mspspec} which describes known MSPs, such a population will be more aptly characterized by one of the phenomenological fitting functions with more parametric freedom considered below.

Because of the relatively low-energy of the exponential cutoff in Eq.~\ref{eq:mspspec} and the relatively robust high-energy tail of the 
GCE as shown in Figs.~\ref{fig:GCE_emission_ALL_models} and \ref{fig:GCE_emission_N_vs_S}, we find that MSPs do not offer a 
good fit to the GCE in any region of interest. This statement holds solely due to the energy spectrum of the GCE 
and particularly to its high-energy tail. This is independent 
of the morphological details related to the MSP spatial distribution. 
Also, our fit is independent of the manner in which MSP pulsars are 
distributed with respect to their luminosity or other astrophysical characteristics. The fact that known 
MSPs 
provide a bad fit to the GCE spectrum
further strengthens the 
results from our morphological analysis which disfavor MSPs as the source of the entirety of the GCE emission.

\begin{figure*}[!t]
\includegraphics[width=0.48\textwidth]{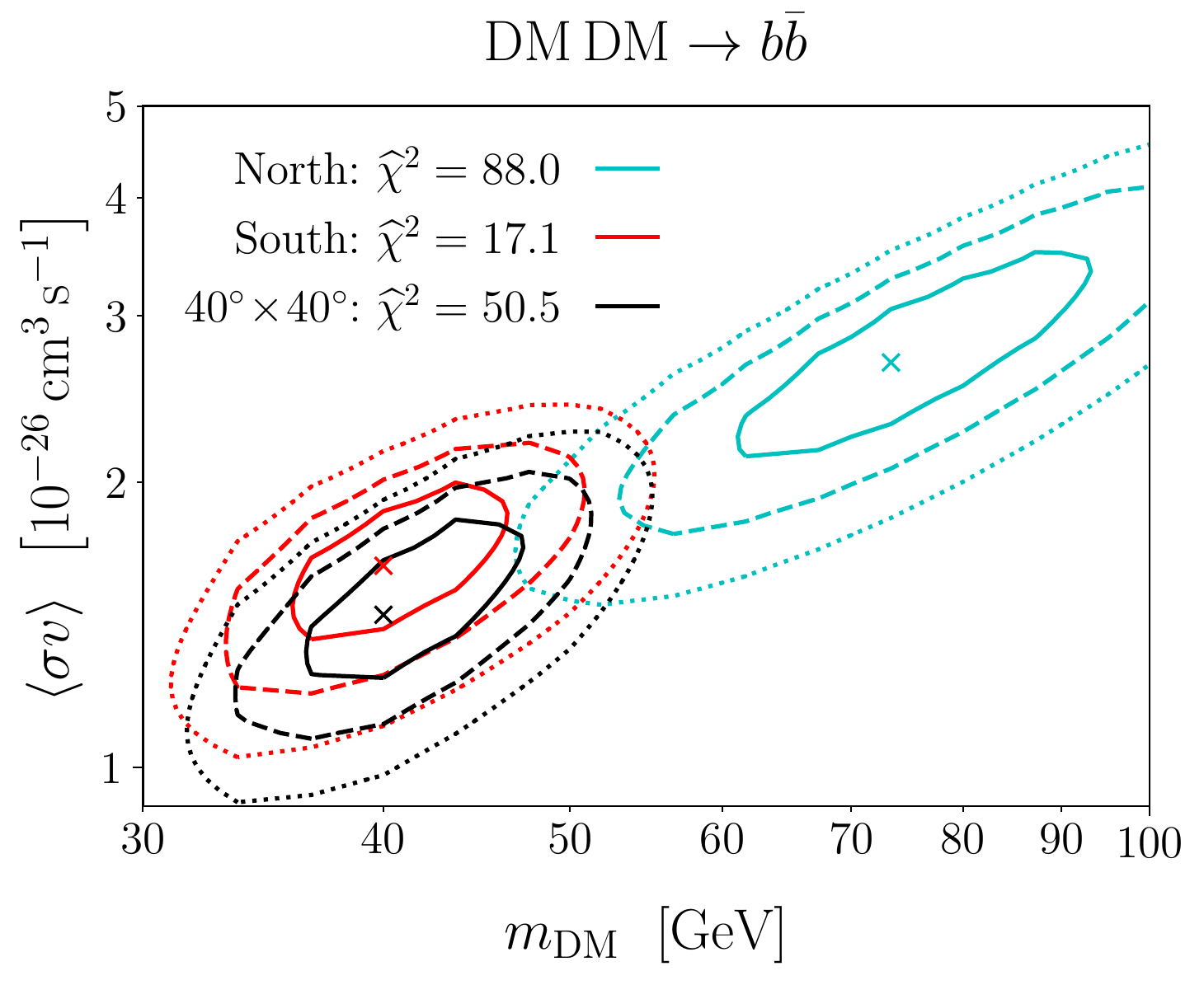}~~~
\includegraphics[width=0.48\textwidth]{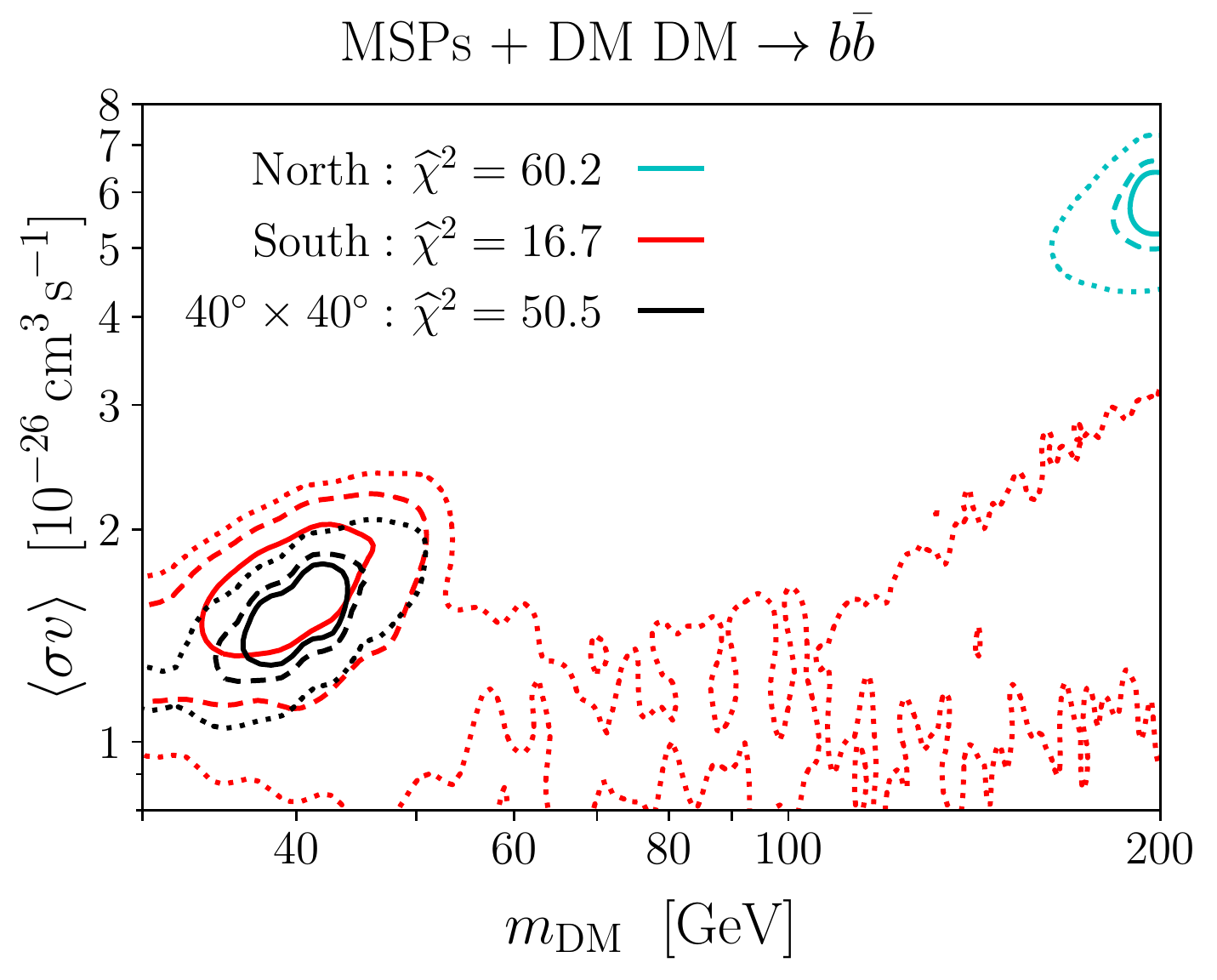}
\centering
\caption{{\it Left:} Preferred parameter space for $\text{DM DM}\to b \bar b$. The $\widehat \chi^2$ value is the best value of Eq.~\ref{eq:chi2} for each model. The solid (dashed) [dotted] contours show values of $\Delta \chi^2=$ 2.30 (6.18) [11.83], suitable for 1(2)[3]$\sigma$ limits for two fit parameters. {\it Right:} Preferred parameter space for $\text{DM DM}\to b \bar b$ 
when allowing a freely floating contribution from MSPs. The $\widehat \chi^2$ value is the 
value of Eq.~\ref{eq:chi2} at the maximum-a-posteriori parameter point for each 
ROI. The solid 
(dashed) [dotted] contours show 
the $50\% (68\%) [95\%]$ credible intervals.}
\label{fig:dmbbonly}
\end{figure*}

\subsection{Dark Matter}

Dark matter annihilation has been previously suggested as an explanation for the GCE \cite{Daylan:2014rsa, Hooper:2011ti, Abazajian:2012pn, Cline:2013gha, Abazajian:2014fta, Calore:2014nla, Hooper:2013rwa, Buckley:2014fba, Berlin:2014tja, Ipek:2014gua, Arhrib:2013ela, Boehm:2014hva, Abdullah:2014lla, Agrawal:2014una, Martin:2014sxa, Alves:2014yha, Agrawal:2014oha, Berlin:2015wwa, Arcadi:2014lta, Karwin:2016tsw}
The qualitative reason that DM annihilation is believed to be a good explanation for the GCE 
is because the spherical morphology at the galactic center is a natural consequence 
of DM structure formation. Moreover the GCE emission is very suggestive of a thermal relic dark matter 
particle. These DM annihilation models have two fit parameters: $m_{\rm DM}$, which determines 
the photon energy spectrum, and the annihilation cross section $\sigmav$, which determines 
the brightness of the signal. Because of the high-energy tail that we detect at high confidence, 
we consider three simple annihilation channels to $b \bar b$, $hh$, and $ZZ$ in this work, leaving 
a more comprehensive analysis on the possible DM particle couplings to known Standard 
Model particles to future work.

First, we consider dark matter annihilation to $b\bar b$. As shown in Tab.~\ref{tab:Interps}, 
the $p$-value of the dark matter models that we test is not good across the $40^{\circ} \times 40^{\circ}$ 
window nor in the northern hemisphere alone. Restricting only to the southern sky, 
the $p$-value is as large as $0.15$, which is among the best fits for any model we have tested. We note again that
the northern sky has more galactic diffuse background 
emission compared to the south. This makes the northern sky a region where characterizing the GCE 
is more challenging. 
We feel that the fit in the cleaner southern sky provides some evidence in favor of the possibility 
that the GCE is due to DM annihilation to $b \bar b$.

\begin{figure}
\begin{centering}
\includegraphics[width=3.65in,angle=0]{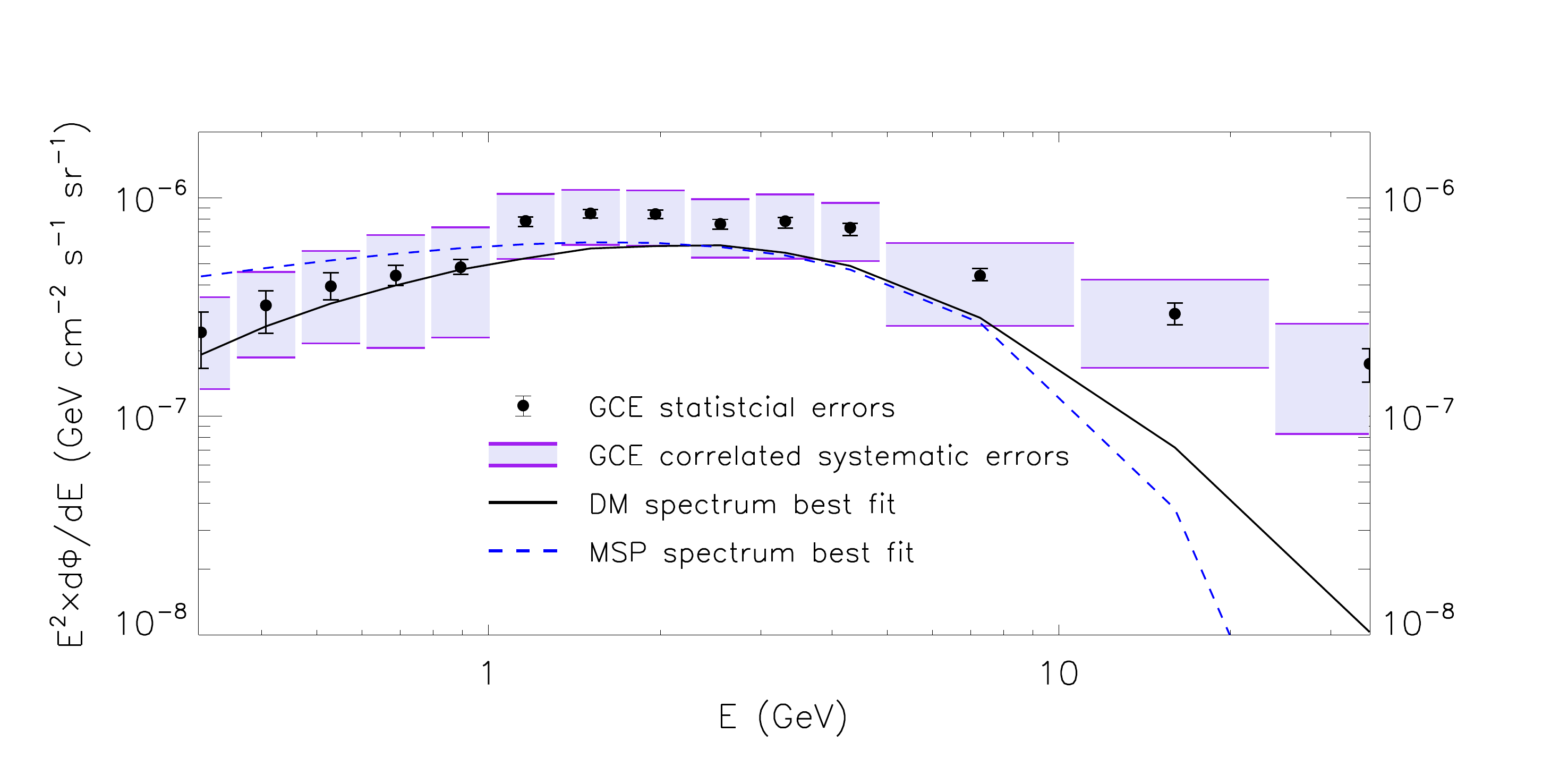}
\end{centering}
\vskip -0.3in
\caption{The GCE statistical and correlated systematic errors for the $40^{\circ} \times 40^{\circ}$ window, 
centered at the best fit normalizations of the best fit background model XLIX. For the correlated systematic 
errors we used a first three principal components of Table~\ref{tab:pcavals}. We also show -in solid back 
and dashed blue respectively- the best fit DM model and MSP spectra for the same window. For the DM 
model the best fit in the $40^{\circ} \times 40^{\circ}$ window is achieved for $m_{\rm DM}$ =40 GeV 
annihilating to $b \bar b$ with $\sigmav = 1.45 \times 10^{-26}\cm^3/\s$, while for the MSP spectrum, we 
take parameterization of Ref.~\cite{Cholis:2014noa}, with a normalization of $6.22 \times 10^{-7}$ 
GeV$\cm^{-2} \s^{-1} \textrm{sr}^{-1}$ at 1.96 GeV.}
\label{fig:GCEErrors}
\end{figure}

We show the results in the DM parameter space for $\text{DM DM}\to b \bar b$ in Fig.~\ref{fig:dmbbonly} (left)
for all three ROIs. As is clear, the preferred parameter space for the $40^{\circ} \times 40^{\circ}$ 
window and the southern-hemisphere analyses are compatible, and prefer dark matter parameter values 
near $m_{\rm DM} \simeq 40 \GeV$ and $\sigmav$ in the range $(1-2) \times 10^{-26}\cm^3/\s$. 
In Fig.~\ref{fig:GCEErrors}, we also show the best fit choices for the DM and MSP spectra from performing
the fit to Eq.~\ref{eq:chi2}. We note that compared to the earlier estimation of the statistical and correlated 
systematic errors of \cite{Calore:2014xka} in the same window, the statistical errors have been reduced by 
about 40$\%$, while the systematic errors have been reduced in the $<2$ GeV range but have not
substantially changed above that energy.  
We also note that the overall GCE spectrum in the $40^{\circ} \times 40^{\circ}$ window is similar to those of 
\cite{Calore:2014xka, DiMauro:2021raz} in the same region. 
The preferred parameter space when analyzing the northern sky only is in tension with these values, 
tending to a larger mass and correspondingly larger cross section. The values in the southern-hemisphere 
and in the $40^{\circ} \times 40^{\circ}$ window  analysis are compatible with previous fits 
\cite{Daylan:2014rsa, Calore:2014nla, Agrawal:2014oha}.

The fits to $hh$ and $ZZ$ alone do not result in large $p$-values, even restricting to the southern sky. 
The reason for this is because these models generically under-predict the gamma-ray flux at low 
energies, where the GCE in fact reaches its peak.

\subsection{Dark Matter Plus Millisecond Pulsars}
\label{sec:Interp_MSPDM}


Here, we explore the possibility that a combination of dark matter and MSPs can explain the GCE. 
MSPs may account for part of the low-energy portion of the GCE spectrum while DM annihilation accounts 
for the emission at higher energies. 

To study this possibility, we have three fit parameters: the dark matter mass and annihilation cross section 
$\sigmav$ as well as the normalization of the MSP spectrum as given in Eq.~\ref{eq:mspspec}. 
We use the dynamic nested sampler {\tt dynesty} \cite{2020MNRAS.493.3132S, 2004AIPC..735..395S, 10.1214/06-BA127, Higson:2019, 2009MNRAS.398.1601F} 
and we visualize results with {\tt getdist} \cite{Lewis:2019xzd}. We take a linear-flat prior on the dark matter mass with maximum value of 200 GeV and minimum value of $30$ GeV, $m_Z$, and $m_h$ for the case of annihilation to $b \bar b$, $ZZ$, and $hh$, respectively; a linear-flat prior on $\sigmav \in 0-20 \times 10^{-26}\cm^3 {\rm s}^{-1}$; and a log-flat prior on the magnitude of the differential flux of the MSP component at a GeV ranging from $10^{-15} - 1 \ {\rm GeV^{-1} \, cm^{-2} \, s^{-1}  \, sr^{-1}}$, which is necessarily positive.
We co-vary all three parameters and present our results in the two-dimensional DM parameter space by marginalizing over the MSP contribution.

We show the results for MSPs$+\text{DM DM} \to b \bar b$ in the right panel of Fig.~\ref{fig:dmbbonly}. We show the corner plot of the full parameter space in Fig.~\ref{fig:corner_mspb} of App.~\ref{sec:cp_mspdm}.
We find that for annihilation to $b \bar b$, the addition of MSPs does not meaningfully improve the 
quality of the fit as measured by the value of $\widehat \chi^2$. 
The addition of one more fit parameter, describing the normalization of the MSP spectrum, 
does not change the $\hat p$-value for the $40^{\circ}\times 40^{\circ}$ window, nor does it have a major impact 
in the southern hemisphere. Only for the northern hemisphere that provided the poorest fit 
is there a substantive change, as is clear from a comparison of the left and right panels of Fig.~\ref{fig:dmbbonly}.
As we show, the preferred mass parameter space in the northern hemisphere shift considerably. 
In the high-mass parameter space, MSPs provide a significant fraction 
of the GCE, particularly at low energy. 
Given the complexity of the northern hemisphere, we believe that 
the additional model complexity is not justified if dark matter annihilates to $b \bar b$.
We take this opportunity to point 
out that, because our analysis only accounts for gamma rays with energies up to 50 GeV, fitting to 
higher-mass DM than the values shown here is not appropriate. 

Because MSPs can ``soak up'' low-energy emission where the excess peaks, the addition of MSPs 
does lead to a significantly improved fit when considering DM annihilation to $hh$ or $ZZ$. The 
largest value of $\hat p$ found in any of our analyses is obtained when fitting the model 
$\text{DM DM}\to ZZ $ plus MSPs to the data in the southern hemisphere. In App.~\ref{sec:cp_mspdm} we show the full results of adding an MSP component to the $\text{DM DM}\to ZZ $ and $\text{DM DM}\to hh $
channels.
Because MSPs are responsible for the low-energy GCE flux, they contribute an order 
one fraction to the total GCE at the best-fit point for a given model: at the best-fit point in the full-sky (southern hemisphere) fit for MSPs 
plus $\text{DM DM}\to ZZ $, the MSPs contribute 69.1\% (75.1\%) of the flux. Very similar values 
are found for the analysis for MSPs plus $\text{DM DM}\to hh$. 
These ratios are calculated after 
finding the maximum-a-posteriori point from our parameter-space exploration. 
In this way, we can define the fraction of the energy flux due to MSPs,
\begin{equation} \label{eq:fhat}
    \hat f_{\rm MSP} = \frac{\int dE \, E \, d\Phi_{\rm MSP}/dE}{\int dE \, E \, d\Phi_{\rm MSP}/dE + \int dE \, E \, d\Phi_{\rm DM}/dE}.
\end{equation}
The integration in energy $E$ is carried between 0.275 GeV to 51.9 GeV since this is the range of gamma-ray energies that we use throughout the analysis.We provide values of $\hat f_{\rm MSP}$ in Tab.~\ref{tab:Interps}.

\subsection{Bursts of Matter from the Inner Galaxy}
The GCE may be the result of relatively recent activity in the central region of the Milky Way, either 
directly related to baryon accretion by the supermassive black hole or a period of stellar activity 
in the inner galaxy environment. 
If so, and the GCE is due to cosmic ray electrons, then its
age is only of the order of Myr \cite{Petrovic:2014uda, Cholis:2015dea}.  We consider this possibility 
by fitting the spectrum with two parametric forms: a power-law plus an exponential cutoff, or a broken 
power-law. The power-law + exponential cut-off
has a spectrum of $dN/dE \propto E^{-\alpha} \cdot \exp(-E/E_{\textrm{cut}})$. The broken power-law has a spectrum of $dN/dE \propto E^{-\alpha_{1}}$ for $E \leq E_{\textrm{br}}$ and $dN/dE \propto E^{-\alpha_{2}}$ for $E > E_{\textrm{br}}$. These parametric forms may also be suitable for describing the spectra of other classes of astrophysical emitters. For example, a central population of young and middle-aged pulsars that are more powerful sources than MSPs and emit significant amounts of high-energy cosmic-ray electrons and positrons, can result in high-energy gamma rays via ICS as well \cite{Hooper:2017rzt, Sudoh:2021avj, Sudoh:2019lav, Sudoh:2020hyu}. As cosmic-ray electrons and positrons diffuse away from the center we expect that ICS emission to be less concentrated than the GCE profile. The exact gamma-ray morphology from such a population would require further study.

These do not provide particularly good global fit to the GCE obtained here, although the broken power 
law spectrum actually provides the lowest $\widehat \chi^2$ and the largest $\hat p$-value for the 
analysis of the $40^{\circ} \times 40^{\circ}$ window, but this $\hat p$-value is only 0.01 which is nevertheless rather small. 
In the southern-sky analysis, the power-law + exponential cut-off and the broken power-law fits 
allow for $\hat p$-values of 0.02 and 0.05 respectively, which are moderately worse than the fit of 
$\text{DM DM}\to b \bar b$ or MSPs combined with DM annihilation.

In Table~\ref{tab:Interps}, we
provide the best fit values for the parametric fits we describe in this section.
If indeed the GCE is the result of cosmic-ray burst activity in the 
past history of the inner galaxy, there is no reason that both hemispheres would have the exact same spectra. 
That could explain why the GCE spectrum is not the same in the north and south. 

\begin{table*}[t]
    \begin{tabular}{l|cccl}
    \hline
            Model & $\widehat \chi^2/{\rm dof}$ & $\hat p$-value & ROI & notes \\
            \hline \hline
            & 76.6/13 & $<10^{-6}$ & $40^{\circ}\times 40^{\circ}$ & - \\
            MSPs & 34.5/13 & $1.0\times 10^{-3}$ & southern sky & - \\
            & 194.5/13 & $<10^{-6}$ & northern sky & - \\
            \hline
            & 50.5/12 & $1.1 \times 10^{-6}$ & $40^{\circ}\times 40^{\circ}$ & see Fig.~\ref{fig:dmbbonly} \\
            $\text{DM DM} \to b \bar b$ & 17.1/12 & 0.15 & southern sky & see Fig.~\ref{fig:dmbbonly}  \\
            & 88.0/12 & $<10^{-6}$ & northern sky & see Fig.~\ref{fig:dmbbonly}  \\
            \hline
            & 107.7/12 & $<10^{-6}$ & $40^{\circ}\times 40^{\circ}$ & - \\
            $\text{DM DM} \to ZZ$ & 62.7/12 & $<10^{-6}$ & southern sky & - \\
            & 80.7/12 & $<10^{-6}$ & northern sky & - \\
            \hline
            & 74.8/12 & $<10^{-6}$ & $40^{\circ}\times 40^{\circ}$ & - \\
            $\text{DM DM} \to hh$ & 39.6/12 & $8.5\times 10^{-5}$ & southern sky & - \\
            & 74.0/12 & $<10^{-6}$ & northern sky & - \\
            \hline
            & 50.5/11 & $<10^{-6}$ & $40^{\circ}\times 40^{\circ}$ & $\hat f_{\rm MSP} < 0$; see Figs.~\ref{fig:dmbbonly} and \ref{fig:corner_mspb}  \\
            MSPs+$\text{DM DM} \to b \bar b$ & 16.7/11 & 0.12 & southern sky & $\hat f_{\rm MSP} =0.82$; see Figs.~\ref{fig:dmbbonly} and \ref{fig:corner_mspb}  \\
            & 60.2/11 & $<10^{-6}$ & northern sky & $\hat f_{\rm MSP} =0.61$; see Figs.~\ref{fig:dmbbonly} and \ref{fig:corner_mspb}  \\
            \hline
            & 53.5/11 & $< 10^{-6}$ & $40^{\circ}\times 40^{\circ}$ & $\hat f_{\rm MSP} =0.69$; see  Fig.~\ref{fig:corner_mspZh}  \\
            MSPs+$\text{DM DM} \to ZZ$ & 15.5/11 & 0.16 & southern sky & $\hat f_{\rm MSP} =0.76$; see  Fig.~\ref{fig:corner_mspZh}  \\
            & 53.3/11 & $< 10^{-6}$ & northern sky & $\hat f_{\rm MSP} =0.53$; see  Fig.~\ref{fig:corner_mspZh}  \\
            \hline
            & 60.9/11 & $<10^{-6}$ & $40^{\circ}\times 40^{\circ}$ & $\hat f_{\rm MSP} =0.61$ see; Fig.~\ref{fig:corner_mspZh}  \\
            MSPs+$\text{DM DM} \to hh$ & 17.5/11 & 0.09 & southern sky & $\hat f_{\rm MSP} =0.74$; see Fig.~\ref{fig:corner_mspZh}  \\
            & 65.9/11 & $<10^{-6}$ & northern sky & $\hat f_{\rm MSP} =0.36$; see Fig.~\ref{fig:corner_mspZh}  \\
            \hline
            & 58.3/11 & $<10^{-6}$ & $40^{\circ}\times 40^{\circ}$ & $\alpha=0.0^{+0.6}_{-0.4}    ,  E_{\rm cut} =1.3^{+0.3}_{-0.4} \textrm{GeV}$ \\
            power-law + exponential & 23.1/11 & 0.02 & southern sky & $\alpha=1.3 \pm 0.2,  E_{\rm cut} =3.5^{+0.6}_{-1.0} \textrm{GeV}$ \\
            & 109.4/11 & $<10^{-6}$ & northern sky & $\alpha=0.1^{+0.4}_{-0.5}    ,  E_{\rm cut} = 1.6\pm 0.4\textrm{GeV}$ \\
            \hline
            & 22.5/10 & 0.01 & $40^{\circ}\times 40^{\circ}$ & $\alpha_{1} = 0.5^{+0.3}_{-0.4}$, $\alpha_{2} = 2.57^{+0.06}_{-0.09} $, $E_{\textrm{br}} = 1.66\pm 0.09$ GeV \\
            broken power-law & 18.0/10 & 0.05 & southern sky & $\alpha_{1} = 1.5^{+0.3}_{-0.1}$, $\alpha_{2} = 2.8^{+0.1}_{-0.2} $, $E_{\textrm{br}} = 2.3\pm 0.6 $ GeV \\
            & 23.7/10 & $8.3 \times 10^{-3}$ & northern sky & $\alpha_{1} = 0.57\pm 0.28  $, $\alpha_{2} = 2.48\pm 0.04 $, $E_{\textrm{br}} = 1.75^{+0.07}_{-0.11} $ GeV \\
            \hline
    \end{tabular}
    \caption{Best fit summary statistics for various hypothetical origins of the GCE. We report the value of the $\hat{\chi}^2$ at the best fit point and the $p$-value of this $\hat{\chi}^2$ value given the number of degrees of freedom in the fit. For the parameters of the ``power-law + exponential'' and ``broken power-law'' models, we report $50\%$ and $68\%$ interval values.}
    \label{tab:Interps}
\end{table*}

\subsection{Summary}

No single explanation provides a good fit (with a $\hat p$-value much greater than 0.1) 
to the GCE across the entire $40^\circ \times 40^\circ$ region we have analyzed. 
This appears to be in contrast with previous works \cite{Calore:2014nla, Daylan:2014rsa}, which found $p$-values closer to 0.5 for some models. 
There are two basic reasons for the degradation in fit. 
One is that we have fewer energy bins than past analyses, which means that our 
statistical errors are smaller throughout and we account for a smaller number of data points 
when we report $\hat{\chi}^2/{\rm dof}$ and its associated $\hat p$-value.
The principal components of the
covariance matrix that we show in Fig.~\ref{fig:pca_cov}, which dominate the combined error budget, are similar to those of \cite{Calore:2014nla}, 
which leads us to believe that with more energy bins the $p$-values would be larger:
the statistical error bars would shrink with fewer energy bins, but the systematic error bars, which dominate the $\chi^2$ value in Eq.~\ref{eq:chi2}, would not change significantly, so the overall $\hat \chi^2$ would remian a similar size but the $\hat p$-value would increase. 
Also, since we use about twice the photons as compared to past analyses, the detection of the high-energy tail 
of the GCE, where the systematic errors are lower, is now much more significant than in the earlier works of Refs.~\cite{Calore:2014xka, TheFermi-LAT:2015kwa}. The high-energy tail is most dominant on the northern hemisphere, though it is still plainly evident in the southern sky despite being somewhat softer. 
Given the brightness of the diffuse emission in the northern sky which makes any fit of a subdominant emission components challenging,  
we endorse fits to the GCE that rely solely on data from the relatively cleaner southern sky.

The best $\widehat \chi^2$ value for the analysis in the $40^\circ \times 40^\circ$ ROI is a broken power law model 
which has four free parameters, but it has a $\widehat \chi^2/{\rm dof} > 2$ and a correspondingly small $\hat p \simeq 0.01$. 
In the southern sky only, $\text{DM DM}\to b \bar b$ can provide a good fit on its own, 
with $\hat p = 0.15$, as can the combination of MSPs with $\text{DM DM}\to ZZ$, 
which has $\hat p = 0.16$ and is the overall best $p$-value found in this analysis. 
Qualitatively, the combination of MSPs with $\text{DM DM}\to hh$ is similarly good, with $\hat p = 0.09$. For these reasons, we believe that our results provide some evidence in favor of the possibility that the GCE is due to DM annihilation. 
We provide a summary of our statistical fit results in Table~\ref{tab:Interps} for all interpretations for the GCE that we tested.

\section{Conclusions}
\label{sec:conclusions}

In this work, we have produced an entirely new set of high-resolution galactic diffuse emission templates.
We started by making use of the  new local cosmic-ray measurements by 
\textit{AMS-02} that inform us about the Milky Way conditions within few kpc from us. We then allowed for additional degrees of freedom to account for the fact that 
the ISM conditions and sources in the inner galaxy may be different. 
This has allowed us to revisit the Galactic center gamma-ray excess, which we explore and characterize in great depth.

The templates generated in this work provide excellent fits to the local cosmic-ray spectra. However, the ``inverse problem'' of understanding the conditions at the center of the galaxy is still much more challenging. Many different global distributions of cosmic-rays are consistent with the high-precision local observations we have access to. Because each of these cosmic-ray distributions will lead to a unique gamma-ray map, cosmic-ray observations alone are not enough to predict gamma-ray observations at high precision.

We test a wide range of galactic diffuse model assumptions. These account for 
the possible spatial distribution of cosmic-ray sources and cosmic-ray injection spectral assumptions in the inner galaxy. 
Our models also account for a 
wide range of assumptions on how these cosmic rays propagate away from their sources through diffusion and convection. 
We test how cosmic rays at low energies can be reaccelerated and how electrons at initially high energies can lose their energy to photons due to 
interactions with the galactic magnetic field, with the ISM gas, and the ISRF. 
We test different profiles for the spatial distribution of the magnetic field, ISM gas and the ISRF. 
All these varying assumptions lead to distinctively different morphologies
for the resulting Pi0, bremsstrahlung and ICS diffuse emission template components and the energy evolution of those morphologies.
We test 80 different models for the conditions of the inner galaxy by fitting the linear combination 
of the Pi0, bremesstrahlung, ICS, isotropic, \textit{Fermi} Bubbles and GCE templates to the \textit{Fermi}-LAT data. 
We use MCMC to explore the likelihood of these models.
A large template normalization deviation from the original values of the
Pi0, bremsstrahlung and ICS templates, would suggest different physical assumptions than those used to produce them.
To avoid that, we restrict the template fits to ranges 
where the best-fit normalization values correspond to minor perturbations to the input spectra.

Here, we have performed these fits across several regions of interest near the galactic center and along the galactic disk. Our primary results, fitting to gamma-ray data in the region defined by galactic latitudes $-20^\circ \leq b \leq 20^\circ$ and longitudes $-20^\circ \leq \ell \leq 20^\circ$, reveal strong evidence in favor of a spherically symmetric excess of gamma rays centered at the galactic center, which rises inversely with galactocentric radius as expected from the annihilation of dark matter particles distributed according to a slightly adiabatically contracted NFW profile. The preference for this specific profile is robust against small perturbations away from spherical symmetry or from the precise inner profile. Moreover, the preference for this specific template is also robust against more dramatic deviations that would follow various stellar-mass and star-forming regions.

We repeated this fitting procedure in a number of different subregions near the galactic center. First, we separately tested the northern galactic hemisphere defined by $2^\circ \leq b \leq 20^\circ$ and $-20^\circ \leq \ell \leq 20^\circ$, and the southern galactic hemisphere defined by $-20^\circ \leq b \leq -2^\circ$ and $-20^\circ \leq \ell \leq 20^\circ$. The excess is robustly identified in both regions, with a very similar normalization but with a subtly different spectrum: compared to the analysis of the full $-20^\circ \leq b \leq 20^\circ$ and $-20^\circ \leq \ell \leq 20^\circ$ region, the fit in the north (south) is spectrally harder (softer).

We also tested 28 additional test regions along the galactic disk, translating the favored excess template in steps of $5^\circ$. This test was designed to give a quantitative estimate of the possibility that the galactic center excess is due to systematic mismodeling, and to give a way of prescribing systematic error bars. In all regions tested, the normalization of the fitted excess was lower by more than a factor of 2, the spectrum in an absolute sense was always substantially softer, and the statistical preference for the GCE-like emission in these other regions was smaller than the statistical preference for the GCE itself as long as we restrict to excesses with positive normalization.

Finally, we tested various possibilities that could be ultimately responsible for the excess. We find that no explanation on its own gives an especially high-quality fit to the excess across the full $-20^\circ \leq b \leq 20^\circ$ and $-20^\circ \leq \ell \leq 20^\circ$ region of interest. Furthermore, we find that millisecond pulars on their own give a very poor fit to the excess centered at the galactic origin, regardless of region of interest. This is largely due to the high-energy tail of the excess detected at high confidence in this work. In contrast, dark matter annihilation to $b \bar b$ is a decent fit to the excess in the southern-hemisphere-only analysis, as is a combination of dark matter annihilation along with millisecond pulsars. In the case that millisecond pulsars are added to the dark matter fits, the preferred dark matter cross section does not change significantly, but higher-mass dark matter becomes an acceptable contributor to the emission. 
We also note that a broken power-law that could be the result of 
recent cosmic-ray burst activity in the inner galaxy provides a relatively good fit to the data.

The galactic center excess has confounded physicists for over a decade. 
Even after more than 12.5 years of \textit{Fermi}-LAT observations 
and several improvements in our
understanding of cosmic rays 
in the Milky Way, it remains true that the GCE has a unique spectrum and amplitude along the galactic disk. 

\acknowledgments{
We would like to thank Patrick J.~Fox for collaboration on related work and in early stages of this project. We thank Edward W.~Kolb for  providing  access to the resources from the University of Chicago’s Research Computing Center, which were extensively used in this work.  We also acknowledge the use of \texttt{GALPROP} \cite{GALPROPSite} and thank Igor V. Moskalenko, Andrew W. Strong, Seth Diegel and Jeff Wade for their support. We thank Kevork Abazajian, Dan Hooper, Rebecca Leane, Tim Linden, Manoj Kaplinghat, Siddharth Mishra-Sharma, Simona Murgia, Nicholas Rodd, Tracy Slatyer, and Tim Tait for comments on a draft.
IC acknowledges support from the Michigan Space Grant Consortium, NASA Grants No. NNX15AJ20H and No. 80NSSC20M0124. 
IC acknowledges that this material is based upon work supported by the U.S. Department of Energy, Office of Science, Office of High Energy Physics, under Award No. DE-SC0022352.
YZ acknowledges the Aspen Center for Physics for hospitality during the completion of this work, which was supported by NSF grant PHY-1607611 and partially supported by a grant from the Simons Foundation. YZ is supported by the Kavli Institute for Cosmological Physics at the University of Chicago through an endowment from the Kavli Foundation and its founder Fred Kavli. 
Fermilab is operated by Fermi Research Alliance, LLC under Contract No. De-AC02-07CH11359 with the United States Department of Energy, Office of High Energy Physics.
}

\begin{appendix}

\section{Testing the Impact of Different Mask Assumptions on the GCE Ellipticity}
\label{app:Ellipticity}

One of the most challenging aspects of our analysis has been to 
constrain the morphology of the GCE
. This 
is because low latitudes where the excess may be brightest are removed by the galactic disk mask of $|b| \leq 2^{\circ}$ and the 4FGL-DR2 point sources.
Moreover, low latitudes are where remaining bright sub-threshold point 
sources may lay, as such latitudes are both where the point source detection threshold is the highest \cite{Fermi-LAT:2019yla} 
and where the galactic background emission peaks.
Of the GCE morphological parameters that we constrain, the ellipticity $\epsilon$ proves to be the most sensitive to the mask choice. 

In this appendix we test the GCE ellipticity parameter for alternative masks to the one we use in the main text in Figs.~\ref{fig:GCE_emission_ALL_models},~\ref{fig:Morphology}, and \ref{fig:GCE_emission_N_vs_S}. We provide the description of these masks in Table~\ref{tab:Masks}.
\begin{table*}[t]
    \begin{tabular}{cccr}
    \hline
            Mask & Galactic Disk Mask & Point Source Catalogue & Masked fraction \\
            \hline\hline
            Standard  & $|b| \leq 2^{\circ}$, $|\ell| \leq 180^{\circ}$ & 4FGL-DR2 & 23.4\%\\ 
            Option 2  & $|b| \leq 2^{\circ}$, $|\ell| \leq 180^{\circ}$ & 4FGL-DR1 & 20.8\%\\     
            Option 3  & $|b| \leq 2^{\circ}$, $|\ell| \leq 8^{\circ}$ & 4FGL-DR2 & 19.7\%\\ 
            Option 4  & $|b| \leq 2^{\circ}$, $|\ell| \leq 5^{\circ}$ & 4FGL-DR2 & 18.8\%\\ 
            \hline
    \end{tabular}
    \caption{The four options for the mask that removes the galactic disk and the known point sources. The last column shows  the fraction of masked pixels for each mask option at $1.02-1.32$ GeV with respect to the total number of pixels in the inner $20^\circ$ Galactic center region ($1.6\times10^5$ pixels).}
    \label{tab:Masks}
\end{table*}

The 4FGL-DR2 \textit{Fermi} point source catalogue \cite{Ballet:2020hze} is an update on 
previous  \textit{Fermi} point-source catalogs. We mask all relevant point sources with a circular disk centered at their best fit
location (provided by Ref.~\cite{Ballet:2020hze}). The radius of that disk depends on energy, which accounts 
for \textit{Fermi}-LAT's energy-dependent PSF, as described in Sec.~\ref{sec:masks} and in 
Table~\ref{tab:PSF_vsE}. The extended sources and bright point sources that have a detection test statistic (TS) of 49 
or more are masked with a circular disk of larger radius as described in Table~\ref{tab:PSF_vsE}.
In Fig.~\ref{fig:Masks}, we compare the standard mask choice to option 2 that uses the 4FGL-DR1 catalogue instead (left versus
right panels). 

\begin{figure*}[t]
\begin{centering}
\hspace{-0.16in}
\includegraphics[width=3.47in,angle=0]{plots/GCE_Ellipticity_New_Oct_v2.pdf} 
\hspace{-0.21in}
\vspace{-0.25in}
\includegraphics[width=3.47in,angle=0]{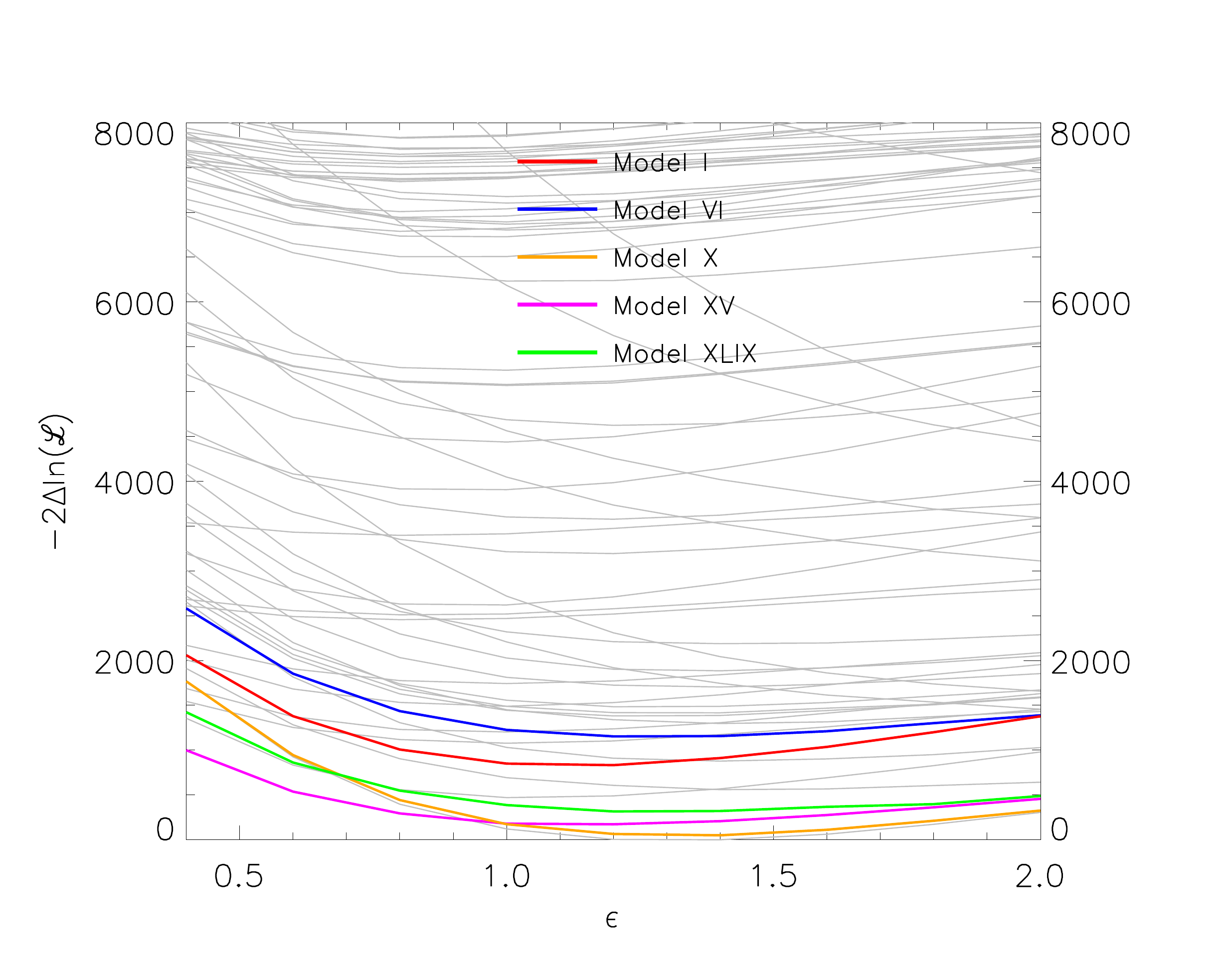} \\
\hspace{-0.16in}
\includegraphics[width=3.47in,angle=0]{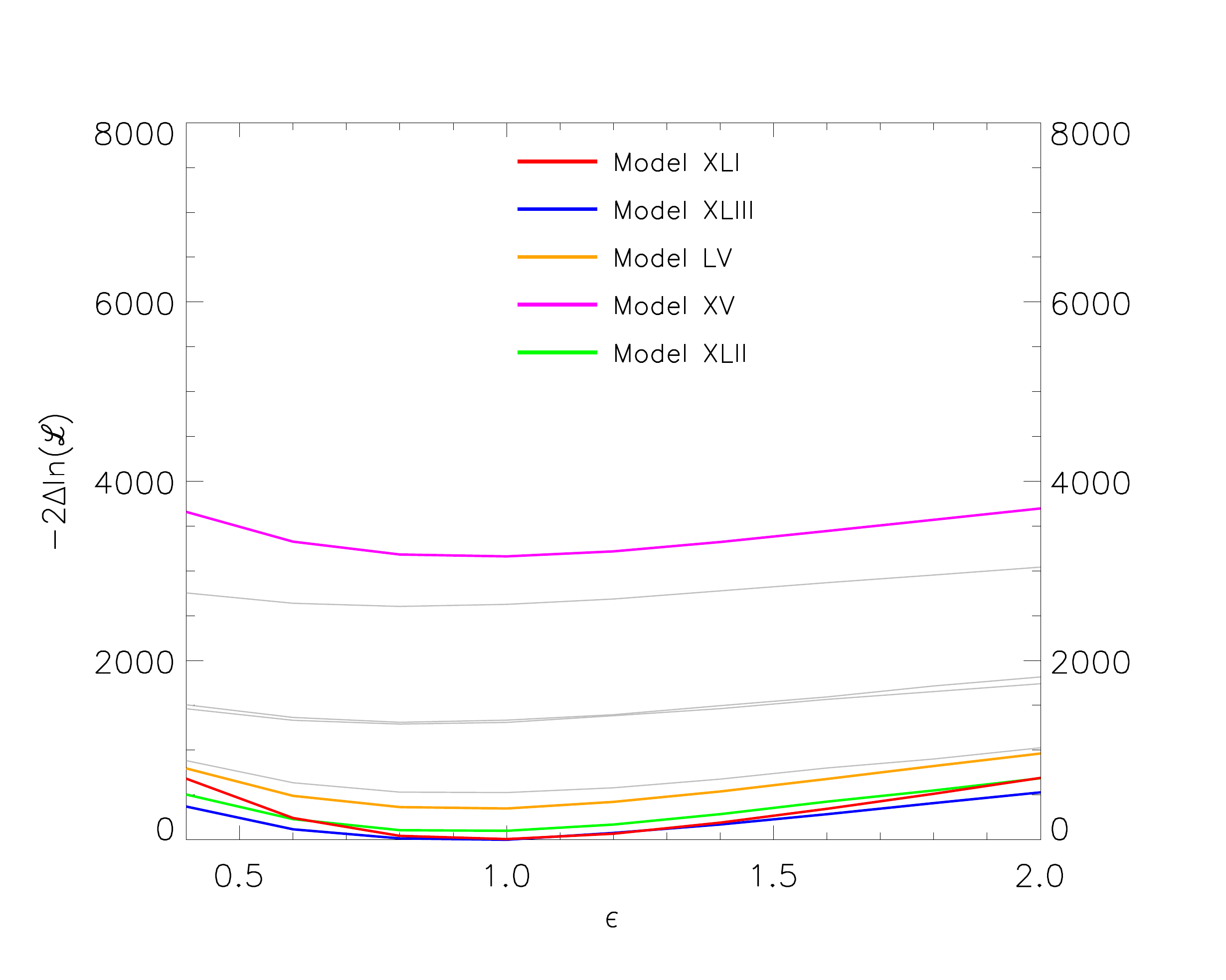}
\hspace{-0.21in}
\vspace{-0.25in}
\includegraphics[width=3.47in,angle=0]{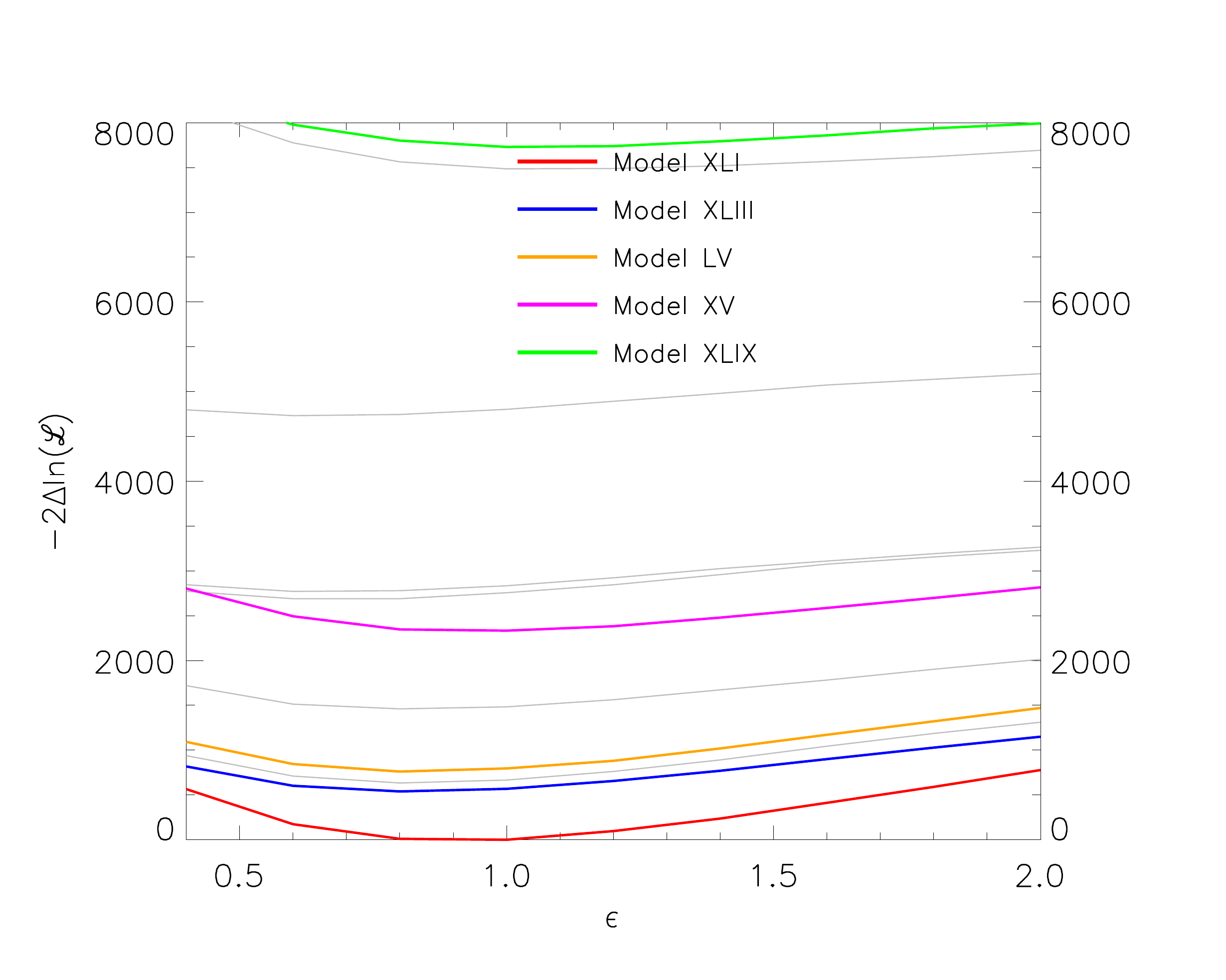} \\
\hspace{-0.12in}
\includegraphics[width=3.47in,angle=0]{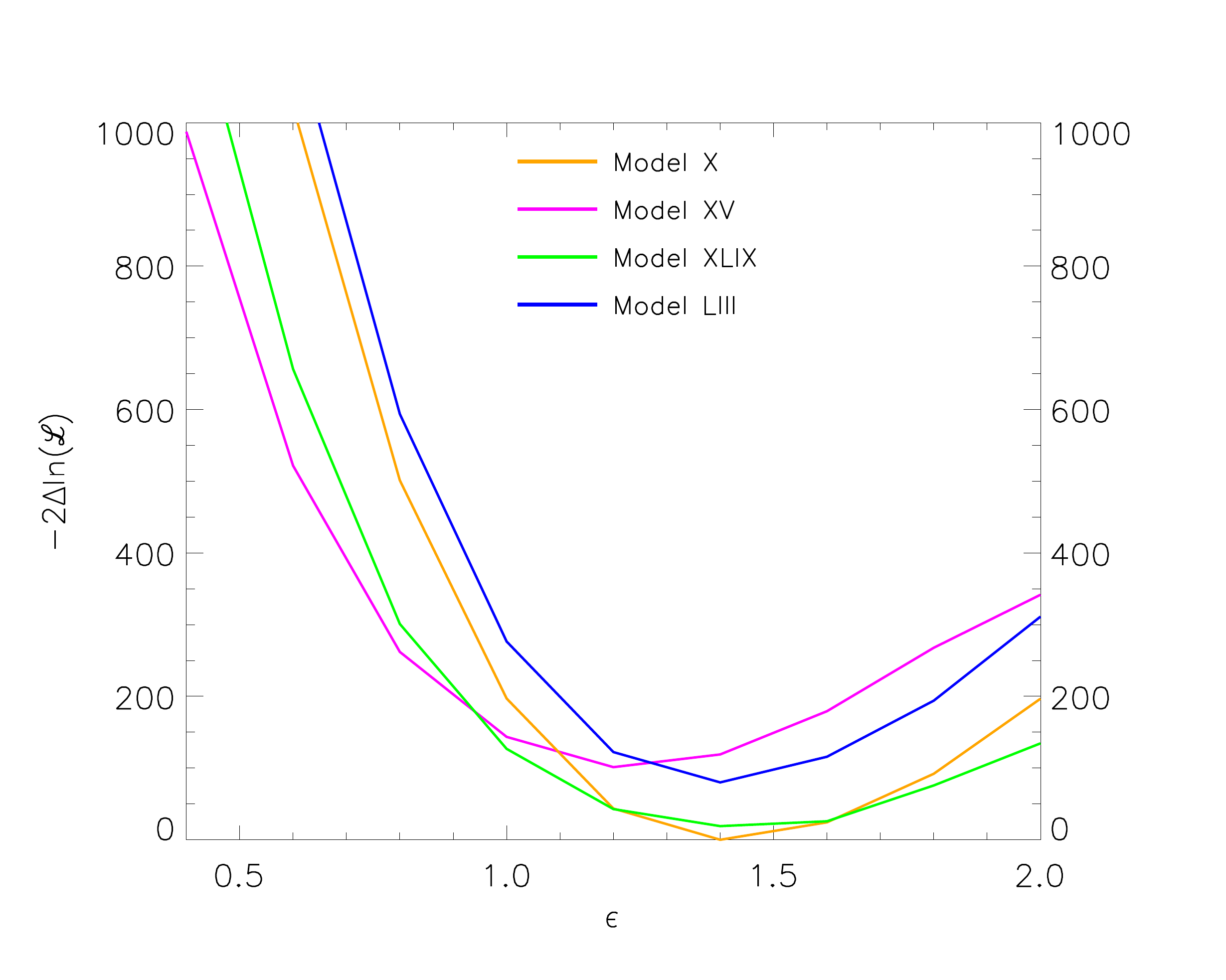}
\hspace{-0.21in}
\includegraphics[width=3.47in,angle=0]{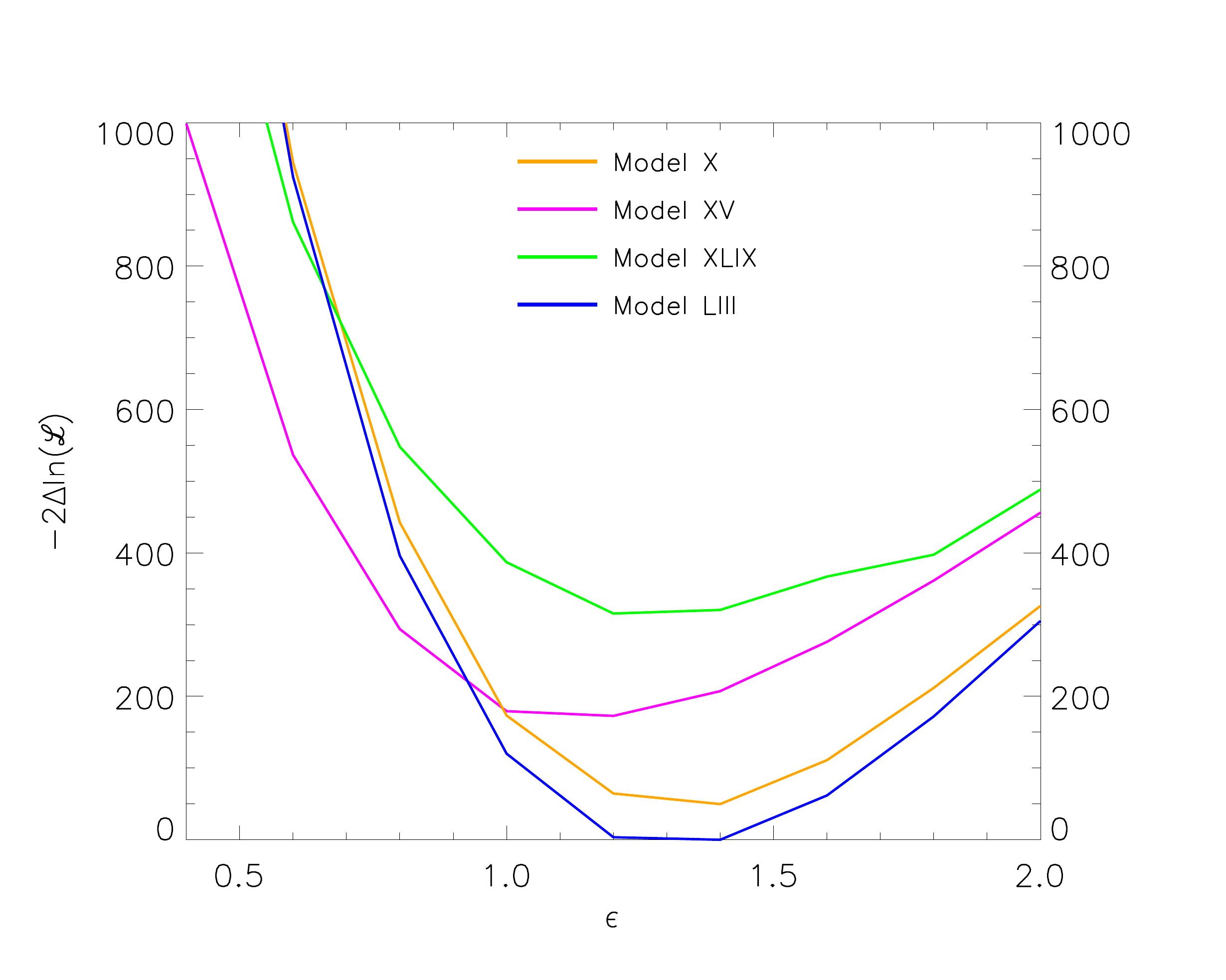}
\end{centering}
\vskip -0.2in
\caption{We show the impact that a changing mask has on the ellipticity of the GCE. Like with 
Fig.~\ref{fig:Morphology}, we show the difference in $-2 \Delta \ln(\mathcal{L})$ between the 
best fit choice and all others. 
\textit{Top Left}: our standard choice using as a mask the 4FGL-DR2 catalogue and masking the entire galactic 
disk within $|b| \le 2^{\circ}$. This is also shown in Figure \ref{fig:Morphology}.
\textit{Top Right}: option 2, using as a mask the 4FGL-DR1 catalogue and masking also the entire galactic disk 
within $|b| \le 2^{\circ}$.  
\textit{Middle Left}: option 3, using as a mask the 4FGL-DR2 catalogue but masking out the disk only within 
$|b| \le 2^{\circ}$ and $|l| \le 8^{\circ}$.
\textit{Middle Right}: option 4, using as a mask the 4FGL-DR2 catalogue and masking out the disk only within 
$|b| \le 2^{\circ}$ and $|l| \le 5^{\circ}$. 
\textit{Bottom Left}: zooming in for our standard choice and plotting our four best models, using as a mask the 
4FGL-DR2 catalogue and masking the entire galactic disk within $|b| \le 2^{\circ}$.
\textit{Bottom Right}:  zooming in for our standard choice and plotting our four best models for option 2 
(4FGL-DR1 catalogue and masking also the entire galactic disk within $|b| \le 2^{\circ}$).  
The exact mask cut has a major impact on the fits. The ellipticity gets its maximum value of $\epsilon = 1.4$ for our standard 
choice, while any mask that includes pixels from the lower latitudes, or even just the 4FGL-DR1 catalogue 
mask prefers values for the ellipticity closer to $\epsilon = 1$.}
\label{fig:Morphology_vs_Masks}
\end{figure*}

In Fig.~\ref{fig:Morphology_vs_Masks} we present the impact of the alternative masks on the ellipticity of the GCE.
Following the main text discussion in Sec.~\ref{sec:MorphologyResults}, we show the difference in $-2 \Delta \ln(\mathcal{L})$ 
between the best fit choice and all others. For the top two rows we keep the range of $-2 \Delta \ln(\mathcal{L})$ values from 0 to $8\times10^{3}$ 
as this is already a very wide one. In the bottom panels, we provide zoomed-in versions for our results with  4FGL-DR2 and 4FGL-DR1, 
showcasing only the change in the likelihood for the four best models.
 For certain choices of masks, many galactic diffuse emission models give fits that fall outside the 
plotted range, which suggests highly excluded options. 

We find that upon removing the galactic disk mask of $|b| \leq 2^{\circ}$, the fit is dominated by low latitudes. 
At these latitudes the GCE is still 
subdominant. The central pixels where the GCE is anticipated to peak are always removed by the presence of the many
point sources there. We also note that the exact mask cut has a major impact on the fits, as the number of 
bright pixels included in the $|b| \leq 2^{\circ}$ region significantly increases. This makes the differences in the 
likelihood between galactic diffuse models more prominent, as can be seen by the middle row panels of 
Fig.~\ref{fig:Morphology_vs_Masks}. 

As a general tendency, we note that for all alternative options where 
we impose less strong mask cuts than the standard mask, the preferred ellipticity decreases, from its maximum
of  $\epsilon = 1.4$ to values closer to $\epsilon = 1.0$. This suggests that the values of $\epsilon = 1.0-1.4$ 
that we get for the best fit models from our standard mask are likely a result of the
fact that once masking the entire disk the templates with $\epsilon = 1.0-1.4$ become difficult to distinguish with
our fit procedure. If the GCE had a genuinely prolate emission morphology, it would have showed up consistently in the 
alternative options that we test here. As a final point we note that are a few models where a slight preference 
for an oblate $\epsilon < 1$ profile exists. This is seen for masks where part of the disk is included, but the 
statistical preference is very weak. Recent cosmic-ray burst activity in the inner galaxy could naturally give 
an oblate GCE. More advanced masking procedures that would allow us to selectively
remove possible point sources below the \textit{Fermi} detection threshold will be interesting to pursue in future work.

\section{Corner Plots for Dark Matter Plus Millisecond Pulsars}
\label{sec:cp_mspdm}

\begin{figure}[t]
    \centering
    \includegraphics[width=0.48\textwidth]{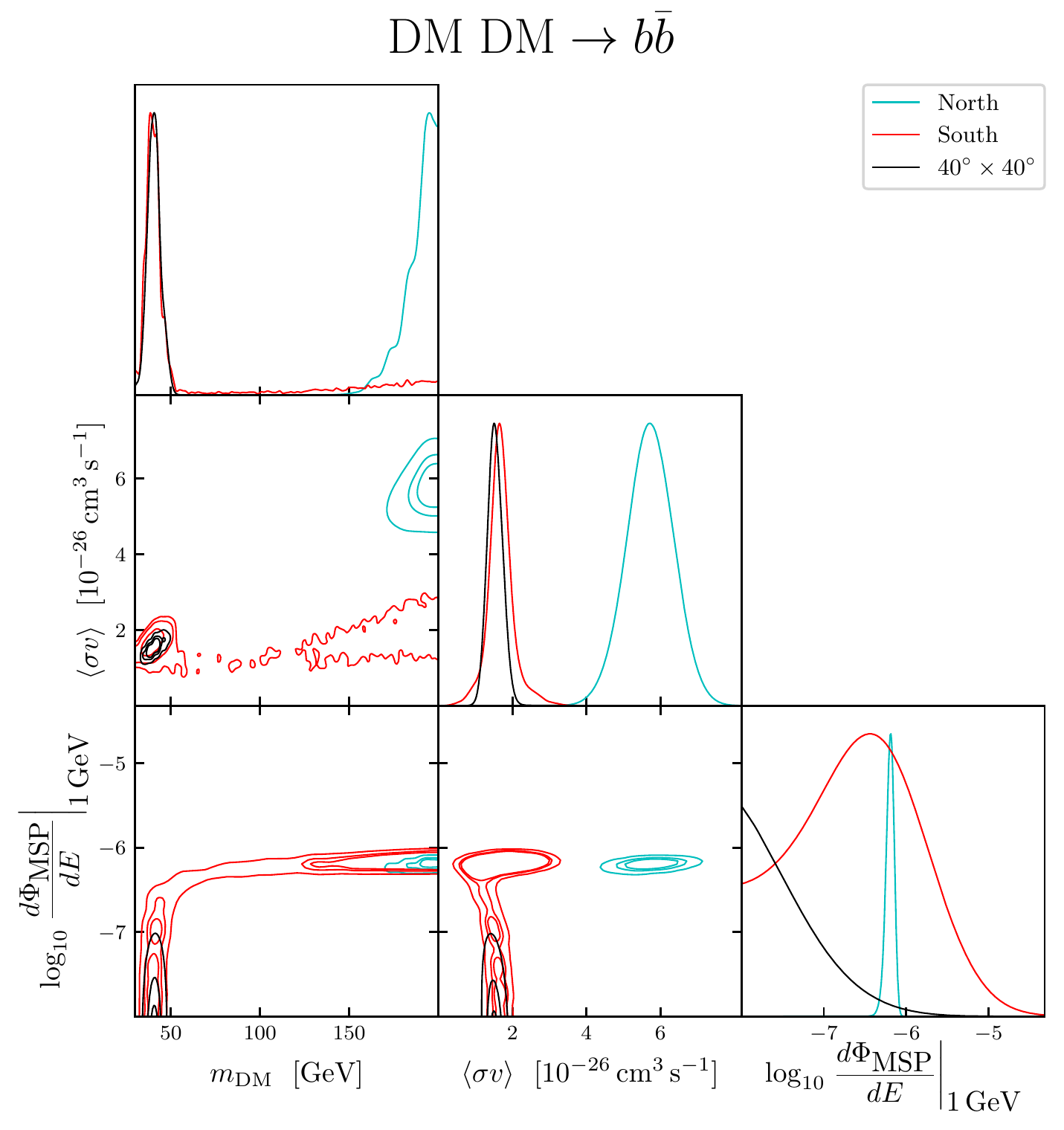}
    \caption{Corner plot for MSPs$+\text{DM DM} \to b \bar b$, as discussed in Sec.~\ref{sec:Interp_MSPDM}. The flux $d\Phi_\text{MSP}/dE$ is in units of $({\rm GeV\, cm^2\, s\, sr})^{-1}$ normalized at 1 GeV.}
    \label{fig:corner_mspb}
\end{figure}

\begin{figure*}[t]
    \centering
    \includegraphics[width=0.48\textwidth]{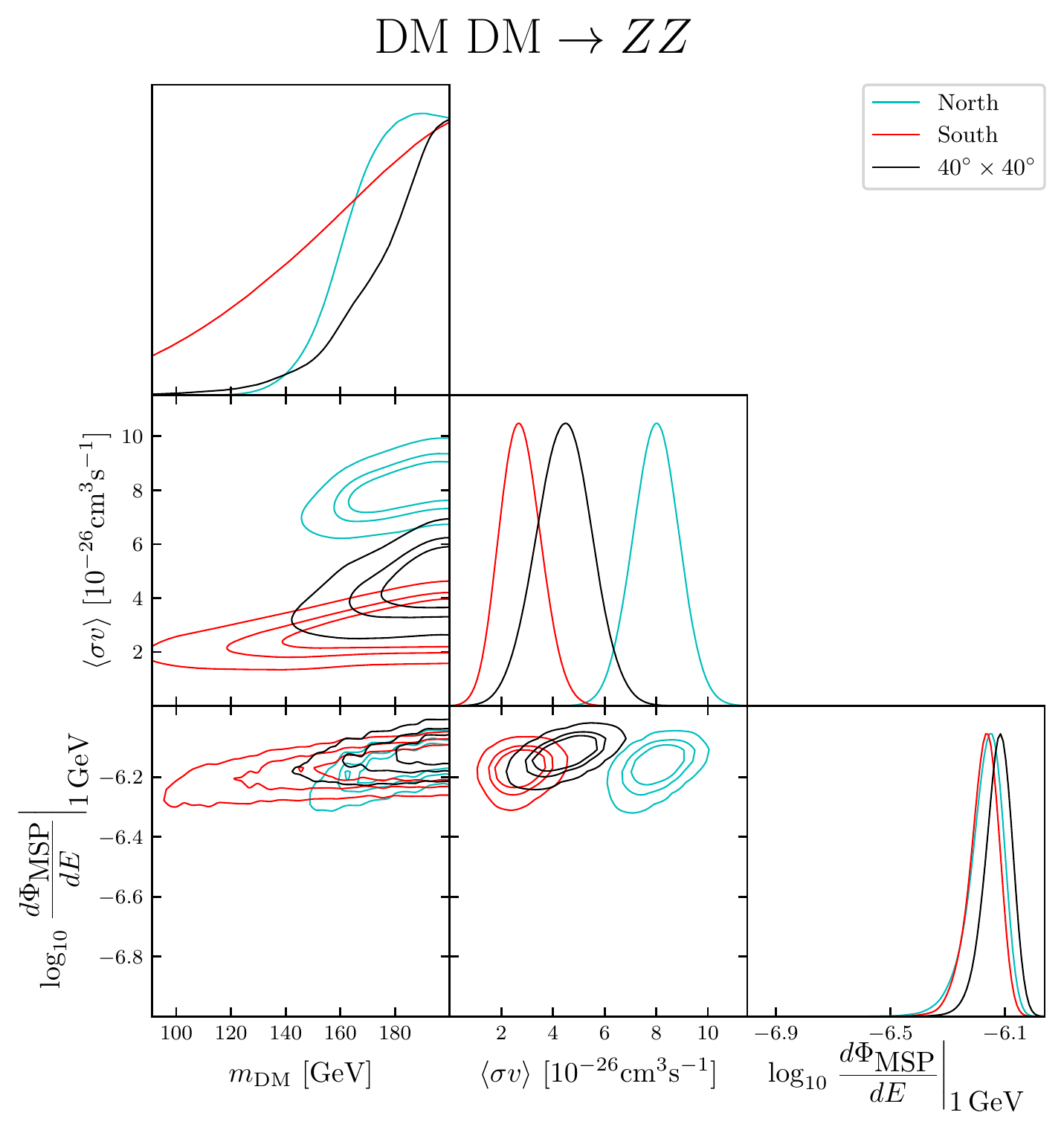}~
    \includegraphics[width=0.48\textwidth]{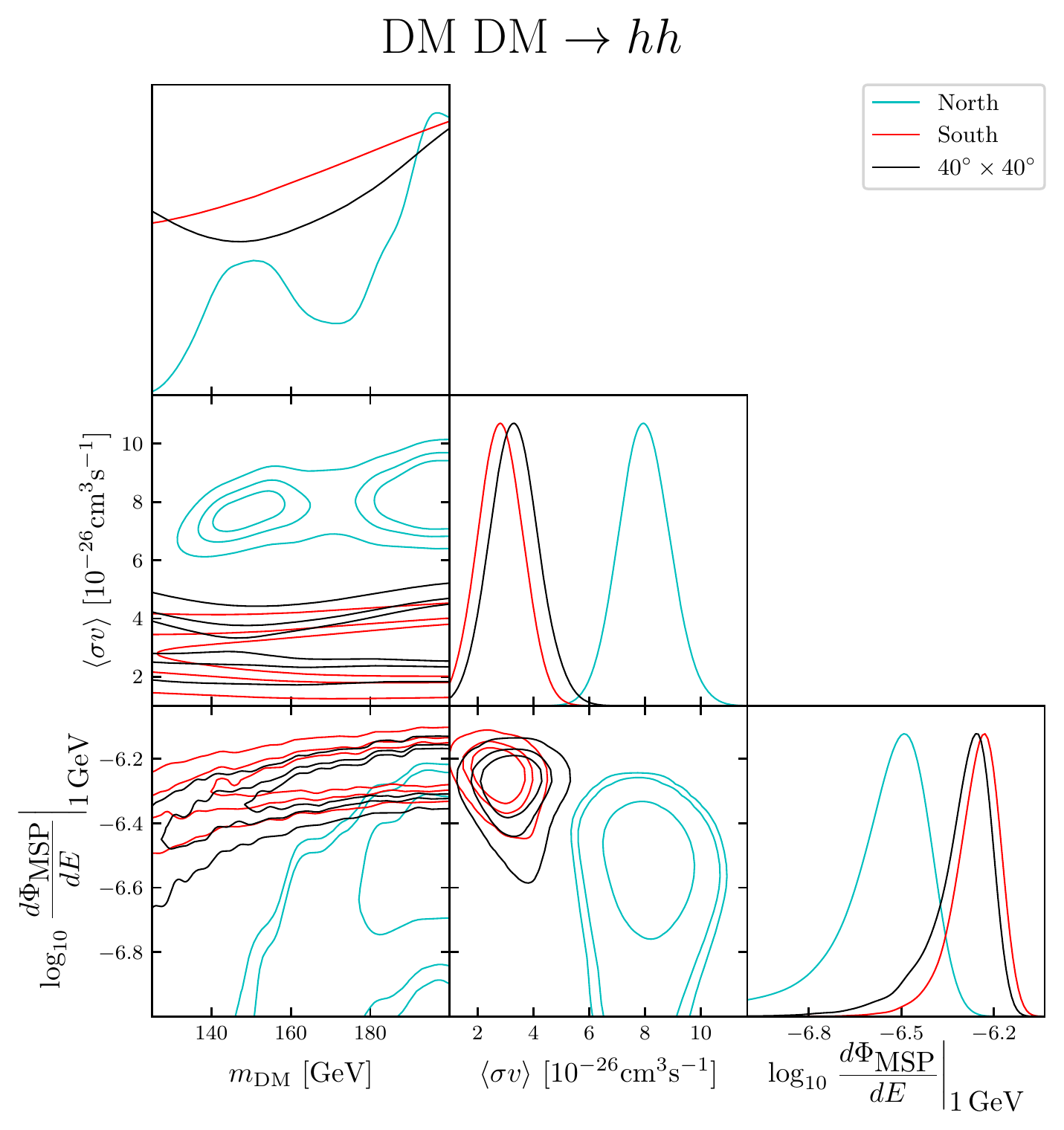}
    \caption{Corner plots for MSPs$+\text{DM DM} \to ZZ$ and MSPs$+\text{DM DM} \to hh$, as discussed in Sec.~\ref{sec:Interp_MSPDM}. The flux $d\Phi_\text{MSP}/dE$ is in units of $({\rm GeV\, cm^2\, s\, sr})^{-1}$ normalized at 1 GeV.}
    \label{fig:corner_mspZh}
\end{figure*}

For completeness, in this appendix we provide corner plots for three models in which MSPs combine with DM annihilation to account for the GCE. This provides a somewhat different perspective on the results in Sec.~\ref{sec:Interp_MSPDM}. 

In Fig.~\ref{fig:corner_mspb}, we show the results for ${\rm MSPs}+ {\rm DM \, DM} \to b \bar b$. The conventional two-dimensional dark matter parameter space is the middle panel of the left column. We see that 
the $40^\circ \times 40^\circ$ analysis is peaked at the smallest allowed values of the MSP flux, as expected from Sec.~\ref{sec:Interp_MSPDM}. The southern-hemisphere analysis results in a bimodal posterior on $\log_{10} \Phi_{\rm MSP}$. The northern-hemisphere analysis strongly prefers a nonzero MSP flux, due to the hard GCE emission there which must be accounted for by relatively heavy DM. We note that we cannot refer to fixed values of the parameter $\hat f$ defined in Eq.~\ref{eq:fhat} since this is a function of both $\sigmav$ and $\Phi_{\rm MSP}$, so it is only well-defined if $\sigmav$ is fixed, which is not true when all parameters are freely varying.

In Fig.~\ref{fig:corner_mspZh}, we show the results for ${\rm MSPs}+{\rm DM \, DM} \to ZZ$ and those for ${\rm MSPs}+ {\rm DM \, DM} \to hh$. The results for ${\rm MSPs}+{\rm DM \, DM} \to ZZ$ are 
as follows: the dark matter mass is peaked towards the high-$m_{\rm DM}$ parameter space, and the favored values of $\sigmav$ are ordered such that the flux due to DM annihilation is brightest in the northern hemisphere and dimmest in the southern hemisphere. The parameter space for ${\rm MSPs}+{\rm DM \, DM} \to hh$ has some interesting features: the preferred values of $m_{\rm DM}$ are bimodal in the northern-hemisphere analysis and the analysis of the full $40^\circ \times 40^\circ$ ROI. The posterior volume is concentrated in the high-$m_{\rm DM}$ limit, with the same ordering of values of $\sigmav$ as in the ${\rm MSPs}+{\rm DM \, DM} \to ZZ$ scenario which was just discussed. For further analysis of these scenarios of DM annihilating to heavy Standard Model particles, additional energy bins may be required in the fit.

A full exploration of these possibilities, as well as more complex dark sectors which may lead to broader photon spectra \cite{Berlin:2014pya}, will be a very interesting topic for future work.

\section{Further Tests for the Contribution of the Stellar Bulge to the GCE}
\label{app:BulgesTests}

Here we expand our discussion on the possible contribution of the stellar bulge to the GCE within the region 
of $2^\circ \leq |b| \leq 20^\circ$, $|\ell| \leq 20^\circ$ that was discussed in the Sec.~\ref{sec:MorphologyResults} and Fig.~\ref{fig:Morphology} (right panel). 
Of special interest is the possible impact of the dense nuclear bulge. The nuclear bulge is composed of the nuclear stellar disk, which has a disk-like morphology 
with a characteristic scale height of 45 pc, and the nuclear stellar cluster, which is spherical and peaks at the Galactic center as observed in radio and microwaves \cite{2002A&A...384..112L} . 
We follow Refs.~\cite{Bartels:2017vsx, 2002A&A...384..112L} in modeling 
these two components of the nuclear bulge and combine them with the boxy bulge, which extends to several degrees in latitude. We normalize the luminosity ratio of the 
boxy bulge relative to the nuclear bulge to a value of 1.5 within the $40^{\circ} \times 40^{\circ}$ window (including the disk). 
We note that because we mask the galactic disk, most of the nuclear bulge's emission is removed especially at the higher energies. The fact that the GCE is seen at 
high latitudes at high energies clearly suggests that the GCE is due to more than just the emission from the nuclear bulge. We also test the boxy bulge without its nuclear bulge component, as it is most 
effectively probed by our analysis. If most of the luminosity of the GCE comes from a less concentrated population of sources as e.g. MSPs from disrupted globular 
clusters \cite{Brandt:2015ula}, then the contribution of the boxy bulge may be more dominant than its mass would suggest and testing it by itself is a worthwhile.  

\begin{figure*}[t]
    \centering
    \includegraphics[width=0.48\textwidth]{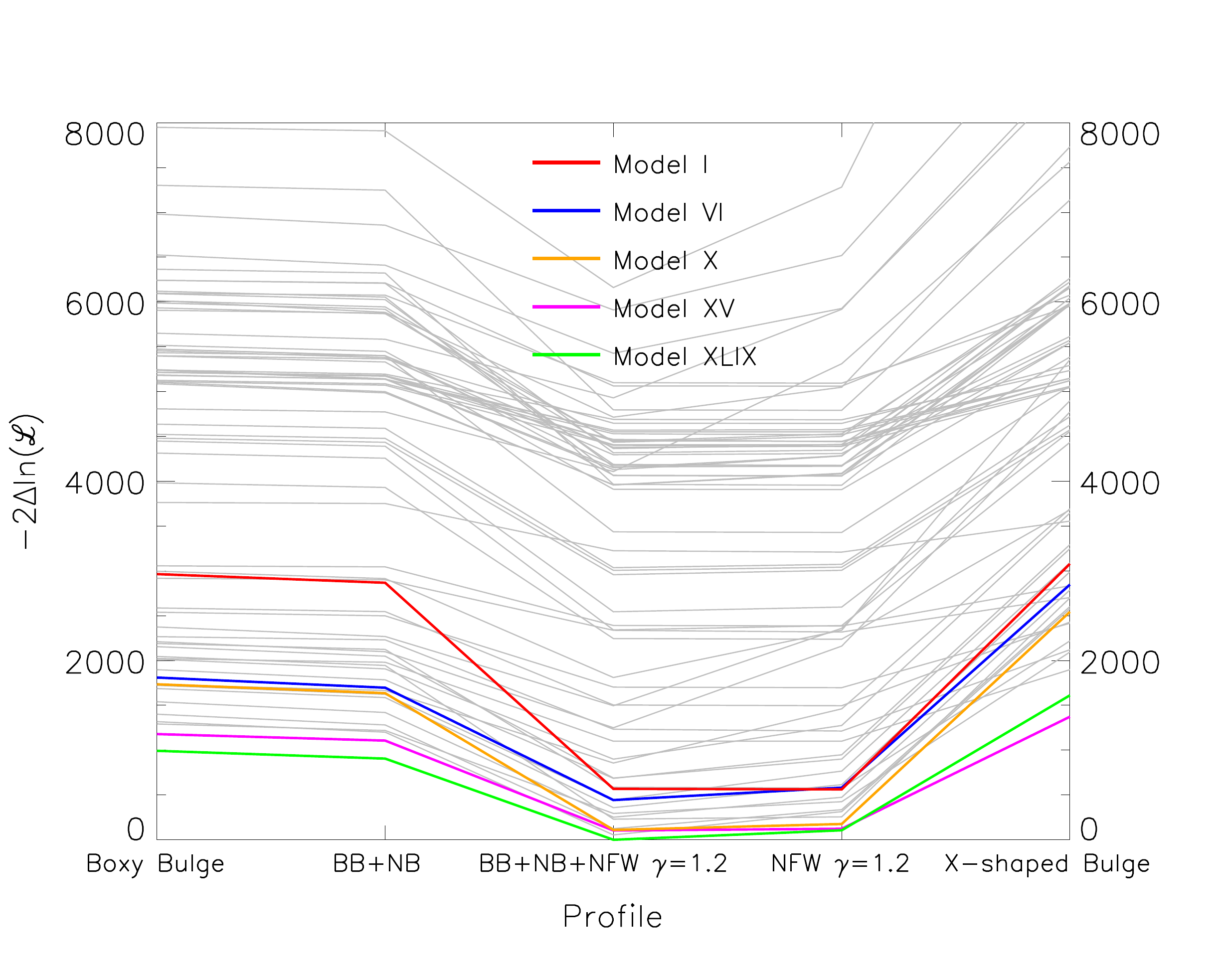}
    \includegraphics[width=0.48\textwidth]{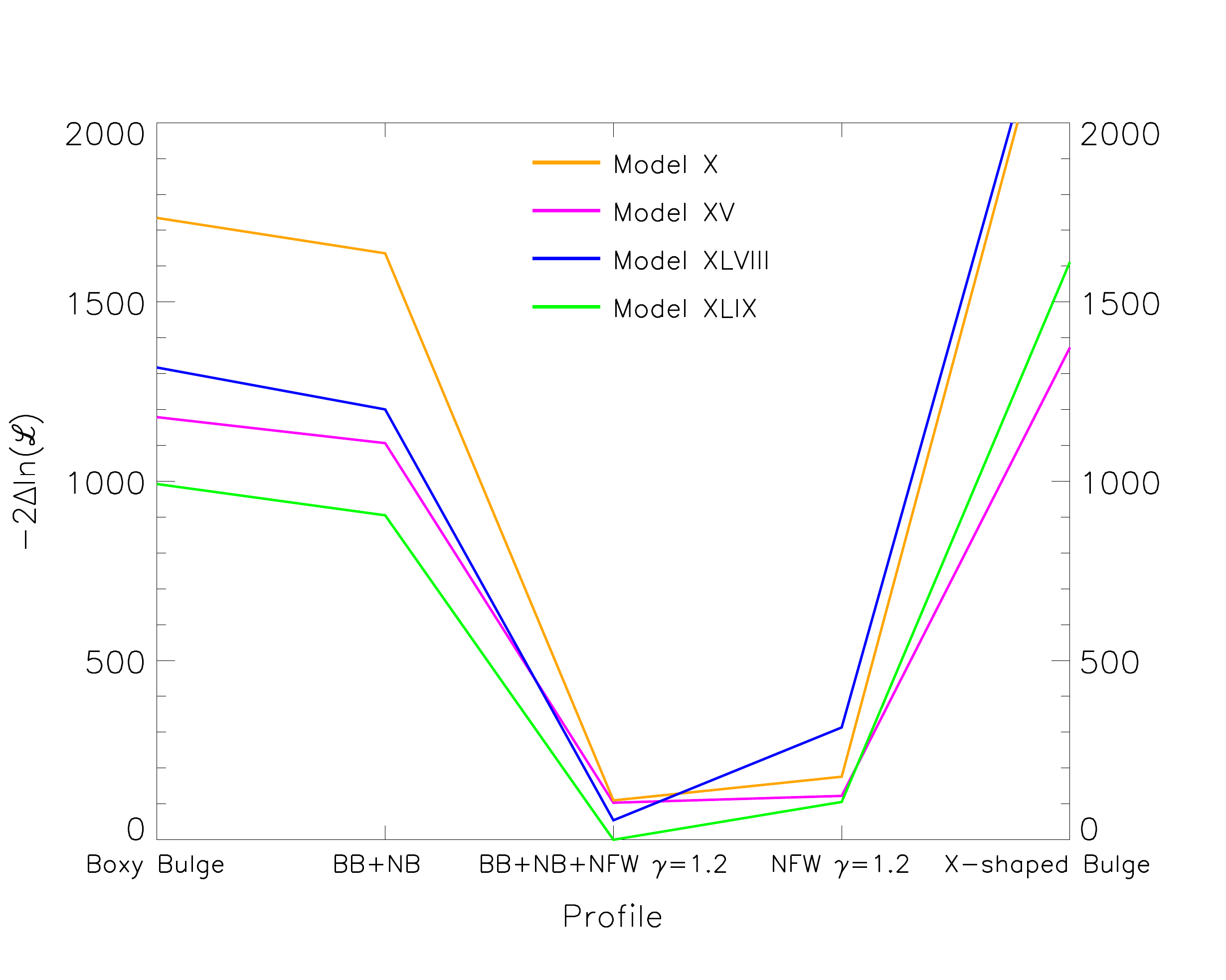}
    \caption{As with Fig~\ref{fig:Morphology}, we show the quality of fit for five different options for the GCE over the 80 galactic diffuse emission models. \textit{Left}: models 
    within a $-2 \Delta \ln(\mathcal{L})$ of $8.0\times 10^{3}$ from the best fit model (XLIX). \textit{Right}: zooming in the four best model assumptions. 
    Allowing for \textit{both} a component of DM annihilation following a profile of NFW with $\gamma=1.2$ and a stellar bulge (BB+NB) provides for the best 
    background models a marginally better fit to the observations than the NFW with $\gamma=1.2$ alone.}
    \label{fig:BulgesExtended}
\end{figure*}

We show in Fig.~\ref{fig:BulgesExtended} a comparison in the quality of fit between the boxy bulge template alone, the template containing the boxy bulge and the 
nuclear bulge (BB+NB), the NFW with $\gamma=1.2$ DM annihilation profile, and the X-shaped bulge. This is an expanded version of the right panel of 
Fig.~\ref{fig:Morphology}. Moreover, we test the combination of the 
NFW with $\gamma=1.2$ and the stellar bulge i.e. BB+NB, where we allow for the relative normalization of the DM-like component and the stellar bulge component to be free. 

\begin{figure}[t]
    \centering
    \includegraphics[width=3.7in, angle=0]{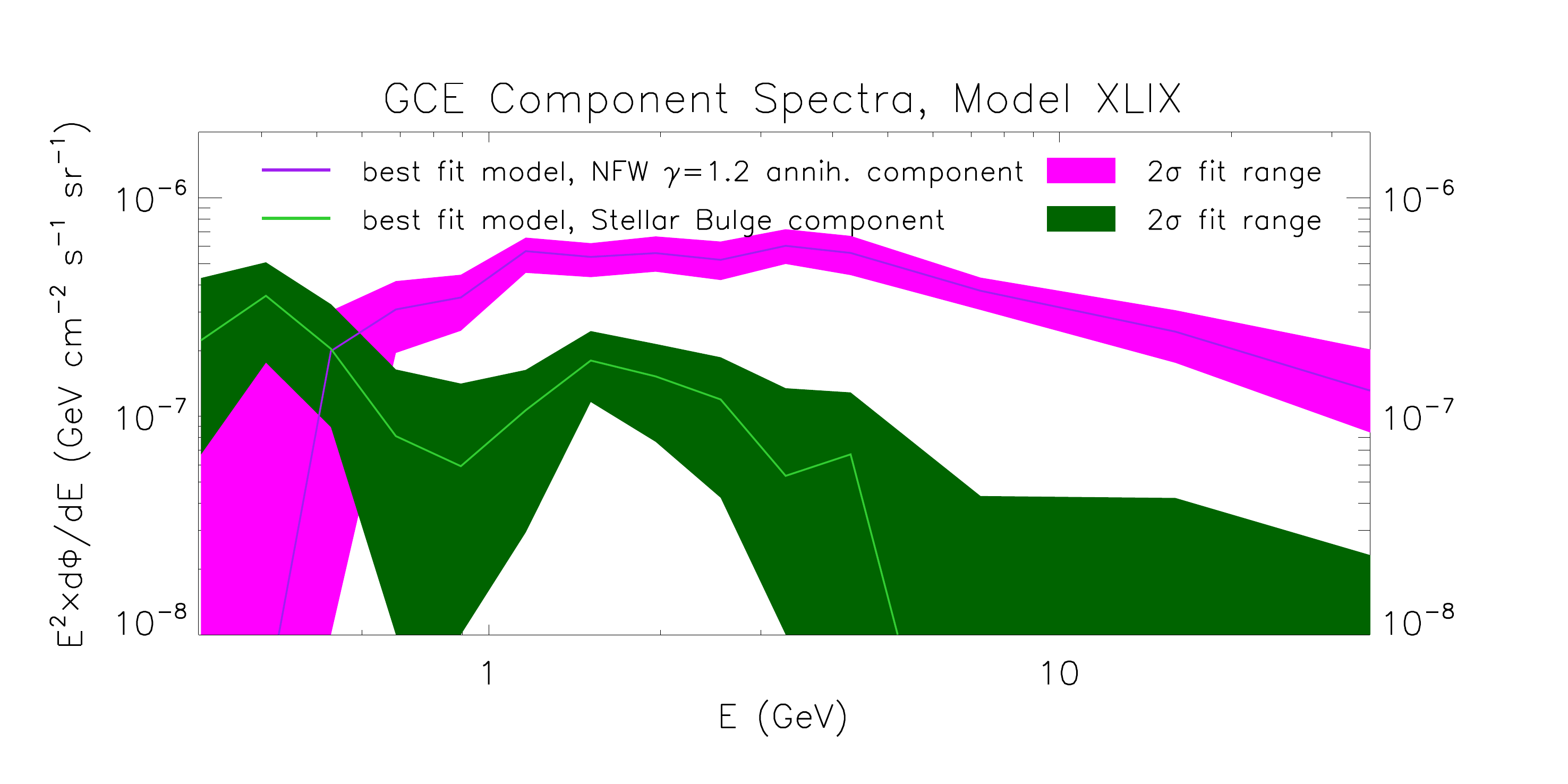}\\
    \includegraphics[width=3.25in, angle=0]{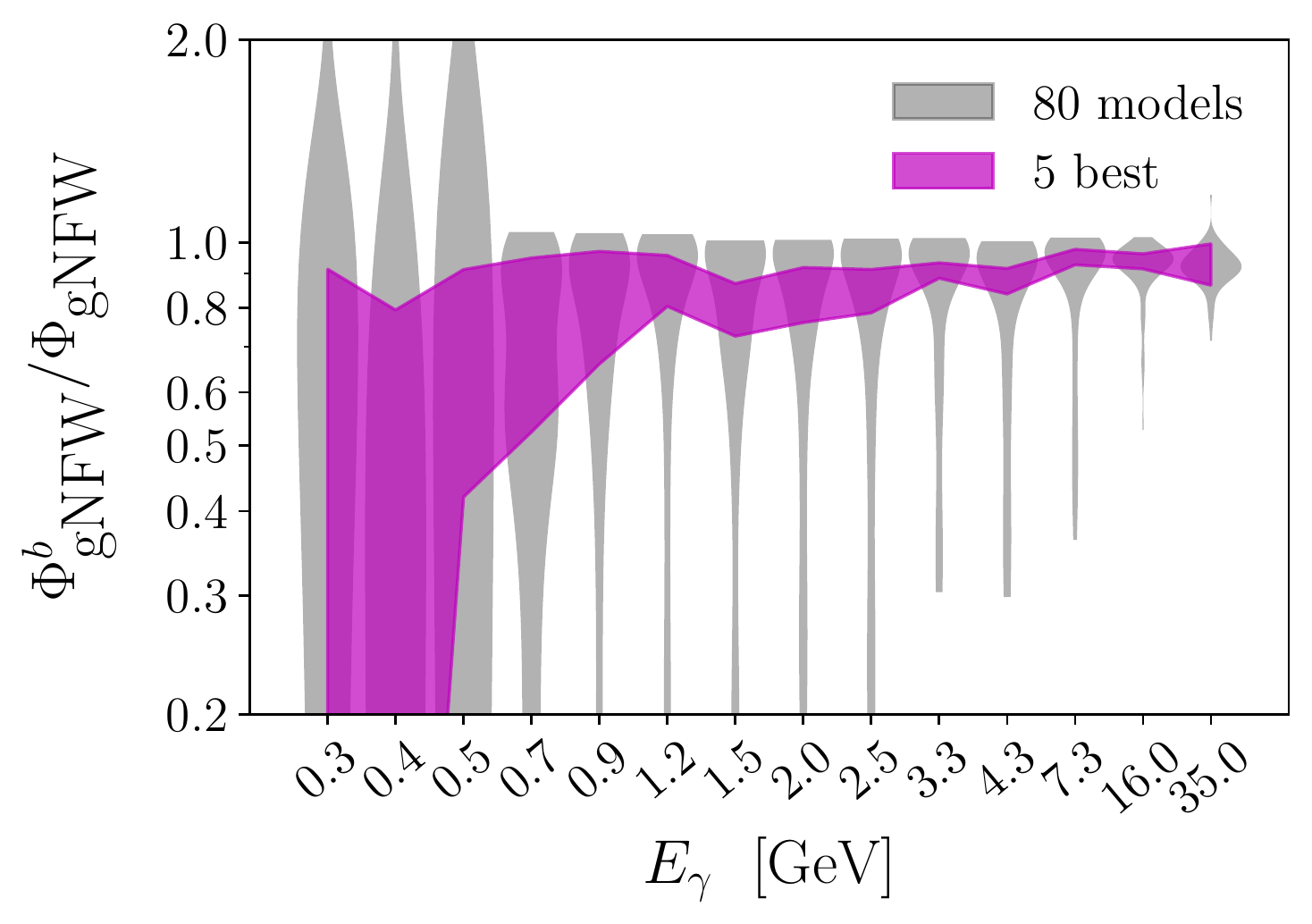} 
    \caption{\textit{Top:} using the best fit model we show the flux spectra from the two separate morphological components averaged 
    over the $40^{\circ} \times 40^{\circ}$ masking just the galactic disk (as we did in Figs.~\ref{fig:GCE_emission_ALL_models} and~\ref{fig:GCE_emission_N_vs_S}).
    \textit{Bottom:} testing all 80 models, we show the relative flux of the GCE absorbed by the NFW with $\gamma=1.2$ annihilation component.}
    \label{fig:GCEcomponentSpectra}
\end{figure}

Allowing for \textit{both} a component of DM annihilation following a profile of NFW with $\gamma=1.2$ and a stellar bulge (BB+NB) provides a marginally 
better fit to the observations than the NFW with $\gamma=1.2$ alone. However, compared to any of the known stellar populations either the NFW 
with $\gamma=1.2$ alone or in combination with the stellar bulge performs much better. Only for background models that provide a poor fit to the \textit{Fermi} data 
we see a significant difference between the NFW with $\gamma=1.2$ alone and its combination with the stellar bulge. This clearly shows that most of the GCE is still described best by the DM 
annihilation morphology coming from a NFW with $\gamma=1.2$ profile (squared). This becomes more clear when we show the relevant spectra from the 
NFW with $\gamma=1.2$ component and that from the stellar bulge in Fig.~\ref{fig:GCEcomponentSpectra} (top panel). We use model XLIX that provides 
the best fit to the data. Only at the low energies does the stellar bulge absorb the GCE emission. Instead above 0.7 GeV the NFW with $\gamma=1.2$  
annihilation component becomes dominant and above 2 GeV it is at least four times brighter to the BB+NB (Bulges). This result is robust to the exact galactic diffuse 
model used as long as we stay within the best fit models. In Fig.~\ref{fig:GCEcomponentSpectra} (bottom panel) we test the relative flux contribution that is in the NFW with $\gamma=1.2$  
annihilation. Above 0.7 GeV most of the flux is in the NFW with $\gamma=1.2$ (gNFW) annihilation component and above 2 GeV at least $80\%$ of the GCE comes from the DM 
annihilation component. 

\section{The Ensemble of Diffuse Gamma-Ray Emission Models for the Inner Galaxy}
\label{app:Allmodels}
In this appendix we give the description for all 80 models that we use in this analysis. 
The model parameters relevant for the production of the galactic diffuse model templates are described in Sec.~\ref{sec:GDGRE}.
In the main paper we only gave the model parameters for a small number of representative models in Table~\ref{tab:ModelsShort}.
Table~\ref{tab:ModelsLong} provides the full information and continues to Table~\ref{tab:ModelsLong2} to include all 80 models. 
These model parameters are the inputs into the \texttt{GALPROP} code \cite{galprop, Moskalenko:2001ya, GALPROPSite}. 
The gamma-ray template maps were created through the ``WebRun'' mode \cite{VLADIMIROV20111156, GALPROPSite} which is based 
on  \texttt{GALPROP} version 54. We remind that the cosmic-ray analysis and comparison to the \textit{AMS-02} data was done using 
the somewhat more recent version 56. Version 54, accessible through the ``WebRun'' mode, is a more efficient way to 
scan the wider parameter space (in number of relevant varying parameters) needed for the gamma-ray analysis. The full description
of each of these parameters and the range covered is given in Sec.~\ref{sec:CRtoGR} and Sec.~\ref{sec:TemplateModels}. We do not 
repeat it here. We describe here the numbering scheme for the last three columns that represents alternative choices for the ISM 
molecular, atomic and ionized hydrogen gases. 
We follow the same numbering scheme as used in the  \texttt{GALPROP}  ``WebRun''. With this our information, our template maps (which we make public)
should be relatively easy to reproduce or perturb around.

The spatial distribution of the molecular hydrogen gas H2 is known to follow well that 
of CO, which is traced though its 2.6 mm rotational transition line \cite{Pohl:2007dz}. The conversion from CO maps to H2 is done 
though the $X_\text{CO}(r)$ parameter, where $r$ is the galactocentric radius and $X_\text{CO} = N_\text{H2}/W_\text{CO}$ with $W_\text{CO}$ the velocity 
integrated radiation temperature \cite{Strong:2004de}. Thus $X_\text{CO}$  does not need to be a constant throughout the Milky Way. For 
the H2 gas the numbers ``0'', ``2'',``3'', ``9'' and ``10'' represent alternative choices for the $X_\text{CO}(r)$ parameter profile. Choice ``0'' is 
the case of a constant $X_\text{CO}(r) = 1.9 \times 10^{20}$  molecules/(K km s$^{-1}$). Choice ``2'' represents the case of $X_\text{CO} = 
X_{0} + A \cdot (r/ 1 \textrm{ kpc}) + B \cdot 10^{C \cdot (r/1 \textrm{ kpc})}$, where we take $X_{0} = 1.8 \times 10^{20}$, 
$A = 1 \times 10^{20}$ molecules/(K km s$^{-1}$) and $B=C=0$. Only for model II we tested $A = 2 \times 10^{20}$ molecules/(K km s$^{-1}$) 
that proved to give a very poor fit to the data. Choice ``3'', uses a tabulated profile for $X_\text{CO}(r)$, while choice ``9'' used the profile of
\cite{2004A&A...422L..47S}. Finally, choice ``10'' used $X_\text{CO} = 1 \times 10^{20} \cdot 10^{(-0.4 + 0.066 \cdot r)}$ molecules/(K km s$^{-1}$)  
for $r$ < 15 kpc and $X_\text{CO} = 1 \times 10^{20} \cdot 10^{(-0.4 + 0.066 \cdot 15)}$ molecules/(K km s$^{-1}$) for $r \geq 15$ kpc. 
For the H2 gas, the intermediate number choices that we did not mention here are not used. 

The spatial distribution of atomic (neutral) hydrogen HI is derived from the 21 cm hyperfine transition line, where assumptions need
to be made on the  spin temperature $T_{S}$ that corrects for the opacity of the Milky Way to that line. The conventional 
assumption is for $T_{S} \simeq 150$ K and the most extreme ones go up to $T_{S} = 10^{5}$ K. Also, dust maps can be used as
alternative tracers of HI. Within \texttt{GALPROP} we tested in our models the two alternative magnitude cuts (2 or 5) on the E(B-V) 
reddening maps of Ref.~\cite{Schlegel:1997yv}. Choice ``1'', takes $T_{S} = 125$ K without using the dust maps.
Choice ``2'', takes $T_{S} = 10^{5}$ K and an  E(B-V) magnitude cut of 2. Choice ``3'', takes $T_{S} = 10^{5}$ K and an  E(B-V) 
magnitude cut of 5. Choice ``4'', takes $T_{S} = 150$ K and an  E(B-V) magnitude cut of 2 and choice ``5'', takes $T_{S} = 150$ K 
and an  E(B-V) magnitude cut of 5.

Finally, the less significant component in terms of total mass is the ionized hydrogen HII gas. However, in the inner galaxy and along the 
line of sight toward this region there may be an appreciable HII contribution. We use three alternatives for the HII gas. Option ``1'', follows the 
gas density model of Ref.~\cite{1991Natur.354..121C}. Option ``2'', follows the model of Ref.~\cite{2002astro.ph..7156C}. Option ``3'', takes 
the most recent HII model of Ref.~\cite{2008PASA...25..184G}. As we show, we test a variety of combinations for the H2, HI and HII gases.
We have tried an even wider range of combinations than those presented here but find that many of them are strongly excluded by the data.

We clarify that the inclusion of all these maps within \texttt{GALPROP} is a standard feature of the  ``WebRun''
\cite{VLADIMIROV20111156, GALPROPSite}, and not our original contribution.

\begin{table*}[t]
    \begin{tabular}{ccccccccccccccc}
    \hline
            Name & $z_{L}$ & $D_{0}$ & $\delta$ & $v_{A}$ & $dv_{c}/d|z|$ & $S^{N}/S^{e}$ & $\alpha_{1}^{p}$/$\alpha_{2}^{p}$ & $\alpha_{1}^{e}$/$\alpha_{2}^{e}$ & $N^{p}$/$N^{e}$ & B-field & ISRF & H2 & HI & HII \\
            \hline\hline
            I  & 4.0 & 5.0 & 0.33 & 32.7 & 55 & Pul/Pul & 1.35/2.33 & 1.5/2.25 & 4.13/3.33 & 200030050 & 1.36,1.36,1.0 & 9 & 5 & 1 \\ 
            II & 6.0 & 7.1 & 0.33 & 50.0 & 0 & Pul/SNR & 1.89/2.30 & 1.40/2.10 & 2.40/2.20 & 050100020 & 1.0,1.0,1.0 & 2$^{*}$ & 1 & 1 \\     
            III  & 5.6 & 4.85 & 0.40 & 40.0 & 0 & Pul/Pul & 1.50/1.90 & 1.5/2.25 & 2.40/1.55 & 200050040 & 1.4,1.4,1.0 & 9 & 4 & 1 \\   
            IV & 6.0 & 6.0 & 0.33 & 10.0 & 50 & Pul/SNR & 1.60/2.30 & 1.6/2.26 & 1.7/4.1 & 200030050 & 1.4,1.4,1.0 & 9 & 5 & 1 \\ 
            V & 10.0 & 10.0 & 0.33 & 32.2 & 0 & Pul/SNR & 1.70/2.39 & 1.6/2.33 & 1.0/2.66 & 200040050 & 1.4,1.4,1.0 & 0 & 5 & 2 \\ 
            VI  & 6.0 & 2.0 & 0.33 & 0 & 200 & Pul/SNR & 1.60/2.10 & 1.6/2.30 & 2.32/5.70 & 200030050 & 1.4,1.4,1.0 & 9 & 5 & 1 \\   
            VII & 10.0 & 8.0 & 0.33 & 0 & 0 & Pul/SNR & 1.40/1.80 & 1.4/2.30 & 1.3/3.33 & 200040050 & 1.4,1.4,1.0 & 0 & 5 & 2 \\
            VIII & 5.6 & 4.85 & 0.40 & 24.0 & 1 & SNR/SNR & 2.00/2.38 & 1.6/2.43 & 5.8/2.00 & 090050020 & 1.36,1.36,1.0 & 9 & 2 & 1 \\  
            IX  & 5.6 & 4.85 & 0.40 & 24.0 & 1 & SNR/SNR & 2.00/2.38 & 1.6/2.43 & 5.8/2.00 & 090050020 & 1.36,1.36,1.0 & 9 & 3 & 1 \\
            X  & 10.0 & 8.0 & 0.33 & 32.2 & 50 & Pul/SNR & 1.40/1.80 & 1.4/2.35 & 1.90/3.20 & 200040050 & 1.4,1.4,1.0 & 0 & 5 & 2 \\       
            XI  & 5.6 & 4.85 & 0.40 & 24.0 & 1 & SNR/SNR & 2.00/2.38 & 1.6/2.43 & 5.8/2.00 & 090050020 & 1.36,1.36,1.0 & 9 & 4 & 3 \\  
            XII  & 5.6 & 4.85 & 0.40 & 24.0 & 1 & SNR/SNR & 2.00/2.38 & 1.6/2.43 & 5.8/2.00 & 090050020 & 1.36,1.36,1.0 & 10 & 4 & 1 \\  
            XIII & 5.6 & 4.85 & 0.40 & 24.0 & 1 & Pul/Pul & 2.00/2.25 & 1.6/2.30 & 5.8/2.00 & 050100020 & 1.5,1.5,1.0 & 9 & 4 & 1 \\   
            XIV & 5.6 & 4.85 & 0.40 & 24.0 & 1 & Pul/Pul & 2.00/2.25 & 1.6/2.30 & 5.8/2.00 & 050100020 & 0.7,0.7,1.0 & 9 & 4 & 1 \\
            XV & 6.0 & 7.1 & 0.33 & 50.0 & 0 & Pul/SNR & 1.89/2.30 & 1.40/2.10 & 2.40/2.20 & 050100020 & 1.0,1.0,1.0 & 0 & 5 & 2 \\
            XVI & 5.6 & 4.85 & 0.40 & 24.0 & 1 & Pul/Pul & 2.00/2.25 & 1.6/2.30 & 5.8/2.00 & 100050020 & 1.0,1.0,1.0 & 9 & 4 & 1 \\   
            XVII  & 5.6 & 4.85 & 0.40 & 24.0 & 1 & Pul/Pul & 2.00/2.25 & 1.6/2.30 & 5.8/2.00 & 025200010 & 0.7,0.7,1.0 & 9 & 4 & 1 \\   
            XVIII & 6.0 & 6.5 & 0.33 & 30.0 & 0 & SNR/Pul & 2.04/2.41 & 1.6/2.43 & 5.8/2.00 & 090050020 & 1.36,1.36,1.0 & 9 & 4 & 1 \\  
            XIX & 6.0 & 6.5 & 0.33 & 30.0 & 0 & Pul/Pul & 2.04/2.41 & 1.6/2.43 & 5.8/2.00 & 025200010 & 0.7,0.7,1.0 & 9 & 4 & 1 \\  
            XX & 5.5 & 5.5 & 0.37 & 30.0 & 2.5 & Pul/Pul & 2.00/2.38 & 1.6/2.43 & 5.8/2.00 & 050100020 & 1.0,1.0,1.0 & 9 & 4 & 1 \\  
            XXI & 5.5 & 5.5 & 0.37 & 30.0 & 2.5 & SNR/Pul & 2.00/2.38 & 1.6/2.43 & 5.8/2.00 & 090050020 & 1.36,1.36,1.0 & 9 & 4 & 1 \\  
            XXII & 5.5 & 5.5 & 0.37 & 30.0 & 2.5 & Pul/Pul & 2.00/2.38 & 1.6/2.43 & 5.8/2.00 & 025200010 & 0.7,0.7,1.0 & 9 & 4 & 1 \\
            XXIII & 5.7 & 3.9 & 0.45 & 25.7 & 6.0 & Pul/Pul & 1.99/2.36 & 1.6/2.43 & 5.8/2.00 & 050100020 & 1.0,1.0,1.0 & 9 & 4 & 1 \\  
            XXIV & 5.7 & 3.9 & 0.45 & 25.7 & 6.0 & SNR/Pul & 1.99/2.36 & 1.6/2.43 & 5.8/2.00 & 090050020 & 1.36,1.36,1.0 & 9 & 4 & 1 \\   
            XXV & 5.7 & 3.9 & 0.45 & 25.7 & 6.0 & Pul/Pul & 1.99/2.36 & 1.6/2.43 & 5.8/2.00 & 025200010 & 0.7,0.7,1.0 & 9 & 4 & 1 \\  
            XXVI & 6.0 & 3.1 & 0.50 & 23.0 & 9.0 & SNR/Pul & 2.02/2.38 & 1.6/2.43 & 5.8/2.00 & 090050020 & 1.36,1.36,1.0 & 9 & 4 & 1 \\ 
            XXVII & 6.0 & 3.1 & 0.50 & 23.0 & 9.0 & Pul/Pul & 2.02/2.38 & 1.6/2.43 & 5.8/2.00 & 025200010 & 0.7,0.7,1.0 & 9 & 4 & 1 \\ 
            XXVIII  & 3.0 & 2.67 & 0.40 & 22.0 & 3.0 & Pul/Pul & 2.08/2.41 & 1.6/2.43 & 5.8/2.00 & 050100020 & 1.0,1.0,1.0 & 9 & 4 & 1 \\ 
            XXIX & 3.0 & 2.67 & 0.40 & 22.0 & 3.0 & SNR/Pul & 2.08/2.41 & 1.6/2.43 & 5.8/2.00 & 090050020 & 1.36,1.36,1.0 & 9 & 4 & 1 \\
            XXX & 3.0 & 2.67 & 0.40 & 22.0 & 3.0 & Pul/Pul & 2.08/2.41 & 1.6/2.43 & 5.8/2.00 & 025200010 & 0.7,0.7,1.0 & 9 & 4 & 1 \\ 
            XXXI & 5.6 & 8.0 & 0.40 & 24.0 & 1.0 & Pul/Pul & 2.00/2.10 & 1.6/2.25 & 5.8/2.00 & 200050040 & 1.4,1.4,1.0 & 9 & 4 & 1 \\  
            XXXII  & 5.6 & 4.85 & 0.40 & 24.0 & 50.0 & Pul/Pul & 2.00/2.10 & 1.6/2.25 & 5.8/2.00 & 200050040 & 1.4,1.4,1.0 & 9 & 4 & 1 \\  
            XXXIII & 5.6 & 3.0 & 0.40 & 24.0 & 0 & Pul/Pul & 1.60/1.80 & 1.6/2.25 & 3.3/1.75 & 200050040 & 1.4,1.4,1.0 & 9 & 4 & 1 \\   
            XXXIV & 6.0 & 3.1 & 0.50 & 10.0 & 9 & Pul/Pul & 1.50/1.80 & 1.6/2.25 & 2.4/3.9 & 200030050 & 1.4,1.4,1.0 & 9 & 4 & 1 \\  
            XXXV & 6.0 & 2.4 & 0.50 & 0 & 9 & Pul/Pul & 1.50/1.80 & 1.6/2.25 & 2.1/3.9 & 200030050 & 1.4,1.4,1.0 & 9 & 4 & 1 \\  
            XXXVI  & 6.0 & 3.1 & 0.50 & 23.0 & 30 & Pul/Pul & 1.50/1.80 & 1.6/2.25 & 2.4/3.9 & 200030050 & 1.4,1.4,1.0 & 9 & 4 & 1 \\    
            XXXVII & 6.0 & 2.4 & 0.50 & 23.0 & 40 & Pul/Pul & 1.50/1.80 & 1.6/2.25 & 2.1/4.2 & 200030050 & 1.4,1.4,1.0 & 9 & 4 & 1 \\  
            XXXVIII & 4.0 & 5.0 & 0.33 & 32.7 & 50 & Pul/Pul & 1.89/2.47 & 1.6/2.42 & 7.52/2.0 & 090050020 & 1.36,1.36,1.0 & 9 & 5 & 1 \\
            XXXIX & 6.0 & 8.3 & 0.33 & 32.7 & 50 & Pul/SNR & 1.89/2.39 & 1.6/2.42 & 4.8/0.49 & 050100020 & 1.0,1.0,1.0 & 9 & 5 & 1 \\  
            XL & 4.0 & 8.0 & 0.33 & 32.7 & 50 & Pul/Pul & 1.89/2.47 & 1.6/2.42 & 7.52/2.0 & 090050020 & 1.36,1.36,1.0 & 9 & 5 & 1 \\   
            XLI  & 10.0 & 12.0 & 0.33 & 32.2 & 0 & Pul/SNR & 1.89/2.39 & 1.6/2.44 & 4.8/0.49 & 090050020 & 1.0,1.0,1.0 & 0 & 5 & 2 \\   
            XLII & 10.0 & 10.3 & 0.33 & 20.0 & 0 & Pul/SNR & 1.89/2.39 & 1.6/2.44 & 4.8/0.49 & 050100020 & 1.0,1.0,1.0 & 0 & 5 & 2 \\ 
            XLIII & 10.0 & 15.0 & 0.33 & 32.2 & 0 & Pul/SNR & 1.89/2.39 & 1.6/2.44 & 4.8/0.49 &  050100020 & 1.0,1.0,1.0 & 0 & 5 & 2 \\   
            XLIV & 6.0 & 7.1 & 0.33 & 31.9 & 0 & Pul/Pul & 1.89/2.39 & 1.6/2.44 & 4.9/0.50 & 050100020 & 0.8,0.8,1.0 & 0 & 5 & 2 \\
            XLV & 6.0 & 7.1 & 0.33 & 31.9 & 0 & Pul/Pul & 1.89/2.39 & 1.6/2.44 & 4.9/0.50 & 050100020 & 1.4,1.4,1.0 & 0 & 5 & 2 \\
            XLVI & 6.0 & 10.0 & 0.33 & 31.9 & 0 & Pul/Pul & 1.89/2.39 & 1.6/2.44 & 4.9/0.50 & 050100020 & 1.0,1.0,1.0 & 0 & 5 & 2 \\
            XLVII  & 6.0 & 8.3 & 0.33 & 32.7 & 50 & Pul/SNR & 1.80/2.39 & 1.6/2.26 & 2.0/4.8 & 200030050 & 1.4,1.4,1.0 & 9 & 5 & 1 \\ 
            XLVIII & 6.0 & 12.0 & 0.33 & 32.7 & 50 & Pul/SNR & 1.60/2.39 & 1.6/2.26 & 2.0/4.8 & 200030050 & 1.4,1.4,1.0 & 9 & 5 & 1 \\  
            XLIX & 10.0 & 7.0 & 0.33 & 32.2 & 0 & Pul/SNR & 1.70/2.39 & 1.6/2.44 & 2.0/1.27 & 050100020 & 1.0,1.0,1.0 & 0 & 5 & 2 \\ 
            L & 10.0 & 8.0 & 0.33 & 40.0 & 0 & Pul/SNR & 1.70/2.39 & 1.6/2.33 & 2.0/1.27 & 050100020 & 1.0,1.0,1.0 & 0 & 5 & 2 \\ 
            LI & 6.0 & 2.0 & 0.33 & 0 & 60 & Pul/SNR & 1.60/2.30 & 1.6/2.26 & 1.5/5.9 & 200030050 & 1.4,1.4,1.0 & 9 & 5 & 1 \\ 
            LII & 6.0 & 2.0 & 0.33 & 0 & 100 & Pul/SNR & 1.60/2.30 & 1.6/2.26 & 1.5/5.9 & 200030050 & 1.4,1.4,1.0 & 9 & 5 & 1 \\ 
            LIII & 10.0 & 8.0 & 0.33 & 32.2 & 100 & Pul/SNR & 1.40/1.80 & 1.4/2.30 & 1.3/3.33 & 200040050 & 1.4,1.4,1.0 & 0 & 5 & 2 \\  
            LIV & 6.0 & 4.0 & 0.33 & 50.0 & 0 & Pul/SNR & 1.89/2.30 & 1.4/2.10 & 2.4/2.20 & 050100020 & 1.0,1.0,1.0 & 0 & 5 & 2 \\
            LV & 6.0 & 12.0 & 0.33 & 0 & 0 & Pul/SNR & 1.89/2.10 & 1.4/2.10 & 2.0/1.10 & 050100020 & 1.0,1.0,1.0 & 0 & 5 & 2 \\
            LVI & 6.0 & 40.0 & 0.33 & 0 & 0 & Pul/SNR & 1.89/2.10 & 1.4/2.10 & 2.0/1.10 & 050100020 & 1.0,1.0,1.0 & 0 & 5 & 2 \\
            \hline
    \end{tabular}
    \caption{Galactic diffuse model parameters $z_{L}$ is in kpc, $D_{0}$ is in $\times 10^{28}$ cm$^{2}$/s, $v_{A}$ is in km/s, $dv_{c}/d|z|$ is in km/s/kpc. $N^{p}$ and $N^{e}$ are the cosmic-ray proton and electron differential flux $dN/dE$ normalizations at the galactocentric distance of 8.5 kpc. They are defined at 100 GeV and 34.5 GeV for the protons and electrons respectively and are in units of $\times 10^{-9}$ cm$^{-2}$s$^{-1}$sr$^{-1}$MeV$^{-1}$. For full details see Sec.~\ref{sec:TemplateModels}.}
    \label{tab:ModelsLong}
\end{table*}

\begin{table*}[t]
    \begin{tabular}{ccccccccccccccc}
    \hline
            Name & $z_{L}$ & $D_{0}$ & $\delta$ & $v_{A}$ & $dv_{c}/d|z|$ & $S^{N}/S^{e}$ & $\alpha_{1}^{p}$/$\alpha_{2}^{p}$ & $\alpha_{1}^{e}$/$\alpha_{2}^{e}$ & $N^{p}$/$N^{e}$ & B-field & ISRF & H2 & HI & HII \\
            \hline\hline
            LVII & 6.0 & 2.0 & 0.33 & 50.0 & 0 & Pul/SNR & 1.89/1.90 & 1.4/2.15 & 2.2/0.80 & 050100020 & 1.0,1.0,1.0 & 0 & 5 & 2 \\ 
            LVIII & 6.0 & 7.1 & 0.33 & 50.0 & 0 & Pul/SNR & 1.89/2.30 & 1.40/2.10 & 2.40/2.20 & 050100020 & 1.0,1.0,1.0 & 9 & 5 & 2 \\
            LIX & 6.0 & 7.1 & 0.33 & 50.0 & 0 & Pul/SNR & 1.89/2.30 & 1.40/2.10 & 2.40/2.20 & 050100020 & 1.0,1.0,1.0 & 10 & 5 & 2 \\
            LX & 6.0 & 7.1 & 0.33 & 50.0 & 0 & Pul/SNR & 1.89/2.30 & 1.40/2.10 & 2.40/2.20 & 050100020 & 1.0,1.0,1.0 & 2 & 5 & 2 \\
            LXI & 6.0 & 7.1 & 0.33 & 50.0 & 0 & Pul/SNR & 1.89/2.30 & 1.40/2.10 & 2.40/2.20 & 050100020 & 1.0,1.0,1.0 & 2 & 4 & 1 \\
            LXII & 6.0 & 7.1 & 0.33 & 50.0 & 0 & Pul/SNR & 1.89/2.30 & 1.40/2.10 & 2.40/2.20 & 050100020 & 1.0,1.0,1.0 & 10 & 1 & 1 \\
            LXIII & 6.0 & 7.1 & 0.33 & 50.0 & 0 & Pul/SNR & 1.89/2.30 & 1.40/2.10 & 2.40/2.20 & 050100020 & 1.0,1.0,1.0 & 9 & 1 & 1 \\
            LXIV & 6.0 & 7.1 & 0.33 & 50.0 & 0 & Pul/SNR & 1.89/2.30 & 1.40/2.10 & 2.40/2.20 & 050100020 & 1.0,1.0,1.0 & 2 & 1 & 1 \\
            LXV & 6.0 & 7.1 & 0.33 & 50.0 & 0 & Pul/SNR & 1.89/2.30 & 1.40/2.10 & 2.40/2.20 & 050100020 & 1.0,1.0,1.0 & 9 & 4 & 3 \\
            LXVI & 6.0 & 7.1 & 0.33 & 50.0 & 0 & Pul/SNR & 1.89/2.30 & 1.40/2.10 & 2.40/2.20 & 050100020 & 1.0,1.0,1.0 & 0 & 3 & 3 \\
            LXVII  & 6.0 & 7.1 & 0.33 & 50.0 & 0 & Pul/SNR & 1.89/2.30 & 1.40/2.10 & 2.40/2.20 & 050100020 & 1.0,1.0,1.0 & 10 & 2 & 3 \\
            LXVIII & 6.0 & 7.1 & 0.33 & 50.0 & 0 & Pul/SNR & 1.89/2.30 & 1.40/2.10 & 2.40/2.20 & 050100020 & 1.0,1.0,1.0 & 9 & 2 & 3 \\
            LXIX & 6.0 & 7.1 & 0.33 & 50.0 & 0 & Pul/SNR & 1.89/2.30 & 1.40/2.10 & 2.40/2.20 & 050100020 & 1.0,1.0,1.0 & 10 & 1 & 3 \\
            LXX & 6.0 & 7.1 & 0.33 & 50.0 & 0 & Pul/SNR & 1.89/2.30 & 1.40/2.10 & 2.40/2.20 & 050100020 & 1.0,1.0,1.0 & 9 & 1 & 3 \\
            LXXI & 6.0 & 7.1 & 0.33 & 50.0 & 0 & Pul/SNR & 1.89/2.30 & 1.40/2.10 & 2.40/2.20 & 050100020 & 1.0,1.0,1.0 & 2 & 1 & 3 \\
            LXXII & 5.6 & 4.85 & 0.40 & 40.0 & 0 & Pul/Pul & 1.50/1.90 & 1.5/2.25 & 2.40/1.55 & 200050040 & 1.4,1.4,1.0 & 10 & 4 & 1 \\   
            LXXIII & 5.6 & 4.85 & 0.40 & 40.0 & 0 & Pul/Pul & 1.50/1.90 & 1.5/2.25 & 2.40/1.55 & 200050040 & 1.4,1.4,1.0 & 0 & 4 & 1 \\ 
            LXXIV & 5.6 & 4.85 & 0.40 & 40.0 & 0 & Pul/Pul & 1.50/1.90 & 1.5/2.25 & 2.40/1.55 & 200050040 & 1.4,1.4,1.0 & 10 & 3 & 1 \\ 
            LXXV & 5.6 & 4.85 & 0.40 & 40.0 & 0 & Pul/Pul & 1.50/1.90 & 1.5/2.25 & 2.40/1.55 & 200050040 & 1.4,1.4,1.0 & 9 & 3 & 1 \\ 
            LXXVI & 5.6 & 4.85 & 0.40 & 40.0 & 0 & Pul/Pul & 1.50/1.90 & 1.5/2.25 & 2.40/1.55 & 200050040 & 1.4,1.4,1.0 & 2 & 3 & 1 \\ 
            LXXVII  & 5.6 & 4.85 & 0.40 & 40.0 & 0 & Pul/Pul & 1.50/1.90 & 1.5/2.25 & 2.40/1.55 & 200050040 & 1.4,1.4,1.0 & 2 & 2 & 1 \\ 
            LXXVIII & 5.6 & 4.85 & 0.40 & 40.0 & 0 & Pul/Pul & 1.50/1.90 & 1.5/2.25 & 2.40/1.55 & 200050040 & 1.4,1.4,1.0 & 10 & 4 & 3 \\ 
            LXXIX & 5.6 & 4.85 & 0.40 & 40.0 & 0 & Pul/Pul & 1.50/1.90 & 1.5/2.25 & 2.40/1.55 & 200050040 & 1.4,1.4,1.0 & 2 & 4 & 3 \\ 
            LXXX & 5.6 & 4.85 & 0.40 & 40.0 & 0 & Pul/Pul & 1.50/1.90 & 1.5/2.25 & 2.40/1.55 & 200050040 & 1.4,1.4,1.0 & 0 & 4 & 3 \\ 
            \hline
        \end{tabular}
    \caption{Continuing Tab.~\ref{tab:ModelsLong}. Galactic diffuse model parameters $z_{L}$ is in kpc, $D_{0}$ is in $\times 10^{28}$ cm$^{2}$/s, $v_{A}$ is in km/s, $dv_{c}/d|z|$ is in km/s/kpc. $N^{p}$ and $N^{e}$ are the cosmic-ray proton and electron differential flux $dN/dE$ normalizations at the galactocentric distance of 8.5 kpc. They are defined at 100 GeV and 34.5 GeV for the protons and electrons respectively and are in units of $\times 10^{-9}$ cm$^{-2}$s$^{-1}$sr$^{-1}$MeV$^{-1}$. For full details see Sec.~\ref{sec:TemplateModels}.}
    \label{tab:ModelsLong2}
\end{table*}

\end{appendix}  
                  
\bibliography{GCE_Templates}
\end{document}